\documentclass[manuscript,screen,nonacm]{acmart}

\AtBeginDocument{%
  \providecommand\BibTeX{{%
    \normalfont B\kern-0.5em{\scshape i\kern-0.25em b}\kern-0.8em\TeX}}}

\usepackage[ruled,vlined]{algorithm2e}
\usepackage{colortbl}
\usepackage{caption}
\usepackage{subcaption}
\usepackage{float}
\usepackage{color}
\usepackage[normalem]{ulem}
\useunder{\uline}{\ul}{}

\usepackage{physics}
\usepackage{amsmath}
\usepackage{tikz}
\usepackage{mathdots}
\usepackage{cancel}
\usepackage{color}
\usepackage{siunitx}
\usepackage{array}
\usepackage{multirow}
\usepackage{gensymb}
\usepackage{tabularx}
\usepackage{booktabs}
\usetikzlibrary{fadings}
\usetikzlibrary{patterns}
\usetikzlibrary{shadows.blur}
\usetikzlibrary{shapes}
\begin{document}

\title{Combating fake news by empowering fact-checked news spread via topology-based interventions}


\author{Ke Wang}
\affiliation{%
  \institution{The University of Sydney}
     \city{Sydney}
  \country{Australia}}
\email{kwan7498@uni.sydney.edu.au}

\author{Waheeb Yaqub}
\affiliation{%
  \institution{The University of Sydney}
  \country{Australia.}
    \institution{New York University}
    \country{USA.}
 }
\email{waheeb.faizmohammad@sydney.edu.au}

\author{Abdallah Lakhdari}
\affiliation{%
  \institution{The University of Sydney}
     \city{Sydney}
  \country{Australia}}
 \email{abdallah.lakhdari@sydney.edu.au}

\author{Basem Suleiman}
\affiliation{%
  \institution{The University of Sydney}
     \city{Sydney}
  \country{Australia}}
 \email{basem.suleiman@sydney.edu.au}

\makeatletter
\let\@authorsaddresses\@empty
\makeatother

\renewcommand{\shortauthors}{Wang and Yaqub, et al.}

\begin{abstract}
Rapid information diffusion and large-scaled information cascades can enable the undesired spread of false information. A small-scaled false information outbreak may potentially lead to an infodemic. We propose a novel information diffusion and intervention technique to combat the spread of false news. As false information is often spreading faster in a social network, the proposed diffusion methodology inhibits the spread of false news by proactively diffusing the fact-checked information. Our methodology mainly relies on defining the potential super-spreaders in a social network based on their centrality metrics. We run an extensive set of experiments on different networks to investigate the impact of centrality metrics on the performance of the proposed diffusion and intervention models. The obtained results demonstrate that empowering the diffusion of fact-checked news combats the spread of false news further and deeper in social networks. 
\end{abstract}




\maketitle


\section{Introduction}
\label{sec:Intro}

A large number of Internet users rely on social networks to view, search, and generate content. By design, online social networks (OSN) promote the cyber-participation culture. Social network users can easily and quickly view and generate content. \textit{Any} user can share \textit{any} sort of information by copying from \textit{any} source (e.g., book, magazine, websites), stating their ideas or opinions, or simply sharing already existing content. The high \textit{connectedness} among OSN users drastically accelerates the spread of information in these online platforms \cite{wardle2017information, yaqub2020credibility}. Indeed,  information diffusion in social networks has attained an unprecedented speed and scale. For instance, multiple nation-wide stories have been triggered by content diffusion on social media, such as the Arab spring 2010, the US election 2016, the Black Lives Matter riots 2020. It is challenging to check all the generated content in a fully decentralized environment like social networks. The freedom to generate \textit{unchecked} content might affect the \textit{veracity} of the shared content in social networks. Veracity is the degree to which the information is accurate and trusted. Detecting and combating the spread of information disorder remains one of the unsolvable issues in social networks.
Fake news persists as one of the most undesired phenomena on social media. For instance, the US presidential elections and the COVID19 vaccine have seen an unprecedented rise of fake news stories on social media. Several studies have recently reported that false news stories spread faster than the actual news \cite{vosoughi2018spread}. Surprisingly, this observation stands correct even when the fake news starts from a small number of a sparse and disconnected set of users ~\cite{schafer2020illusion, vosoughi2018spread,zeng2016rumours,gupta2013faking,shao2016hoaxy} and across different platforms \cite{zannettou2017theweb}. A growing body of research has been interested in \textit{analyzing} and \textit{understanding} the social and cognitive mechanisms leading to this exponential spread of false news, even their flagrant implausibility  ~\cite{zhou2019fake, sharma2019combating, Pennycook2019Fighting, yaqub2020credibility, micallef2020role, yaqub2020bias,pennycook2020cognitive,pennycook2021psyschology}. One analysis indicated that Facebook engagement (likes, comments, shares) towards the top 20 fake news stories about US presidential elections 2016 was far greater than the top 20 real news stories~\cite{silverman2016hyperpartisan}. To understand and analyze the fake news phenomenon in social networks, researchers are focusing on \textit{detecting}, and \textit{tracking} the spread of false news over a social network ~\cite{zhou2019fake, sharma2019combating, Zhou2019Network,karimi2019learning,Shi2016Discriminative,Gupta2014TweetCred, volkova2017separating, Wei2017Learning, Zhao2015Enquiring,Tschiatschek2018Fake, Sethi2017Crowdsourcing, mosleh2021perverse, yaqub2020credibility, micallef2020role, graves2016aThe}. This group of researchers attempts, mainly, to define the social structure of the affected communities by false news and understand the common traits among these communities.

In this paper, we proceed beyond understanding the spread mechanism over social media users. We design an intervention mechanism to insert and expand the verified true news (i.e., fact-checked news conflicting with the corresponding false news)\footnote{In the rest of the paper, we will be using true news, verified true news and fact-checked news interchangeably}. Typically, the verified true news and highly credible information do not spread at the same scale as false news~\cite{shao2018spread}. Furthermore, it has been shown that fact-checked news is less engaging compared to false news, especially if it is received after the false news~\cite{Pennycook2019Fighting}. At the same time, users might associate perceived credibility with high user engagement of news\cite{avram2020exposure}.  It is an uphill battle to overcome or combat the spread of false news which has high user engagement and spreadability compared to the corresponding verified true news that comes later with a lower user engagement and spreadability as shown in few examples in the appendix Table \ref{table:engagement}. The false news user engagements shown in Table \ref{table:engagement} have a median of 4461 and mean of 191316, which are respectively 2 and 70 times higher than that of the verified true news. As the p-value of the one-tailed Wilcoxon signed-rank test being $4.62*10^{-12}$, it suggests that under a 99\% confidence interval the verified true news has less user engagement compared with the corresponding false news. Motivated by the aforementioned observations, we aim to proactively protect social media users from the fast spread of false news. We empower the diffusion of true news by leveraging the \textit{centrality metrics} which, consequently, increase the engagement of users towards the verified true news. 

In this work, we only consider the topology of online social networks to investigate and combat the spread of false news. A social network is modeled by a set of nodes (social media users) connected by edges (relationships). Information cascade occurs when several nodes share the same content sequentially over time~\cite{leskovec2007patterns, watts2002simple}. This phenomenon can be critical, if false news gets sequentially re-shared multiple times, the originally shared information gets amplified and possibly reaches an exponential number of users. We intend to combat the false news by empowering the diffusion process of the true news. We claim that starting from a central node to diffuse the fact-checked news in a social network would significantly expand the coverage of true news, i.e., the number of nodes where verified true news has reached.  According to our best knowledge, none of the proposed methods have investigated the spread of true news through influential nodes to combat false information in social networks.

The contribution of this paper is \textit{a novel centrality-based information diffusion model for true news to combat conflicting false news} in large social networks. Our main goal is to propose an effective and efficient method to share fact-checked news based on central nodes. The proposed method considers the conflicting nature of the true and false news, competing against each other over the topology of the online social network. It is worth mentioning that the adopted diffusion method would strictly obey the three conventional assumptions for any topology-based information diffusion model \cite{najar2012predicting}; (i) no external source of information in the network, apart from the information originator, (ii) information can only diffuse on the edges of the network, (iii) only one piece of the information is diffusing in the model at a time. We first show the impact of the network topology on the diffusion process. 
We then shed light on the impact of central nodes on the information diffusion process. Presumably, the network-based centrality metrics reflect the social aspect that impacts the users' engagement, thereby determine the effectiveness and efficiency of the information diffusion process \cite{bakshy2012role}. For instance, an average person with 300 followers \cite{kempe2003maximizing} would, at best,  have 300 of their followers will share their piece of news (i.e.engage with news). In contrast, a central person with 300K followers would reach much higher user engagement on average\cite{bakshy2012role,harrigan2012influentials,shoroye2015exploring}. Finally, we assess the false news combating effectiveness by measuring the verified true news spread resulting from the proposed diffusion method against the spread of the false news. We define the following research questions to lead our investigation:
\begin{itemize}
    \item What is the effect of the network structure and sparsity on the efficiency of the information diffusion process? 
    
    \item What will be the optimal centrality measurements to start a diffusion process under different network structures?
    
    \item How to measure the impact of the verified true news spread on the network against the already spread false news?

\end{itemize}

\begin{table*}
\caption{Fact-checking process}

\tikzset{every picture/.style={line width=0.75pt}} 

\begin{tikzpicture}[x=0.75pt,y=0.75pt,yscale=-1,xscale=1]

\draw   (115,75) -- (503,75) -- (503,122) -- (115,122) -- cycle ;
\draw    (71,98) -- (110,98) ;
\draw [shift={(112,98)}, rotate = 180] [color={rgb, 255:red, 0; green, 0; blue, 0 }  ][line width=0.75]    (10.93,-3.29) .. controls (6.95,-1.4) and (3.31,-0.3) .. (0,0) .. controls (3.31,0.3) and (6.95,1.4) .. (10.93,3.29)   ;
\draw   (2,75) -- (67,75) -- (67,122) -- (2,122) -- cycle ;
\draw   (549,74) -- (659,74) -- (659,121) -- (549,121) -- cycle ;
\draw   (115,122) -- (231,122) -- (231,162) -- (115,162) -- cycle ;
\draw   (237,122) -- (371,122) -- (371,162) -- (237,162) -- cycle ;
\draw   (378,122) -- (503,122) -- (503,162) -- (378,162) -- cycle ;
\draw    (115,162) -- (115,179) ;
\draw    (133,162) -- (133,179) ;
\draw    (150,162) -- (150,179) ;
\draw    (166,162) -- (166,179) ;
\draw    (281,162) -- (281,179) ;
\draw    (298,162) -- (298,179) ;
\draw    (315,162) -- (315,179) ;
\draw    (436,162) -- (436,179) ;
\draw    (453,162) -- (453,179) ;
\draw    (471,162) -- (471,179) ;
\draw    (503,162) -- (503,179) ;
\draw    (246,162) -- (246,179) ;
\draw    (264,162) -- (264,179) ;
\draw    (331,162) -- (331,179) ;
\draw    (349,162) -- (349,179) ;
\draw    (487,162) -- (487,179) ;
\draw    (503,98) -- (542,98) ;
\draw [shift={(544,98)}, rotate = 180] [color={rgb, 255:red, 0; green, 0; blue, 0 }  ][line width=0.75]    (10.93,-3.29) .. controls (6.95,-1.4) and (3.31,-0.3) .. (0,0) .. controls (3.31,0.3) and (6.95,1.4) .. (10.93,3.29)   ;
\draw    (182,162) -- (182,179) ;
\draw    (199,162) -- (199,179) ;
\draw    (215,162) -- (215,179) ;
\draw    (231,162) -- (231,179) ;
\draw    (366,162) -- (366,179) ;
\draw    (383,162) -- (383,179) ;
\draw    (399,162) -- (399,179) ;
\draw    (418,162) -- (418,179) ;

\draw (249,90) node [anchor=north west][inner sep=0.75pt]   [align=left] {Fact-Checking Methods};
\draw (5,90) node [anchor=north west][inner sep=0.05pt]   [align=left] {$\displaystyle False News_{i}$};
\draw (506,75) node [anchor=north west][inner sep=0.75pt]   [align=left] {Fake};
\draw (550,89) node [anchor=north west][inner sep=0.75pt]   [align=left] {$\displaystyle Fact-check of News_{i}$};
\draw (3,124) node [anchor=north west][inner sep=0.75pt]   [align=left] {{\scriptsize \textcolor[rgb]{0.82,0.01,0.11}{High user engagement}}\\{\scriptsize Share: 5491}\\{\scriptsize Like: 22973}\\{\scriptsize Comments: 9500}};
\draw (551,124) node [anchor=north west][inner sep=0.75pt]   [align=left] {{\scriptsize \textcolor[rgb]{0.82,0.01,0.11}{Low user engagement}}\\{\scriptsize Share: 44}\\{\scriptsize Like: 389}\\{\scriptsize Comments: 170}};
\draw (150,133) node [anchor=north west][inner sep=0.75pt]   [align=left] {Manual};
\draw (270,133) node [anchor=north west][inner sep=0.75pt]   [align=left] {Crowdsourced};
\draw (414,134) node [anchor=north west][inner sep=0.75pt]   [align=left] {Automated};
\draw (106,176) node [anchor=north west][inner sep=0.75pt]   [align=left] {\cite{vlachos2014fact,paul2019principles,moran2018deciding,beers2020examining,nguyen2018therise,ccomlekcci2021combating,mantzarlis2018fact,coddinton2014fact}};
\draw (123,176) node [anchor=north west][inner sep=0.75pt]   [align=left] {};
\draw (140,176) node [anchor=north west][inner sep=0.75pt]   [align=left] {};
\draw (156,176) node [anchor=north west][inner sep=0.75pt]   [align=left] {};
\draw (272,176) node [anchor=north west][inner sep=0.75pt]   [align=left] {};
\draw (288,176) node [anchor=north west][inner sep=0.75pt]   [align=left] {};
\draw (305,176) node [anchor=north west][inner sep=0.75pt]   [align=left] {};
\draw (427,176) node [anchor=north west][inner sep=0.75pt]   [align=left] {};
\draw (444,176) node [anchor=north west][inner sep=0.75pt]   [align=left] {};
\draw (461,176) node [anchor=north west][inner sep=0.75pt]   [align=left] {};
\draw (494,176) node [anchor=north west][inner sep=0.75pt]   [align=left] {};
\draw (237,176) node [anchor=north west][inner sep=0.75pt]   [align=left] {\cite{haque2020combating,pinto2019towards,zubiaga2014tweet,jooyeon2018leveraging,shan2018lingusitic,roitero2020crowd,kriplean2014integrating,cerone2020watch}};
\draw (255,176) node [anchor=north west][inner sep=0.75pt]   [align=left] {};
\draw (322,176) node [anchor=north west][inner sep=0.75pt]   [align=left] {};
\draw (340,176) node [anchor=north west][inner sep=0.75pt]   [align=left] {};
\draw (478,176) node [anchor=north west][inner sep=0.75pt]   [align=left] {};
\draw (172,176) node [anchor=north west][inner sep=0.75pt]   [align=left] {};
\draw (189,176) node [anchor=north west][inner sep=0.75pt]   [align=left] {};
\draw (205,176) node [anchor=north west][inner sep=0.75pt]   [align=left] {};
\draw (221,176) node [anchor=north west][inner sep=0.75pt]   [align=left] {};
\draw (356,176) node [anchor=north west][inner sep=0.75pt]   [align=left] {};
\draw (373,176) node [anchor=north west][inner sep=0.75pt]   [align=left] {\cite{naeemul2017claimbuster,thorne2018fever,hanselowski2017framework,wu2018early,karadzhov2017fully, nguyen2018believe,rony2018baitbuster,natali2017csi}};
\draw (390,176) node [anchor=north west][inner sep=0.75pt]   [align=left] {};
\draw (408,176) node [anchor=north west][inner sep=0.75pt]   [align=left] {};

\end{tikzpicture}

\end{table*}



\section{Related Work}
\label{sec:related}
The background of our work comes from three different
areas, i.e., spread of misinformation, information diffusion, and combating misinformation. We describe the
related work to our research in each of these domains.
\subsection{The spread of misinformation}
The spread of misinformation aims to influence the public and decision-makers in various domains, including politics and economics. Recently, in the light of the outbreak of COVID-19, several research papers have studied the propagation of misinformation related to the coronavirus on social media ~\cite{cinelli2020covid19, kouzy2020coronavirus, gallotti2020assessing, singh2020first, yang2020prevalence, shahi2020exploratory}. It has been proven that false information spreads significantly faster and more broadly than truthful information in various categories ~\cite{vosoughi2018spread}. It is challenging to detect potential initiators and spreaders of false information. Usually, social media users who attempt to spread false information often have significantly fewer activities, followers, and followees~\cite{vosoughi2018spread}. In the case of the initial understanding of social media conversations about COVID-19, evidence also shows that information from low-quality sources transmits faster than information from high-quality sources ~\cite{singh2020first}. One alternative explanation emerges from the Bayesian decision theory and the information theory. Novelty can easily attract the attention of human beings ~\cite{itti2008bayesian} and encourage information sharing easily  \cite{aral2011diversity} when false information are usually containing novel and surprising contents.

Cinelli has proposed an information diffusion analysis about COVID-19 with massive data analysis, introducing an exploratory study into the COVID-19 misinformation diffusion to get an early insight~\cite{cinelli2020covid19}. One important conclusion found is that information diffusion from both reliable and questionable source do not present with different spreading patterns ~\cite{cinelli2020covid19}, which convincingly support the assumptions in a universal independent information diffusion model for both true and false information. Besides, the fraction of re-posting in the first propagation layer of false information is found to be significantly smaller than that of the true information. In contrast, fractions in the subsequent propagation layers of false information are more significant ~\cite{zhao2020false}. Thus, the number of layers in false information is typically larger than true information, while the creator usually has a smaller degree in the information diffusion process ~\cite{zhao2020false}.

Additional studies applied to false information diffusion provided some common identifiable characteristics of false information. Information diffusion analysis plays a vital role in the misinformation detection process. Kouzy provided an early quantification of the misinformation spreading and analyzed some of the key characteristics associated with the diffusion on a small subset of Twitter data manually annotated ~\cite{kouzy2020coronavirus}. Gallotti et al. developed an Infodemic risk index to capture the exposure of misinformation across countries and provided findings of competing misinformation and reliable information ~\cite{gallotti2020assessing}. Yang proposed estimation of the prevalence of links to misinformation on Twitter during the COVID-19 outbreak and the role of bots in the propagation process ~\cite{yang2020prevalence}. Shahi conducted an exploratory analysis into the accounts involved, the propagation, and the content for COVID-19 misinformation ~\cite{shahi2020exploratory}. However, the observed and summarized findings would be limited to a specific situation of misinformation diffusion, while no modeling is involved for further and deeper analysis about the diffusion mechanism. Nevertheless, the characteristics findings provide great inspiration to parameterize the simulation of false information diffusion.


\subsection{Information diffusion}
Several models have been proposed for information diffusion in a social network. Typically, a diffusion model is associated with a directed graph, a set of nodes initially carrying the information, and a mechanism for information propagation. \textbf{Independent Cascade} model is one of the most widely studied models for influence analysis on social networks. Nodes are distinguished as active or inactive. In the independent cascade model, the active nodes will recursively infect their neighbors with a certain probability. The influence maximization in the independent cascade model is a well-established research problem in social networks~\cite{kempe2003maximizing, d2016influence, wu2015efficient, sheldon2010maximizing, bogunovic2012robust,kimura2009blocking}. The influence expectation is estimated through multiple simulation processes. This simulation makes the maximization strategy update computationally expensive ~\cite{kempe2003maximizing}. It is worth mentioning that the optimization techniques to solve the influence maximization problem are mostly domain-dependent. Differently put, the application of the maximization strategy can only be limited to a particular situation or domain~\cite{d2016influence}~\cite{bogunovic2012robust,kimura2009blocking}.

In the \textbf{linear threshold model} model, a random threshold in the range [0,1] is determined for each node based on the weighted sum of its infected neighbors. As the determination of infection on a network is a random process without any predefined hyper-parameters, the modeling itself might differ from the actual information diffusion probability. \textbf{Susceptible-Infected-Recovered} (SIR) model is the most fundamental epidemic model. Nodes are dynamically converted from one of the three stages with a certain probability and finally reach an equilibrium. However, the stage conversion in the SIR model does not reflect the information disorder in a combating way, which could be less applicable in modeling an intervention. 

Several studies have been analyzing and predicting the information cascades differently. Mostly, they claim that a prediction model may be achieved by observing the information cascade for a fixed given time ~\cite{kupavskii2012prediction, ma2013predicting, tsur2012what}.  Cheng et al. have proposed a novel methodology for predicting the growth of a cascade ~\cite{cheng2014cascades}. They consider the cascade as time-series data, which should be tracked over time. Using the cascade's current information, they propose a sequential prediction, which does not suffer from skew biases ~\cite{cheng2014cascades}. Three main advantages have been achieved in Cheng's research~\cite{cheng2014cascades}. Firstly, the prediction problem becomes balanced in classes rather than highly unbalanced in previous modeling methods. Secondly, variation in the cascade's predictability over the range of its growth from small to large now becomes analyzable. Lastly, the prediction is more closely connected to real-world tasks, which often need the management of viral content.

\subsection{Combating misinformation}
\textit{Sznajd model} is the most fundamental for the evolution of combating opinion ~\cite{SZNAJD2000opinion}. In the evolution process, nodes of a pair having the same opinion will make their nearest neighbors agree. On the contrary, nodes of a pair having different opinions would make their nearest neighbors disagree. The Sznajd model has been firstly applied in the study of combating information disorder ~\cite{Bernardes2021Information}. However, due to the simplicity of the diffusion model, the relatively influential strength of the two information is not modeled. Therefore, the Sznajd model cannot explain the origins of very complicated phenomena observed in complex systems ~\cite{SZNAJD2000opinion}. 

\textit{Self-excited Hawkes process} states that the rate of spread in a homogeneous graph depends on two factors, namely, an external source and self-excitation~\cite{hawkes1974cluster}. \textbf{Time-Dependent Hawkes process with false information correction}  extends the state of art models for popularity dynamics prediction~\cite{kobayashi2016tideh}. The proposed model is a two-stage process. The first stage consists of cascading the original information as an ordinary information item. 
The second stage is another cascade that discloses and rectifies the falsehood of the original information~\cite{murayama2020modeling}. The text-mining results indicate that some users would realize the falsehood of the misinformation~\cite{murayama2020modeling}, which provides sight evidence of competing for true and false information. Inspired by these studies, the proposed diffusion algorithm in this project considers the cascade prediction as a sequential task to analyze the spread of misinformation and devise a strategy to combat it.
%


\section{Methodology}
\label{sec:method}

\begin{figure*}[!htb]
    \centering
    \includegraphics[width=13cm]{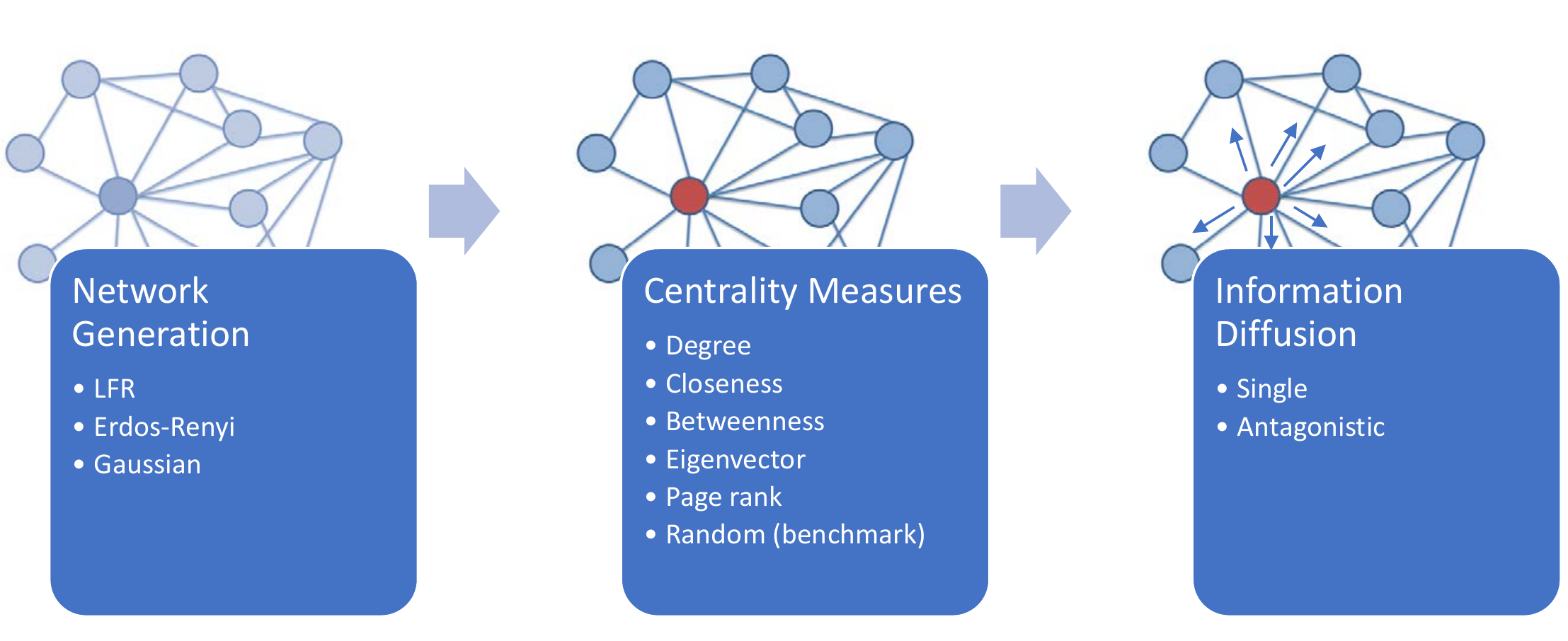}
     \caption{High-level overview of our methodology}
    \label{fig:methodology}
\end{figure*}

In this section, we present our investigation methodology to define an efficient and effective method to spread the true news. We first present the adopted single and combating topology-based information diffusion models, We then examine the effect of centrality metrics on the diffusion models. We conduct our investigation on different social network models, namely, random networks (i.e., \textit{Erdos-Renyi} and \textit{Gaussian} networks) and complex networks (i.e., large scale networks, \textit{Lancichinetti–Fortunato–Radicchi benchmark}). In our current work, we represent an anline social network by a undirected graph, nodes represent the users. The edges represent the connections between users as shown in Figure \ref{fig:methodology} 

\subsection{Single information diffusion model}
\label{sec:SingleModel}
We model the information diffusion as an independent transmitting process across a network. The parameters involved in our model are shown in Table \ref{table:data}. The proposed model first applies the BFS algorithm to define the diffusion \textit{layers}, where the source node is the information creator $IC$ and the sink is the group of nodes that have the longest distance from the $IC$. Our model is based on a number of assumptions. First, the \textit{information transmission process} in a social network occurs only on the edges. Furthermore, the \textit{information transmission process} on each edge is considered independent from the other edges. Second, we use a \textit{universal transmission probability} for any two directly connected nodes. It is the statistical general probability of a node (user) to transmit a piece of information to a connected node (a friend). Third, when updating the $P_I$ of $node_i$ from a source node, an edge of $node_i$ within its layer is considered to be effective for information transmission only when the connected neighbor node links to the same source node in the previous layer. That is because a friend who is forwarding information from a known source would especially increase the information reliability.



\begin{table*}
  \caption{Parameter description of the proposed single information diffusion model}
  \label{table:data}
  \begin{tabular}{lll}
    \toprule
      Symbol & Parameter & Description \\
    \midrule
    $P_I$ & Information gain probability	& The probability of one node (user) to believe that a piece of information is true. \\
    $P$ & Transmission probability & The diffusion probability on a edge.\\
    $IC$ & Information creator	& The initial creator of a piece of information with $P_I$ = 1.\\
    $T$ & Probability threshold &  Node is defined as believer if its $P_I  \geq T$. \\
  \bottomrule
\end{tabular}
\end{table*}



Algorithm ~\ref{alg:update_PI} shows the logic for updating the probability of information gain $P_I$ of a current node from a source node based on the definition of number of effective edges discussed above. The definition of the number of effective edges for $node_u$ from source $node_v$ is the number of $node_i$ in the same layer of $node_u$, where $node_u$, $node_v$, $node_i$ are a closed triplet.

Our proposed information transmission process consists of \textit{n} iterations as shown in Algorithm ~\ref{alg:info_tran}. In each iteration, the $P_I$ of the nodes within one layer are updated from the previous layer simultaneously and independently. Furthermore, we introduce a \textit{probability threshold} $T$ to determine final status of a node; i.e., whether a node (user) accepts in the transmitted information or not.  



\begin{algorithm}
\SetAlgoLined
 initialization \;
$N_{edge}$ = compute effective number of edges using its definition\;
 
$p = P_I(node_{cur})*P$ \;
 \For{n = 1 to $N_{edge}$}{
$p = p + P_I(node_{cur})*P^n*(1-P)^{N_{edge}+1-n}*C(N_{edge}, n)*(1-(1-P)^n)$
}
 \KwRet{$p$}
 \caption{$P_I$ update from one source node}
 \label{alg:update_PI}
\end{algorithm}

\begin{algorithm}
\SetAlgoLined
 initialization\;
   $P_I = 1 \ and \ \overline{P_I} = 0$ for IC\;
   $P_I = 0 \ and \ \overline{P_I} = 1$ for nodes not in IC\;
   $T$ \; 
   rearrange the network into layered form\;
   L = 0\;
   
 \For{iteration = 1 to n}{
 current layer = list of nodes in layer L\;
 next layer = list of nodes in layer L+1\;
\For {$node_{next}$ in the next layer}{
\For{$node_{cur}$ in the current layer connecting to $node_{next}$}{

$\hat p$ = compute using Algorithm ~\ref{alg:update_PI} \;
$\overline{P_I(node_{next})} = \overline{P_I(node_{next})} * (1-\hat p)$\;
$P_I(node_{next}) = 1 - \overline{P_I(node_{next})} $\;
}
}
L = L + 1\;
}
\For {$node$ in the network}{
\uIf{$P_I(node) \geq T$}{Node is labelled "infected" //believing in the piece of information\;}
\Else{Node is labelled "susceptible" //not believing in the piece of information\;}

}
 
 \caption{Information transmission algorithm}
 \label{alg:info_tran}
\end{algorithm}

\subsection{Combating information diffusion model}



\begin{table}
  \caption{Parameter description of combating information diffusion model}
  \label{table:data2}
  \begin{tabular}{ll}
    \toprule
   Symbol & Parameter\\
    \midrule
    $P_{IT}$\ \& \ $P_{IF}$ & Information gain probability for true \&  false \\ & information of a node\\
    $P_T$ \ \& \ $P_F$ & Transmission probability for true \&  false \\ &  information  of all the edges\\
    $IC_T$  \ \& \ $IC_F$ & Information creator for true \&  false  information\\
    $T_D$ & Decisive probability threshold \\
    $T_C$ & Comparative probability threshold \\
    \bottomrule
  \end{tabular}
\end{table}


We model the spread of two pieces of information in a social network. Particularly, we model the diffusion of two conflicting pieces of information, representing false news and the corresponding true news following the similar independent probability update process presented in section ~\ref{sec:SingleModel}. Table ~\ref{table:data2} describes the used parameters in our proposed combating information diffusion model. Particularly, there are two sets of parameters for each of the information gain probability, information transmission probability, and information creator used in the diffusion model for false and true news. Consequently, there are two diffusion processes; one for true news and the other for false news. These two diffusion processes are simultaneously and independently permeating all the nodes on the social network. However, the false news diffusion process always has one iteration ahead of the true news process.

When the two diffusion processes encounter each other, the \textit{decisive probability threshold} determines the availability of the nodes for true information diffusion. For nodes whose $P_{IF}$ reach the decisive probability threshold, they are no longer available for true information diffusion. In addition, for the final determination of the status of a node accepting in either piece of information, the comparative probability threshold is the critical point for the difference of two values of accumulated information gain probabilities. The pseudo-code for belief determination is presented in Algorithm ~\ref{alg:determination}.


\begin{algorithm}
\SetAlgoLined
 initialization\;
 \uIf{$P_{IF} - P_{IT} \geq T_C$}{
   Node is labelled "infected" //believing in false information\;}
 \uElseIf{$P_{IF} \geq P_{IT}$}{
   Node is labelled "susceptible" //potentially believing in false information\;}
 \Else{ 
   Node is labelled “protected” // believing in true information.}
 
 \caption{Node status determination algorithm for combating true and false information }
 \label{alg:determination}
\end{algorithm}

\begin{figure}[!htb]
    \centering
    \includegraphics[width=4.5cm]{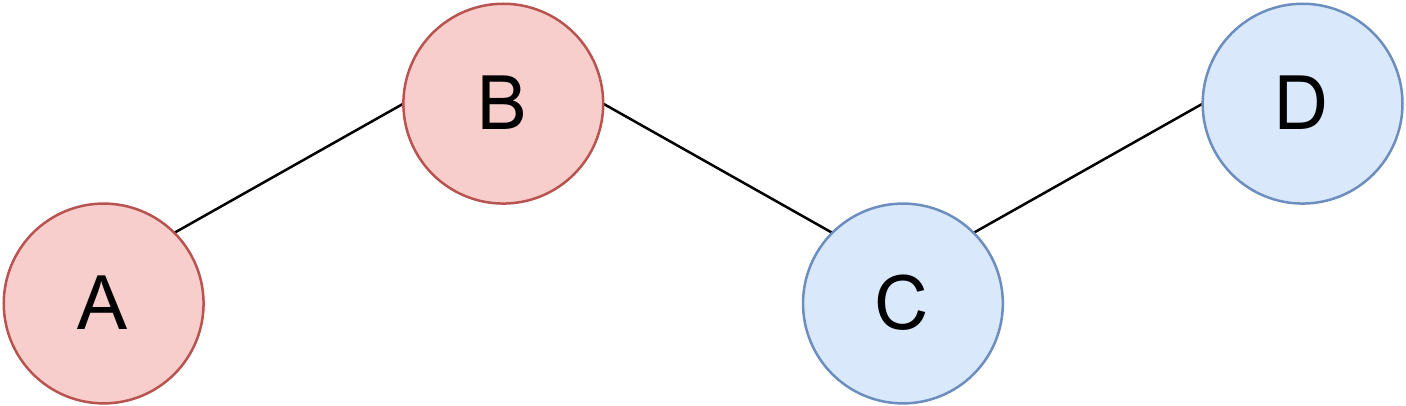}
     \caption{Network example for combating information diffusion model}
    \label{fig:Figure45}
\end{figure}

For instance, consider a network structure as shown in Figure \ref{fig:Figure45} with the set of parameters $IC_F$ = node A, $IC_T$ = node D, $P_F$ = 0.5, $P_T$ = 0.4, $T_D$ = 0.5, $T_C$ = 0.1. Then, the information diffusion progress works as follows. First, the $P_{IF}$ for node A is initialized as 1 and for node B, C, D are initialized as 0. The $P_{IT}$ for node D is initialized as 1 and for node A, B, C are initialized as 0. In the first iteration, the $P_{IF}$ of node B will be updated to $1*0.5 = 0.5$. After that, in the second iteration, the $P_{IF}$ of node C will be updated to $0.5*0.5 = 0.25$, which is less than the $T_D$. Then, the $P_{IT}$ of node C will be updated to $1*0.4 = 0.4$. In the third iteration, the $P_{IF}$ of node D will be updated to $0.25*0.5 = 0.125$. As $P_{IF}$ of node B reaches to the $T_D$, the $P_{IT}$ of node B will be kept to 0, which makes node A and node B both unavailable for true information diffusion. As a result, node A and B are labelled as "false" (red color). Node C and D are labelled as "true" (green color).

\section{Experimental Results}
\label{sec:Eval}
In this section, we present our systematic investigation of our proposed diffusion model. Our goal is to highlight the impact of different centrality metrics on the performance of our proposed diffusion and intervention models to combat false news in online social networks. We investigate the combating methodology on two different random network models (namely, \textit{Erdos-Renyi} and \textit{Gaussian} partition graph) and a large scale complex network model. We generate the complex network based on \textit{LFR benchmark}. We used Wilcoxon signed-rank test to compare the effectiveness of centrality-based strategies information diffusion with the random selection benchmark ~\cite{wilcoxon1992individual,rey2011international}.      
\subsection{Network generation algorithms}


For the data analysis, we use the \textit{Erdos-Renyi random}, \textit{Gaussian random partition} and \textit{LFR benchmark} graphs to generate random networks’ structure. The Erdos-Renyi random graph generates the network with each pair of nodes having a probability of $p$ to be connected. The advantages of the ER method are the simplicity and efficiency of the process. The Gaussian random partition graph creates $k$ partitions with a size drawn from a normal distribution. Unlike the ER method, the probability of an edge that exists is not universal. This can lead to a more complicated graph generation. The LFR benchmark graph has pre-known communities. It accounts for heterogeneity in the node's degree of distribution and community size distribution. Community structure is one of the most important features of real social networks \cite{fortunato2010community,lakhdari2016link}, where the LFR benchmark network reflects the real properties of nodes and communities ~\cite{lancichinetti2008benchmark}.




\subsection{Topology-based single information diffusion}

\begin{figure*}
    \centering
    \includegraphics[width=15cm]{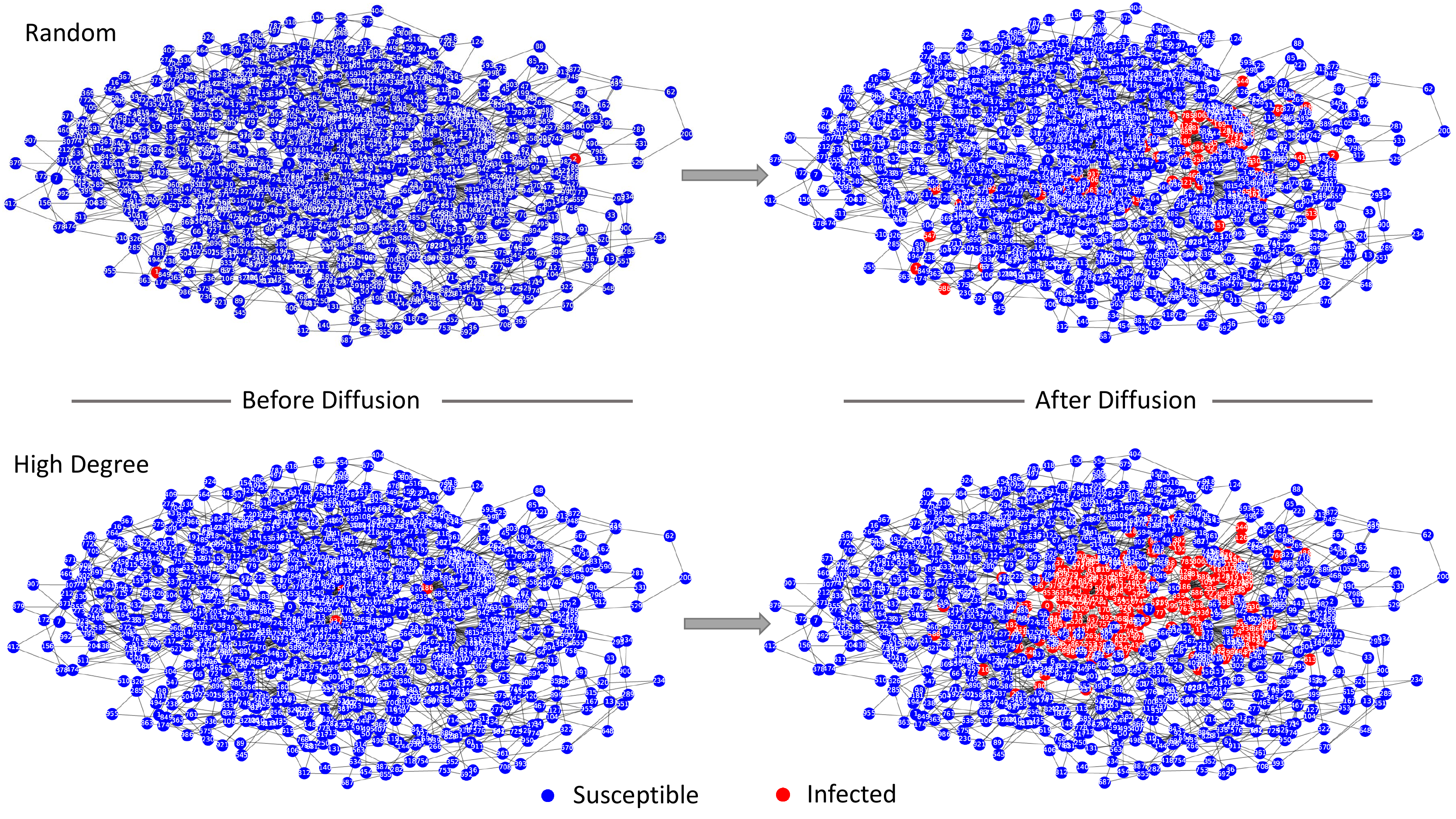}
     \caption{Information diffusion of the simulated social network}
    \label{fig:Figure1234}
\end{figure*}

As explained in the methodology, we generate information diffusion with randomly selected creator, and the information diffusion of creator with highest-centrality degree (see Figure \ref{fig:Figure1234}). A key observation is that the centrality-based strategy has a deeper information influence on the social network. The nodes with higher centrality are expected to have a greater influential impact for information diffusion. Further discussion about the generalization of this observation is presented in section ~\ref{sec:P_info_diff} and ~\ref{sec:network_info_diff}.

\subsubsection{\textbf{Impact of transmission probability on single information diffusion}}
\label{sec:P_info_diff}

\begin{figure*}[!htbp]
\begin{subfigure}{.45\textwidth}
  \centering
  \includegraphics[width=.9\linewidth]{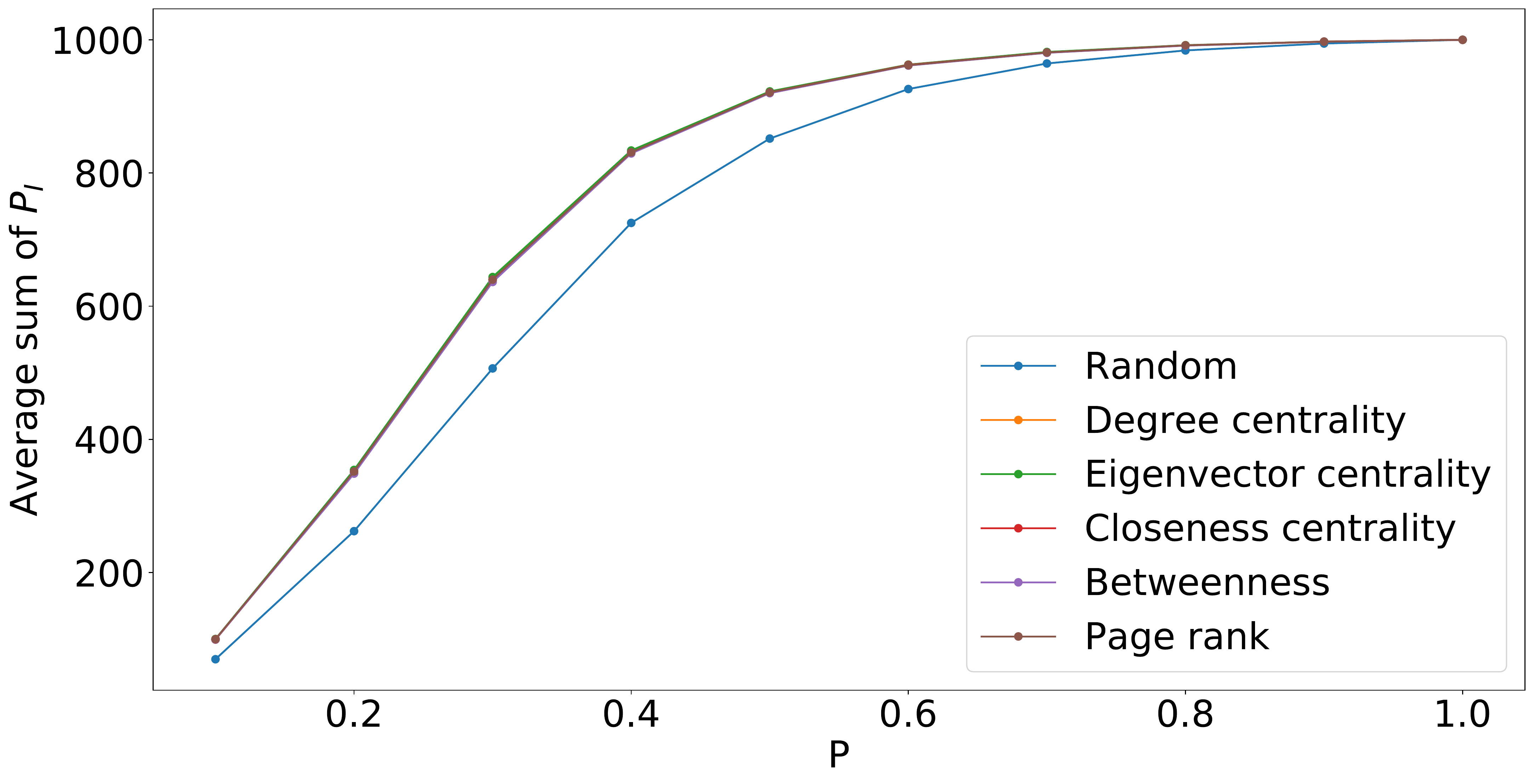}  
  \caption{ }
  \label{fig5:sub-first}
\end{subfigure}
\begin{subfigure}{.45\textwidth}
  \centering
  \includegraphics[width=.9\linewidth]{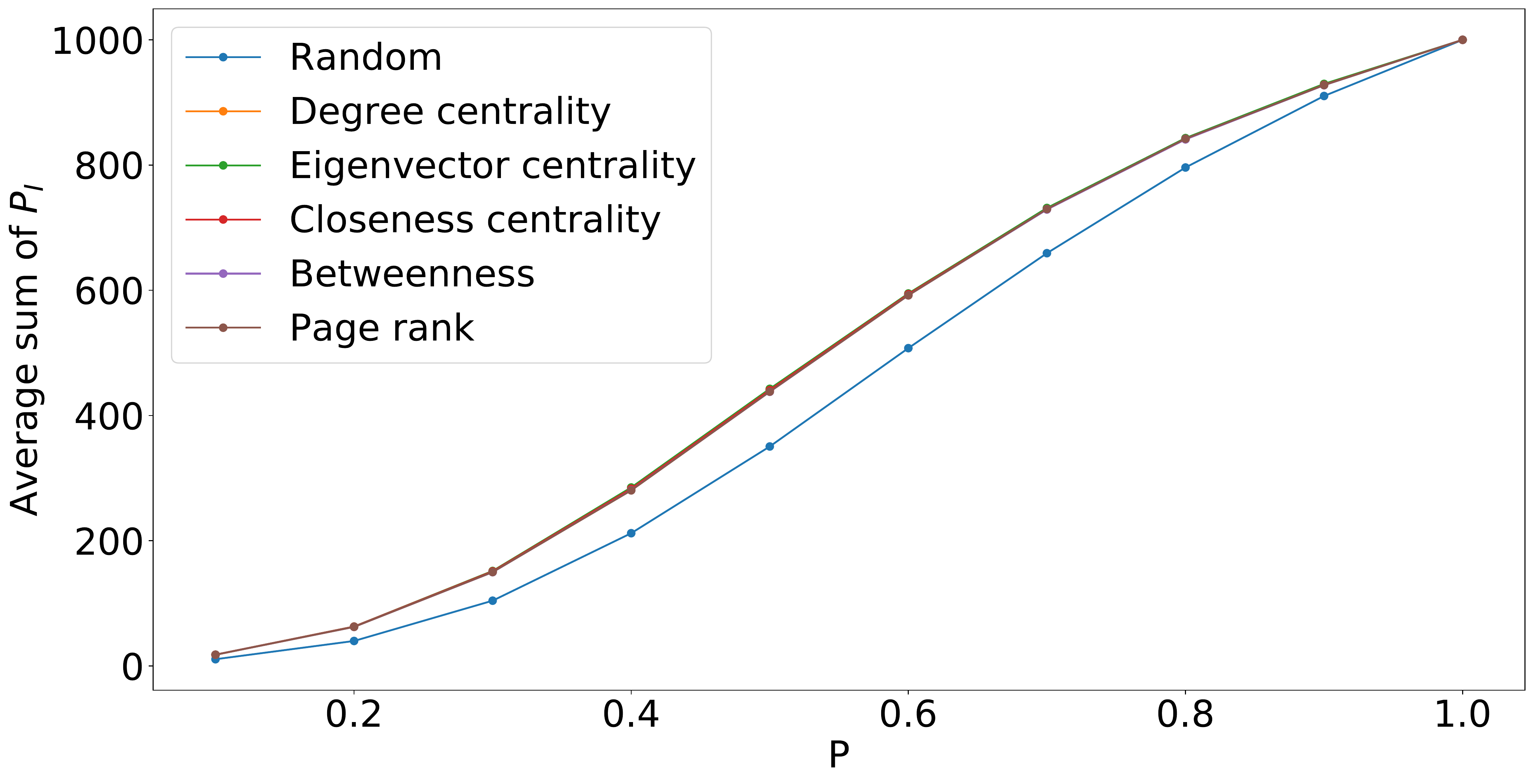}  
  \caption{ }
  \label{fig5:sub-second}
\end{subfigure}


\begin{subfigure}{.45\textwidth}
  \centering
  \includegraphics[width=.9\linewidth]{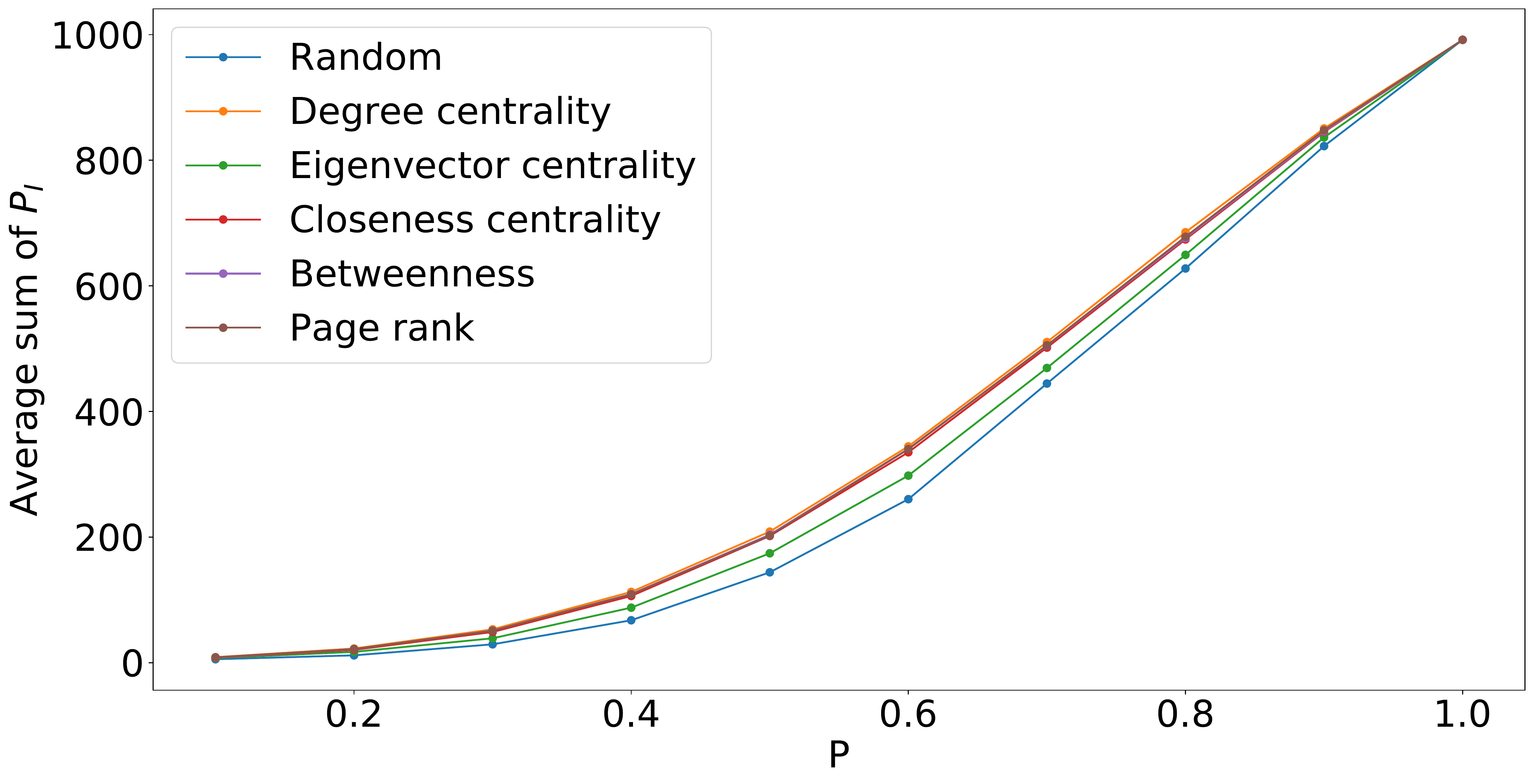}  
  \caption{ }
  \label{fig5:sub-third}
\end{subfigure}
\begin{subfigure}{.45\textwidth}
  \centering
  \includegraphics[width=.9\linewidth]{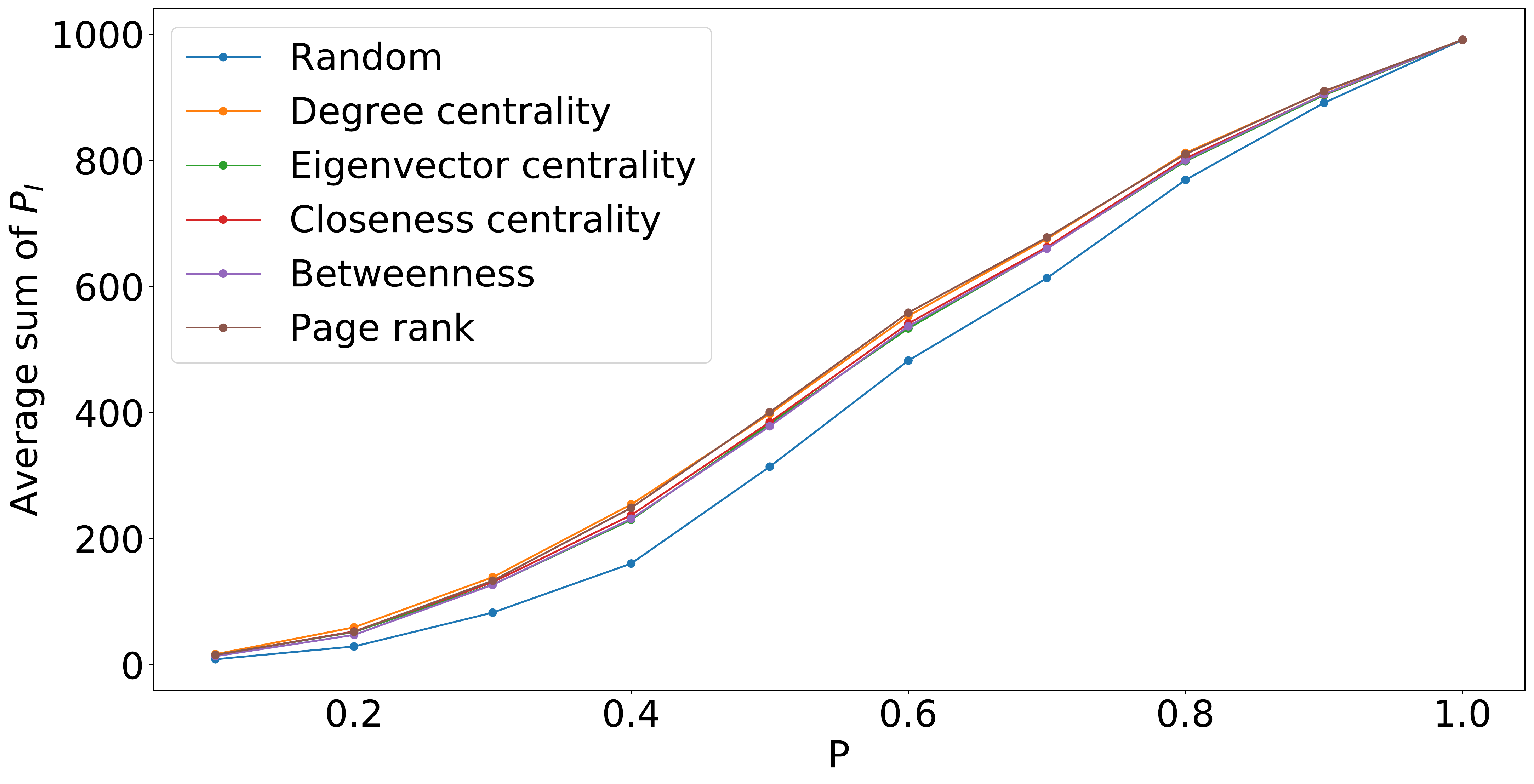}  
  \caption{ }
  \label{fig5:sub-fourth}
\end{subfigure}

\caption{Sum of $P_I$ versus $P$ on (a) 50 dense ER random graph; (b) 50 sparse ER random graph; (c) 50 Gaussian random graph with similar community size; (d) 50 Gaussian random graph with varying community size.}
\label{fig:fig5}
\end{figure*}

One of the key aspects under investigation of the topology-based information diffusion is the transmission probability ($P$), which refers to the infection power of one piece of information. To further analyze that, we compare all the different centrality measures (presented in section ~\ref{sec:method}) with the random selection while changing $P$ from 0.1 to 0.9. The results of these comparisons are illustrated in Figure \ref{fig:fig5}. We observe that the Creators with higher centrality can consistently diffuse information on a social network faster, deeper, and stronger compared to the random selecting strategy. When the infecting ability of a piece of information increases, the relative advantage of a centrality-based strategy decreases. We can conclude that a different creator would be less effective when the transmission probability is very high. This is because a larger transmission probability will result in a higher expectation of information coverage. Also, studying the extreme cases would be less valuable, as it would be almost impossible to intervene in false information with a $P$ close to 1. False information with insignificant impact power should not prioritised. Therefore, the parameter setting in the following analysis would be focused on moderate $P$ in similar-size networks.  

\subsubsection{\textbf{Impact of network structure on single information diffusion}}
\label{sec:network_info_diff}

In addition to the transmission probability, we study the network structure that is a key aspect for generalizing the analysis. As a huge network structure can be broken into a smaller structure, we analyse each network structure separately, which can lead to a universal conclusion about the whole network structure. The two network structures we will investigate are the network of one-community (sparse and dense) and the network of multi-communities. 

\paragraph{\textbf{Information diffusion analysis on one-community networks}}

We use the ER random graph algorithm to generate one-community network. We generate this one-community network fifty times using the same parameter values ($n = 1000$, $ info\_starter = 3$, $P = 0.5$, $T = 0.5$, $edge\_exist\_prob = 0.04$). This results in fifty different network structures that represents the one-community (dense) environment, where users are highly connected. Similarly, we generate one-community (sparse) environment using the following parameters ($n = 1000$, $ info\_starter = 3$, $P = 0.5$, $T = 0.5$, $edge\_exist\_prob = 0.0005$) fifty times. The parameter $info\_starter$ refers to the number of $IC$. In each centrality-based strategy, the corresponding centrality measure are used to select the most important nodes as $IC$.

The results of one-community (dense and sparse) structures are shown in Figure \ref{fig:fig06_07_08_09} and Table \ref{table:table3}. For all centrality-based strategies, the diffusion speed (number of iterations) of the information is outperforming the one based on random selection of $IC$. Information diffusion on a highly-connected network is reasonably fast. In a network where all the nodes are directly connected, based on any selecting strategy of $IC$ all nodes can be reached in one iteration. Therefore, the topology-based diffusion strategy makes very small advantage on a dense social network structure. For the one-community (sparse) environment, topology-based diffusion strategy has greater iteration-wise advantages. 

Based on all the centrality measure strategies, the sum of $P_I$ outperforms corresponding ones from the random selection strategy on dense and sparse one-community networks. It can be concluded that the $IC$ selection strategy of the centrality-based measures contributes to intensifying the information diffusion. The information being diffused intuitively has a decreasing influence on the nodes with longer distance to the $IC$. Thus, as nodes with higher centrality play an important role in the network, the resulting information accumulation will be higher than a random node on the network.

Table \ref{table:table3} shows the $p$ values of the Wilcoxon signed-rank test between five different centrality-based strategies and the random selection strategy on the dense one-community networks. The null hypothesis is that the iteration of the centrality-based strategy and the random selection strategy is the same. Thus, the alternative hypothesis is that the iteration of a centrality-based strategy is lower than that of the random selection. Furthermore, the null hypothesis for $P_I$ is that the sum of $P_I$ is the same, while the alternative hypothesis is that the sum of $P_I$ based on centrality is higher than that of the random selection.

With 99\% confidence intervals, all the $p$ values suggest that based on each centrality measure, the iteration is lower than the random selection and the the sum of $P_I$ is higher than the random selection. Therefore, we can reject the null hypotheses of the iteration and the $P_I$. Based on that it can be concluded that centrality-based $IC$ selection strategies can accelerate and deepen the information diffusion on one-community (dense and sparse) network.

Among the centrality measures, all the strategies are performing similarly compared with each other, while eigenvector and closeness centrality strategies have slightly lower iteration needed for information diffusion on sparse one-community network.

\begin{figure*}[!htb]

\begin{subfigure}{.49\textwidth}
  \centering
  \includegraphics[width=.9\linewidth]{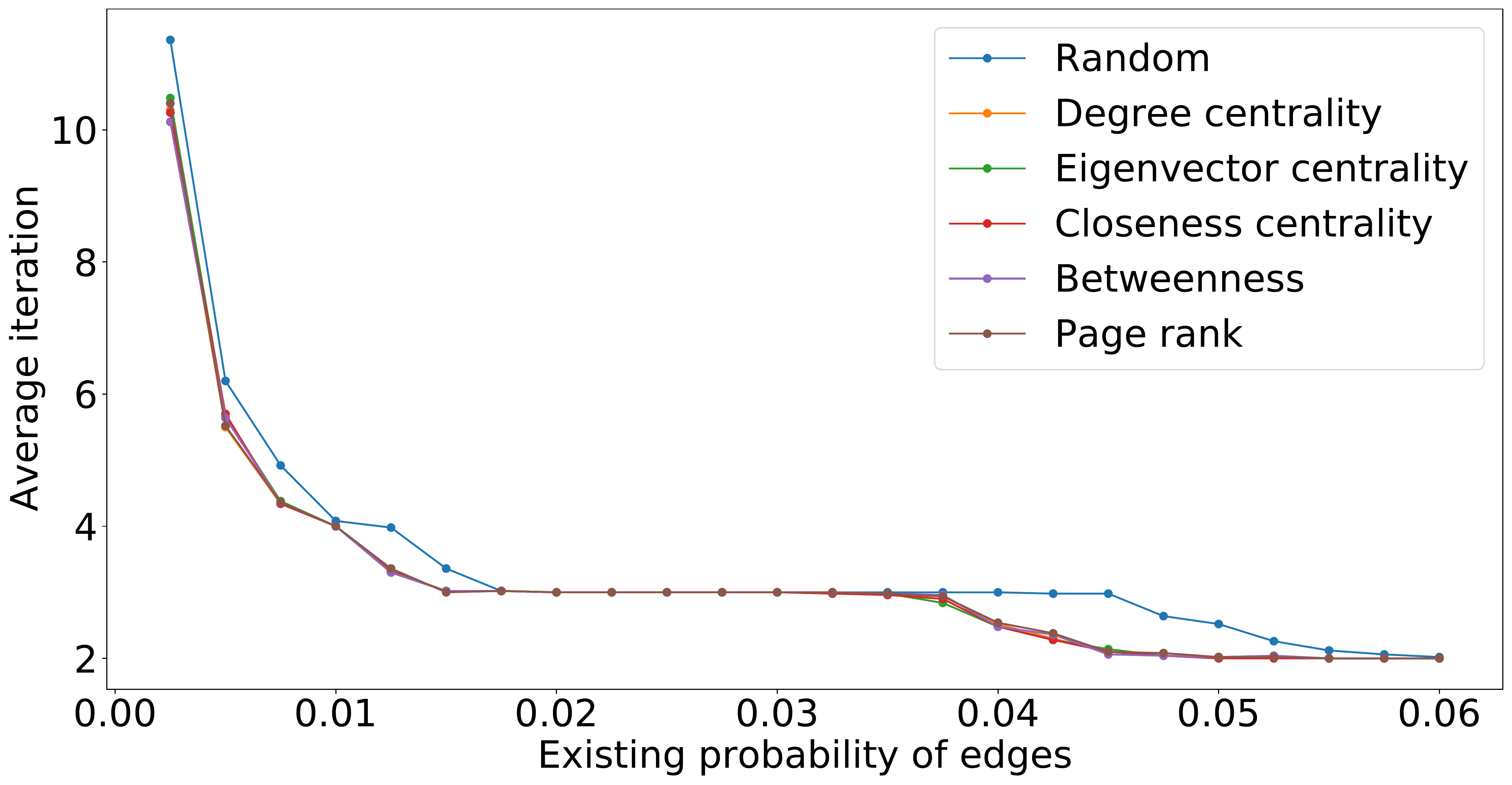}  
  \caption{ }
  \label{fig:sub-first}
\end{subfigure}
\begin{subfigure}{.49\textwidth}
  \centering
  \includegraphics[width=.9\linewidth]{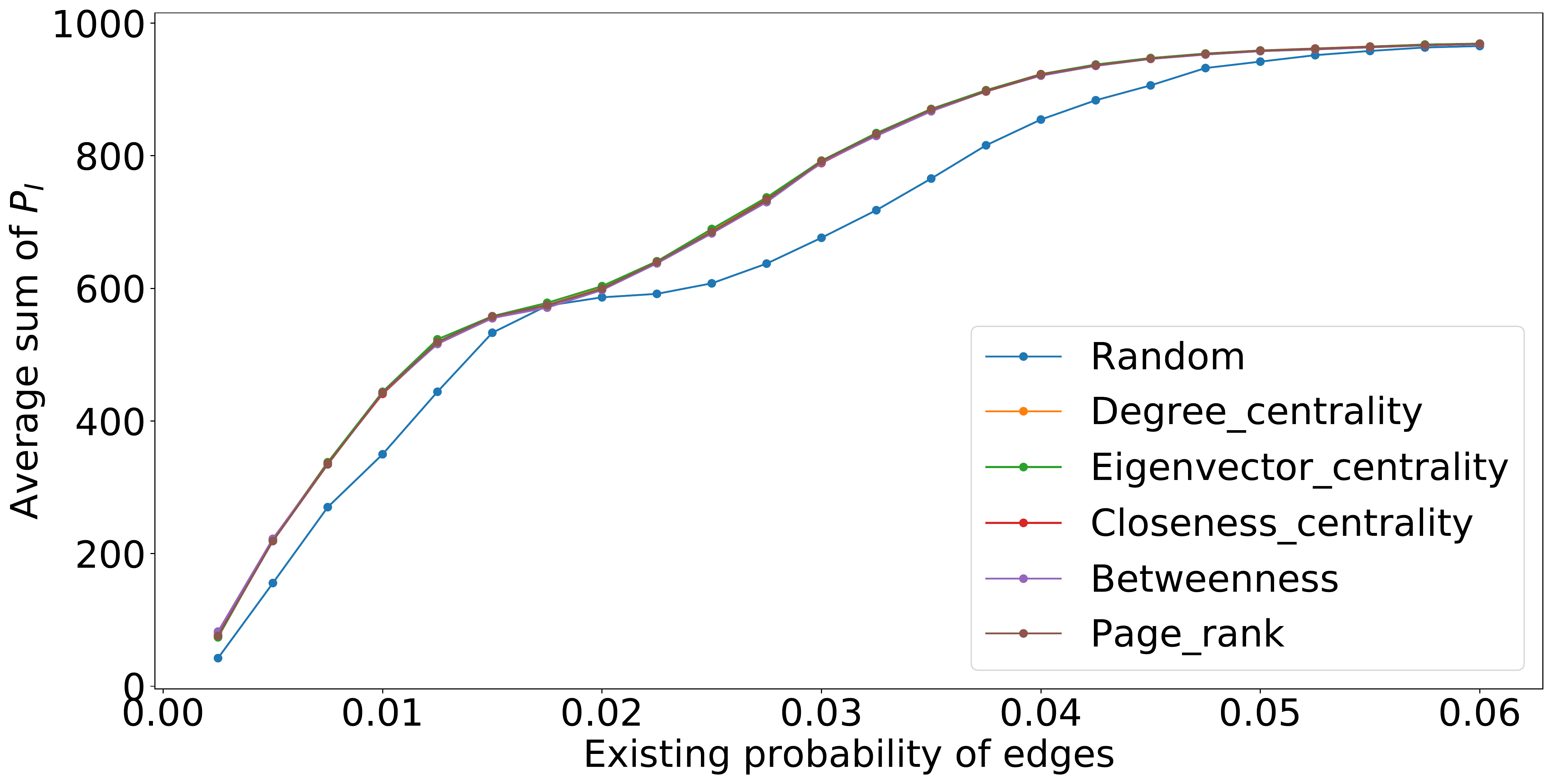}  
  \caption{ }
  \label{fig:sub-second}
\end{subfigure}
\caption{Effect of network density analysis for information diffusion on (a) iteration; (b) sum of $P_I$.}
\label{fig:fig_density}
\end{figure*}

Based on our previous findings, as the density of a one-community social network increases, the advantage of information diffusion based on five centrality decreases. We further analyze the effect on centrality-based information diffusion strategies in a one-community network, targeting at the micro-scope analysis of information diffusion in a social network ($n = 1000$, $ info\_starter = 3$, $P = 0.5$), with the result shown in Figure \ref{fig:fig_density}.

%


When the density is close to zero, the centrality-based strategies have slight advantages compared with random selection in diffusion speed and effectiveness. When the density of a network increases, the topological structure plays a more important role in the diffusion of information. After the density reaches a certain threshold, the network becomes dense enough where random selection can also pick an information creator with high centrality. Therefore, in extremely dense and sparse networks, the topology-based information diffusion model works poorly. At the same time, centrality contributes significantly to a community with moderate density.

\paragraph{\textbf{Information diffusion analysis on multi-communities networks}}

We use the \textit{Gaussian random partition graph} to generate two set of multi-communities networks. First, we generate fifty multi-communities networks with similar community size using the parameter values ($n = 1000$, $num\_info\_starter = 3$, $P = 0.5$, $T = 0.5$, $s = 40$, $v = 40$, $p\_in = 0.1$, $p\_out = 0.001$). Second, we generate another fifty multi-community networks with varying community sizes using the parameters ($n = 1000$, $num\_info\_starter = 3$, $P = 0.5$, $T = 0.5$, $s = 40$, $v = 1$, $p\_in = 0.1$, $p\_out = 0.001$). The expected community size of the generated networks is 40. The variance of community-size is 1 for similar-sized communities and 40 for the varying-sized communities. The edge existing probability within and between communities are 0.1 and 0.001 respectively. Therefore, each node is expected to have four neighbors within its community, and every node in one community are expected to have one cross-community connection. 

The results on multi-community networks are shown in Figure \ref{fig:fig10_11_12_13} and Table \ref{table:table5}. It can be concluded that the iteration distribution of multi-communities networks significantly differs from that of one-community (dense) network. When the networks are having similar-sized-multi-communities, the centrality-based stragies are generally outperforming the benchmark. However, in the simulated multi-communities networks with varying community size, the distribution of the iteration by strategies based on degree, eigenvector and closeness centrality does not show any advantage when compared with the random selection strategy. Apart from that, the iteration distribution based on the other two centrality measures only slightly outperform the random selection strategy. It can be noted that the two better-performing measures consider the global structure of the network, while the other three mainly take the local neighbors into consideration. In a multi-communities networks, local-focusing measures may have $IC$ trapped in a closely connected community, leading to a poor reaching-speed from the whole network. 


Similarly, the sum of $P_I$ distribution also validates the benefits of topology-based diffusion. Degree centrality-based diffusion has overall optimal performance. A high degree $IC$ can pass the high accumulation advantage regardless of the drawback in diffusion iteration. Therefore, a small clustered community close to $IC$ would have a high sum of $P_I$, which contributes to the well-performance. The other centrality measures also outperform the benchmark. It could be noticed that the eigenvector centrality has a significantly weak performance on similar-sized-communities networks, when compared with the other four centrality-based strategies.


The $p$ values (as shown in Table \ref{table:table5}) are consistent with our primary observation on the box plot. $p$ values of all the centrality measures suggest that they outperform the random selection in a 99\% confidence interval in terms of iteration and sum of $P_I$. However, eigenvector centrality strategy has a less significant advantage compared with the other centrality measures.  

When the community size varies sharply, the resulting social network structure will have a less-clustered effect. After that, the generated social networks will be closer to a dense one-community structure, where the information diffusion are easier. Therefore, the performance of information diffusion based on iteration or sum of $P_I$ generally are better in varying-sized multi-communities structure (as shown in Figure \ref{fig:fig10_11_12_13}).

\begin{figure*}[!htb]

\begin{subfigure}{.49\textwidth}
  \centering
  \includegraphics[width=.99\linewidth]{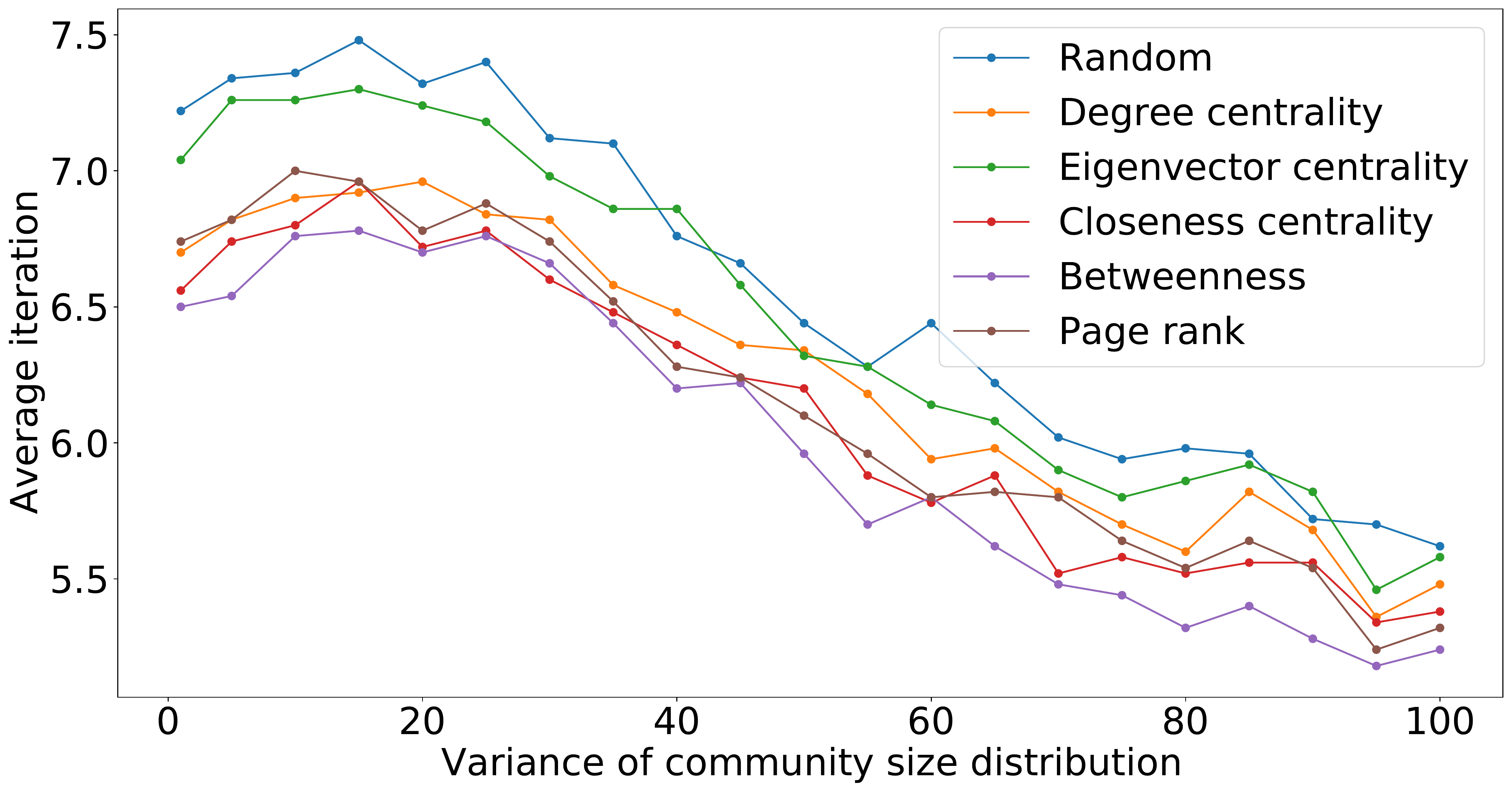}  
  \caption{ }
  \label{fig16_17:sub-first}
\end{subfigure}
\begin{subfigure}{.49\textwidth}
  \centering
  \includegraphics[width=.99\linewidth]{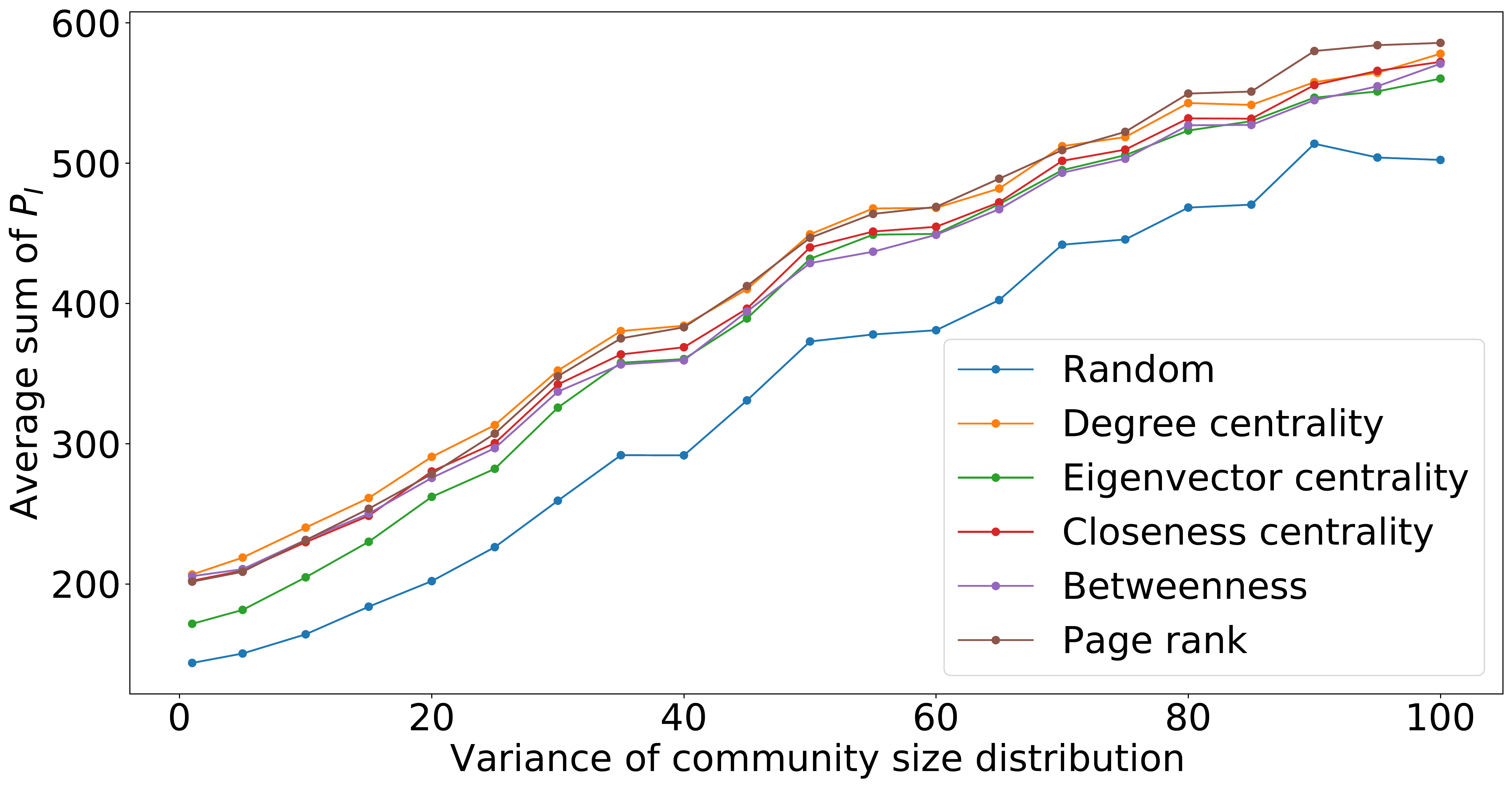}  
  \caption{ }
  \label{fig16_17:sub-second}
\end{subfigure}
\caption{Effect of variance in community size analysis for information diffusion on (a) iteration; (b) sum of $P_I$.}
\label{fig:fig16_17}
\end{figure*}

Based on the previous results, the variance of the community size in a multi-community network can be negatively correlated with the significance of the $p$ values. Further analysis with parameters of ($n = 1000$, $num\_info\_starter = 3$, $P = 0.5$, $T = 0.5$, $s = 40$, $p\_in = 0.1$, $p\_out = 0.001$) provides no clue that variance in community size is related to the absolute difference in iteration or sum of $P_I$ (shown in Figure \ref{fig:fig16_17}). However, we can see a descending trend with increasing community size variance. As the connection between nodes within a community is higher than across communities, the increasing variance in the community size would lead to a gradual transformation from a clustered network structure towards a one-community network structure. Therefore, the information diffusion in a one-community structure is faster and more effective than that in a clustered multi-community structure. 

%


\paragraph{\textbf{Information diffusion analysis on LFR benchmark networks}}

Lastly, we apply our proposed information diffusion model to 50 LFR benchmark networks with parameters ($n = 1000$, $num\_info\_starter = 3$, $P = 0.5$, $T = 0.5$, $tau1 = 3$, $tau2 = 1.5$, $\mu = 0.1$,  $average\_degree = 5$, $min\_community=50$). The LFR benchmark networks are reflecting real-world properties of nodes and communities, which makes it a good simulated environment for modelling real-world network structure. 

Based on result in Figure \ref{fig:LFRsingle} and Table \ref{table:LFRsingle}, all the centrality based strategies outperform random selection. The result is consistent with our previous finding, where only eigenvector centrality is the weaker measures of the topology-based information diffusion compared with the others. The topology-based strategy is validated to be effectively useful to empower information diffusion.

\subsection{Topological-based interventions for combating information disorder}
\begin{figure*}[!htbp]
    \centering
    \includegraphics[width=15cm]{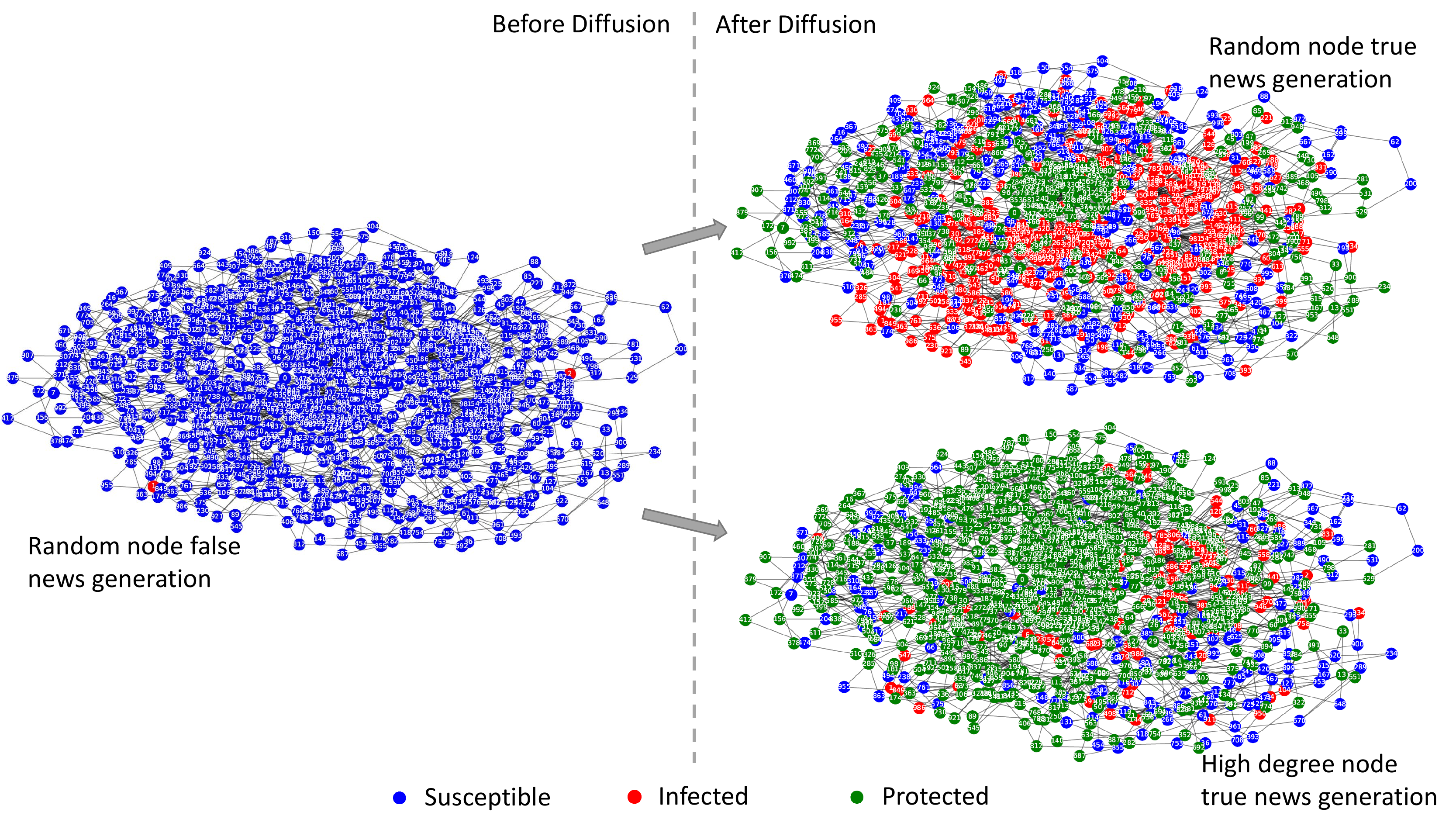}
    \caption{Information interventions of the simulated social network}
    \label{fig:Figure18}
\end{figure*}

Apart from modelling and analysing information diffusion, we have also extended our model for combating information disorder. The topology-based interventions for combating information is depicted in Figure \ref{fig:Figure18}. If the initial spreader of true information has the highest degree centrality, the diffusion of false information can be significantly retardant. Compared with the random selection strategy, the number of people being “infected” and “susceptible” are distinctly lower, while more people are “protected”.

\subsubsection{\textbf{Impact of network structure on information disorder interventions}}
To be conservative, we randomly select the information creator of false information, which would be a more challenging condition compared with real world situation. 

\paragraph{\textbf{Information interventions analysis on one-community networks}}
We apply further statistical analysis for structural impact on topology-based interventions on a generated dense ER random graph. Similarly, the dense and sparse ER random graphs are used to model one-community social networks with high and low connection between the users respectively.

We define a complete intervention as the expectation of number of protected nodes is higher than that of the infected nodes. Given that the number of $IC_F$ is three, the minimum number of $IC_T$ which lead to a complete intervention under averaged result of 50 ER random graph is shown in the  Table \ref{table:tableIC}, using parameter ($n = 1000$, $ false\_info\_starter = 3$, $ false\_info\_starter = 10$, $P_F = 0.5$, $P_T = 0.4$, $T_D = 0.4$,  $T_C = 0.1$, $edge\_exist\_prob = 0.03$). Centrality measure can significantly reduce the minimum number of $IC_T$ for a complete intervention, which means the topology-based interventions are more efficient than the benchmark. 

We generate one-community (dense) network fifty times using parameter ($n = 1000$, $ false\_info\_starter = 3$, $ true\_info\_starter = 10$, $P_F = 0.5$, $P_T = 0.4$, $T_D = 0.4$,  $T_C = 0.1$, $edge\_exist\_prob = 0.04$) and one-community (sparse) network fifty times using ($n = 1000$, $ false\_info\_starter = 3$, $ true\_info\_starter = 10$, $P_F = 0.5$, $P_T = 0.4$, $T_D = 0.4$,  $T_C = 0.1$, $edge\_exist\_prob = 0.005$) for information intervention analysis. As mentioned before, the false information diffuse faster and deeper on a social network. Therefore we set the transmission probability of false information $P\_F$ as 0.5, while $P\_T$ is set as 0.4. Besides, to simulate the fastness of false information, the diffusion of false information is one iteration ahead compared with true information. The $T\_D$ is set as 0.4, which means when the sum of $P\_{IF}$ for one node has reached 0.4, the node is no longer available for true information diffusion. In addition, $T\_C$ is set as 0.1, which is the initial difference in $P\_F$ and $P\_T$. Therefore, when the difference in sum of $P\_{IF}$ and $P\_{IT}$ for one node is larger than 0.1, the node is labeled as false information believer or “infected” here. Lastly, the potential believer of false information is labeled as “susceptible” and the believer of true information is labeled as “protected” in the result figures.

The box plots in Figure \ref{fig:fig21_22_23_24} and \ref{fig:fig25_26_27_28} show the detailed performance based on different centrality measures. Topology-based interventions generally have a better performance on sparse one-community networks compared with dense ones. Within the five centrality intervention strategies, there is a tiny difference in the effectiveness of diffusion on dense ER random graph. 

As shown in Table \ref{table:table6} and \ref{table:table7}, all the centrality measures achieved a sum of $P_{IT}$ under 99\% confidence interval higher than those in the random selection benchmark. Apart from that, more nodes are  “protected”, and fewer are “infected” because of topology-based strategies. As for “susceptible” nodes, all centrality measures show no advantage compared with random selection. After all, the centrality-based strategies all out-perform random selection on one-community networks.

We further analyze the effect of community density on combating information diffusion in a one-community network, targeting the micro-scope analysis of information diffusion strategy in a social network  ($n = 1000$, $ false\_info\_starter = 3$, $ true\_info\_starter = 10$, $P_F = 0.5$, $P_T = 0.4$).

\begin{figure*}[!htbp]
\begin{subfigure}{.49\textwidth}
  \centering
  \includegraphics[width=.9\linewidth]{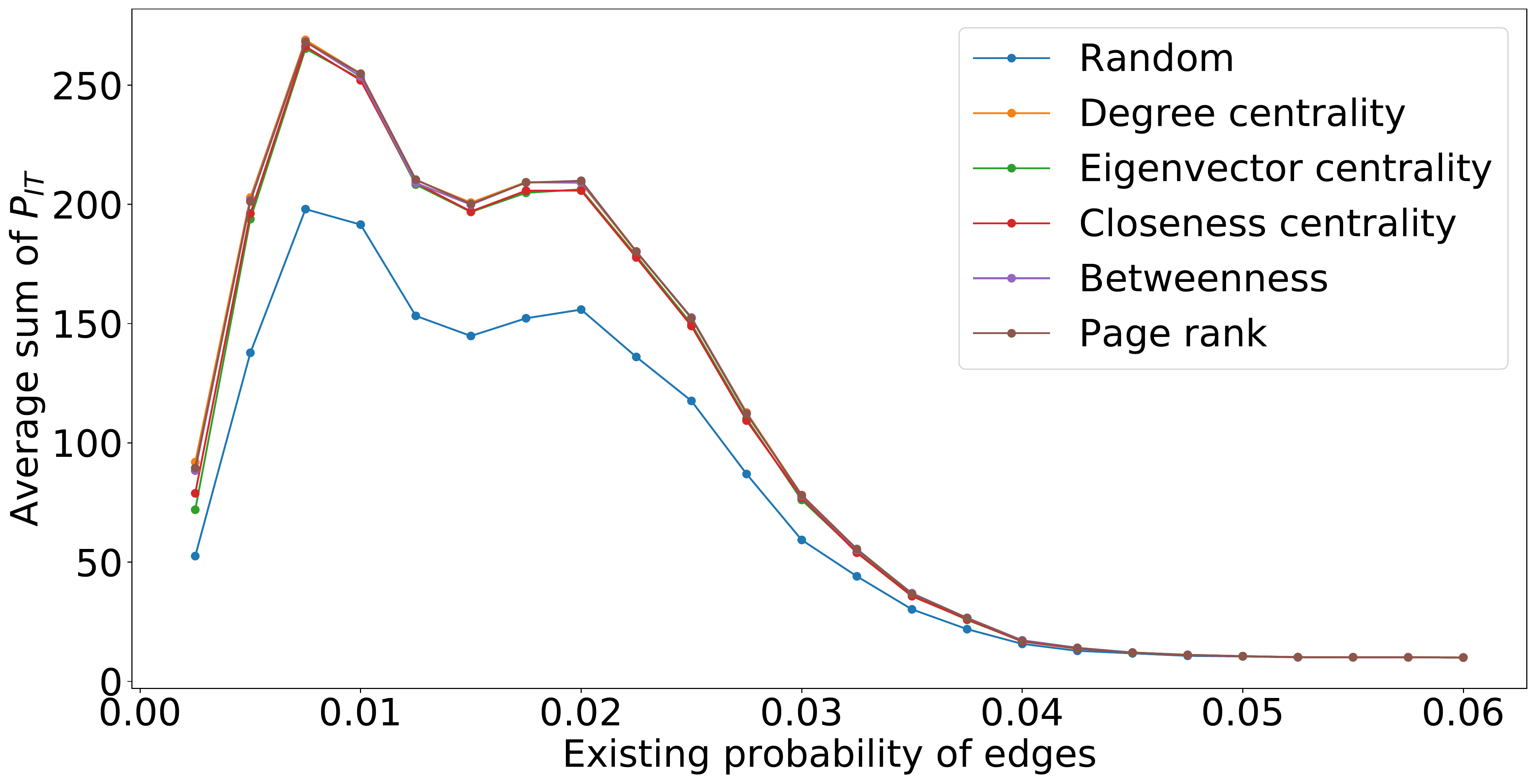}  
  \caption{ }
  \label{fig37_38_39_40:sub-first}
\end{subfigure}
\begin{subfigure}{.49\textwidth}
  \centering
  \includegraphics[width=.9\linewidth]{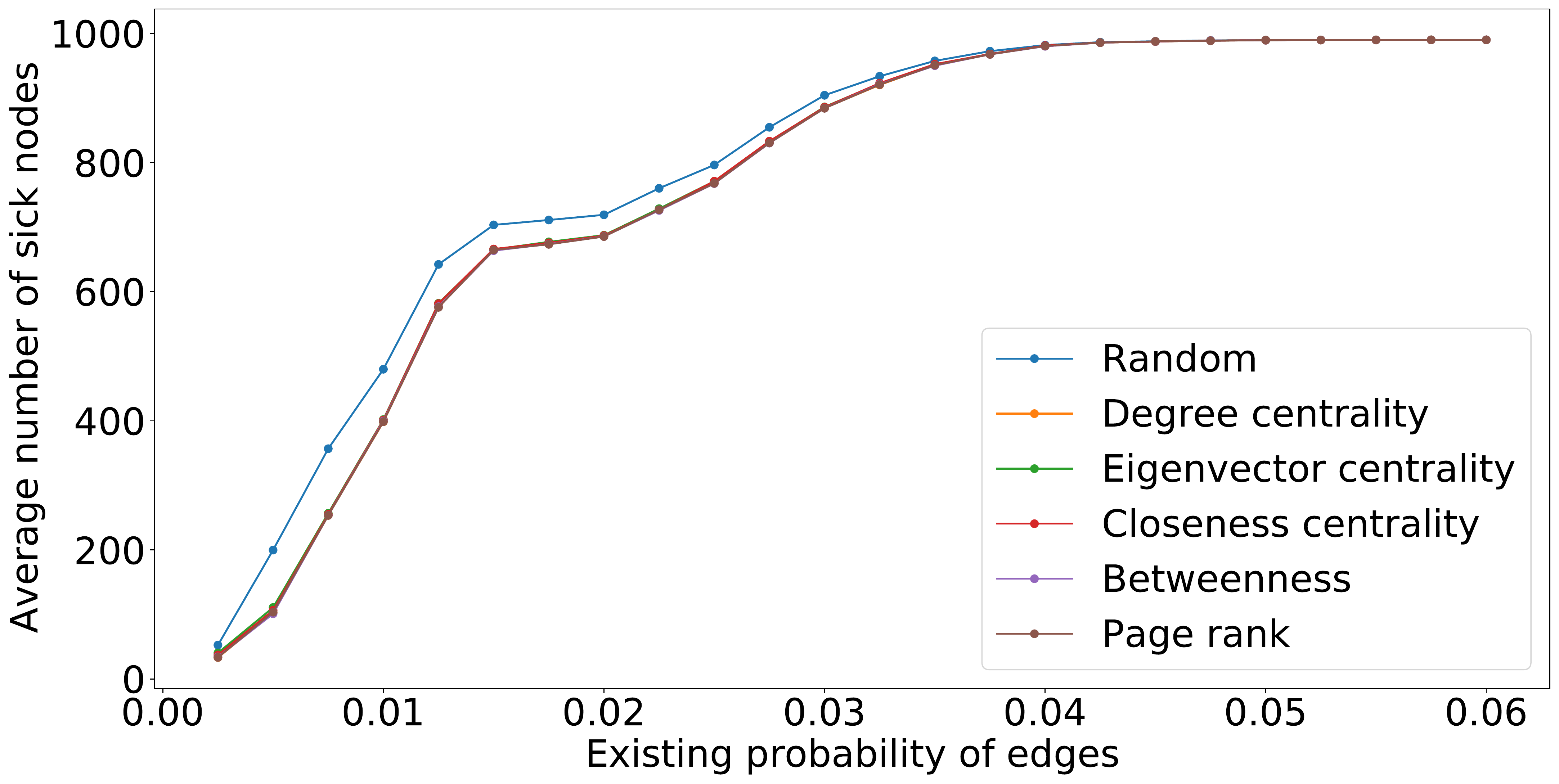}  
  \caption{ }
  \label{fig37_38_39_40:sub-second}
\end{subfigure}


\begin{subfigure}{.49\textwidth}
  \centering
  \includegraphics[width=.9\linewidth]{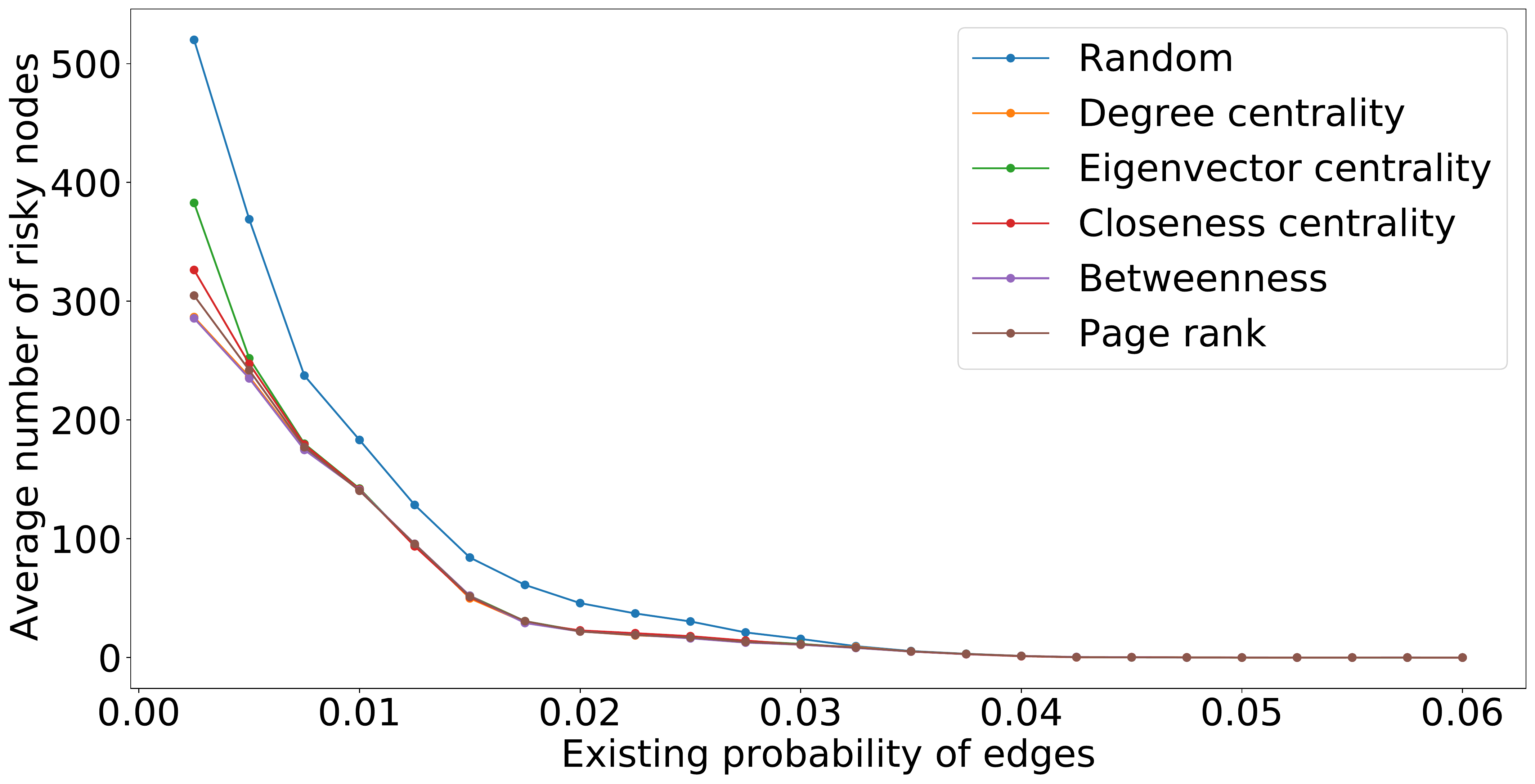}  
  \caption{ }
  \label{fig37_38_39_40:sub-third}
\end{subfigure}
\begin{subfigure}{.49\textwidth}
  \centering
  \includegraphics[width=.9\linewidth]{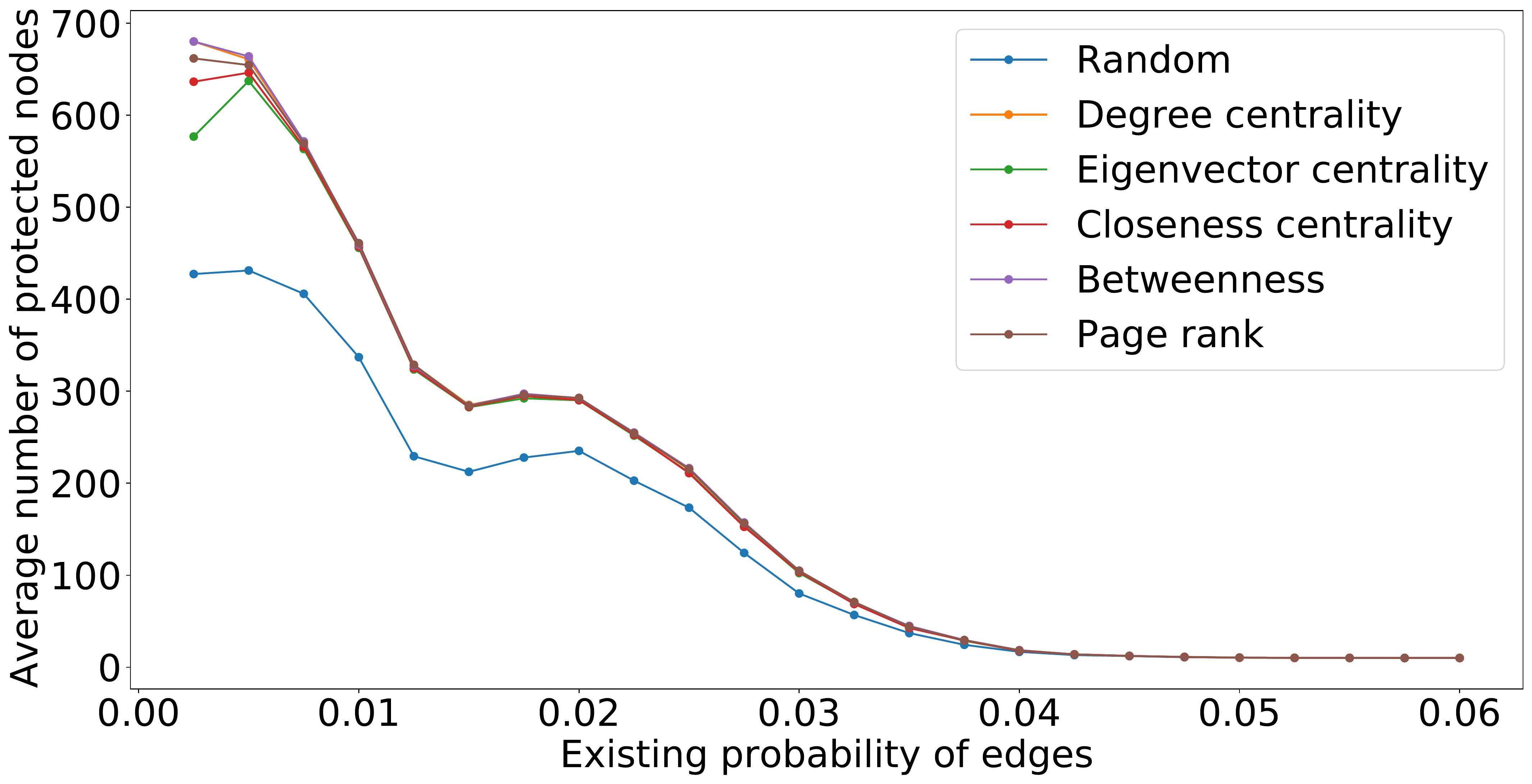}  
  \caption{ }
  \label{fig37_38_39_40:sub-fourth}
\end{subfigure}

\caption{Effect of network density analysis for information interventions on (a) sum of $P_{IT}$; (b) number of infected nodes; (c) number of susceptible nodes; (d) number of protected nodes.}
\label{fig:fig37_38_39_40}
\end{figure*}

With increasing density observed in Figure \ref{fig37_38_39_40:sub-first}, \ref{fig37_38_39_40:sub-second}, \ref{fig37_38_39_40:sub-third}, and \ref{fig37_38_39_40:sub-fourth}, the centrality-based strategies perform optimally on a sweet point. After that, the increasing density will lead to an excessively dense network where false information dominates the nodes. Therefore, on networks with high density, centrality-based strategies will have no advantages compared with random selection.

%


\paragraph{\textbf{Information interventions analysis on multi-communities networks}}
Then we use a clustered Gaussian random partition graph to model a multi-communities social networks. Given that the number of $IC_F$ is three, the minimum number of $IC_T$ which lead to a complete intervention under averaged result of 50 multi-communities networks is shown in the Table \ref{table:tableIC}, using parameter ($n = 1000$, $false\_info\_starter = 3$, $P_F = 0.5$, $P_T = 0.4$, $T_D = 0.4$, $T_C = 0.1$, $s = 40$, $v = 40$, $p\_in = 0.1$, $p\_out = 0.001$). Based on the results shown, the topology-based interventions are significantly more efficient than the benchmark. On multi-communities networks, all the centrality-based strategies except the eigenvector centrality-based one outperform the benchmark.

\begin{table}[!htbp]
  \caption{minimum number of $IC_T$ for a complete intervention on 50 ER random graph and on 50 50 multi-communities networks}
  \label{table:tableIC}
\begin{tabular}{lrc}
\hline
\multirow{2}{*}{\begin{tabular}[c]{@{}l@{}}\textbf{Centrality}\\ \textbf{measure}\end{tabular}} & \multicolumn{2}{r}{\textbf{minimum number of $IC_T$}} \\ \cline{2-3} 
            & \multicolumn{1}{c}{ER} & multi-communities \\ \hline
Degree      & $10$                   & $3$               \\
Eigenvector & $12$                   & $16$              \\
Closeness   & $10$                   & $3$               \\
Betweenness & $10$                   & $3$               \\
Page rank   & $10$                   & $3$               \\
Random      & $19$                   & $6$               \\ \hline
\end{tabular}
\end{table}

Networks are generated with parameter ($n = 1000$, $false\_info\_starter = 3$, $true\_info\_starter = 10$, $P_F = 0.5$, $P_T = 0.4$, $T_D = 0.4$, $T_C = 0.1$, $s = 40$, $v = 40$, $p\_in = 0.1$, $p\_out = 0.001$) fifty times for information intervention analysis on similar-sized multi-communities networks. Besides, networks with parameter($n = 1000$, $false\_info\_starter = 3$, $true\_info\_starter = 10$, $P_F = 0.5$, $P_T = 0.4$, $T_D = 0.4$, $T_C = 0.1$, $s = 40$, $v = 1$, $p\_in = 0.1$, $p\_out = 0.001$) are generated fifty times for information intervention analysis on varying-sized multi-communities networks. The set of parameter will result in a network with average community size of 40. The variance of community size will be 1 for similar-sized and 40 for varying-sized. The edge existence probability within and between community is 0.1 and 0.001 respectively. Therefore, each node will have a expectation of four neighbours within its community, and every node in one community are expected to have one cross-community connection.

On the Gaussian random partition graphs, only betweenness and page rank are significantly better than the benchmark on both similar-sized and varying-sized multi-communities networks. Eigenvector centrality on multi-communities networks shows no advantage compared with random selection. The different performance of degree centrality based intervention strategy on similar-sized and varying-sized results from that of information diffusion analysis similarly.

For analysis of variance in community size, the parameters are set as ($n = 1000$, $false\_info\_starter = 3$, $true\_info\_starter = 10$, $P_F = 0.5$, $P_T = 0.4$, $T_D = 0.5$, $T_C = 0.1$, $s = 40$, $p\_in = 0.1$, $p\_out = 0.001$). When the two combating information are diffused on a network with multi-communities where the variance is moderate, the centrality-based strategies for dissemination of true information are significantly more effective. Page rank, Betweenness, and closeness centrality are the leading measures for a deeper information influence.



\begin{figure*}[!htbp]

\begin{subfigure}{.49\textwidth}
  \centering
  \includegraphics[width=.9\linewidth]{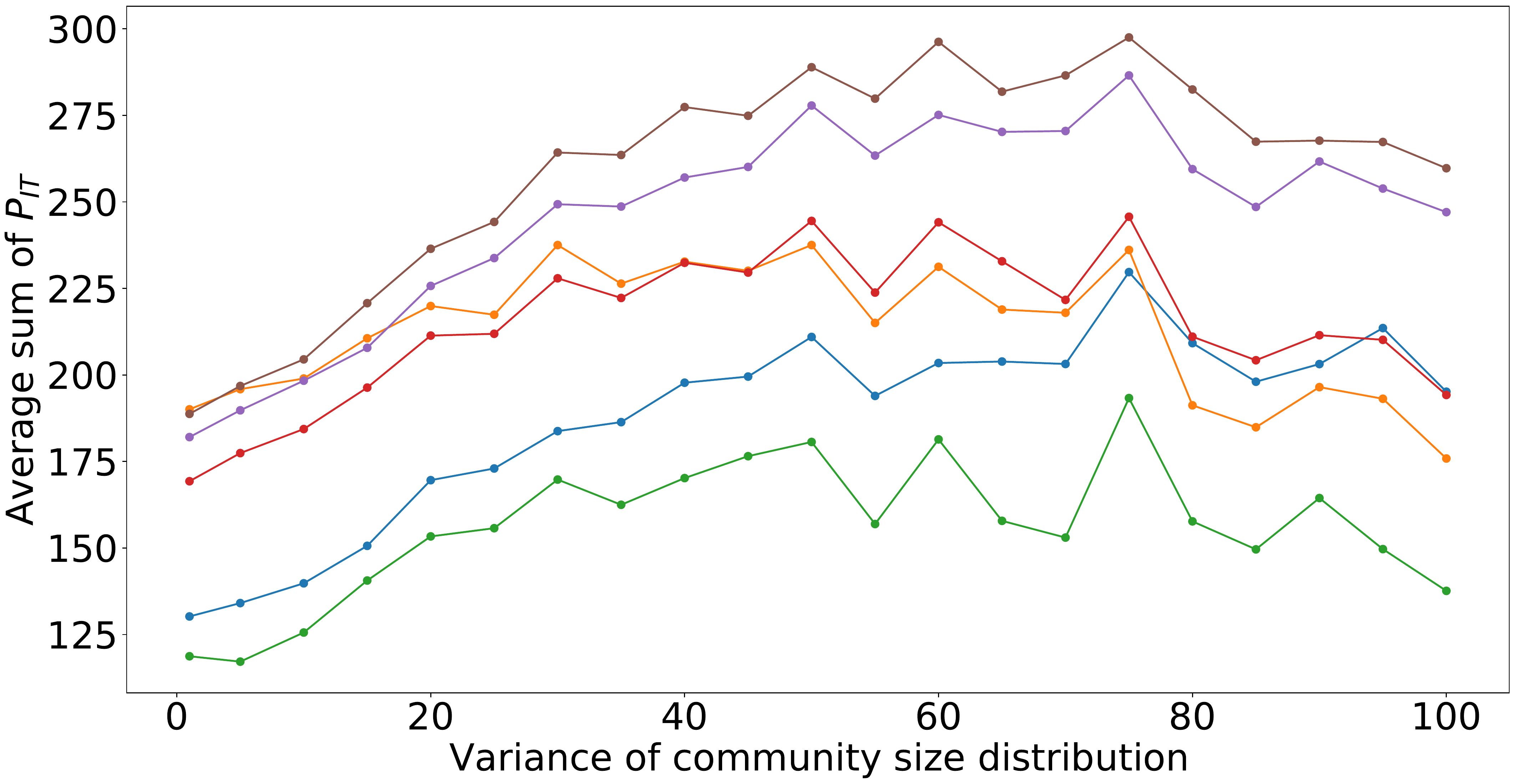}  
  \caption{ }
  \label{fig41_42_43_44:sub-first}
\end{subfigure}
\begin{subfigure}{.49\textwidth}
  \centering
  \includegraphics[width=.9\linewidth]{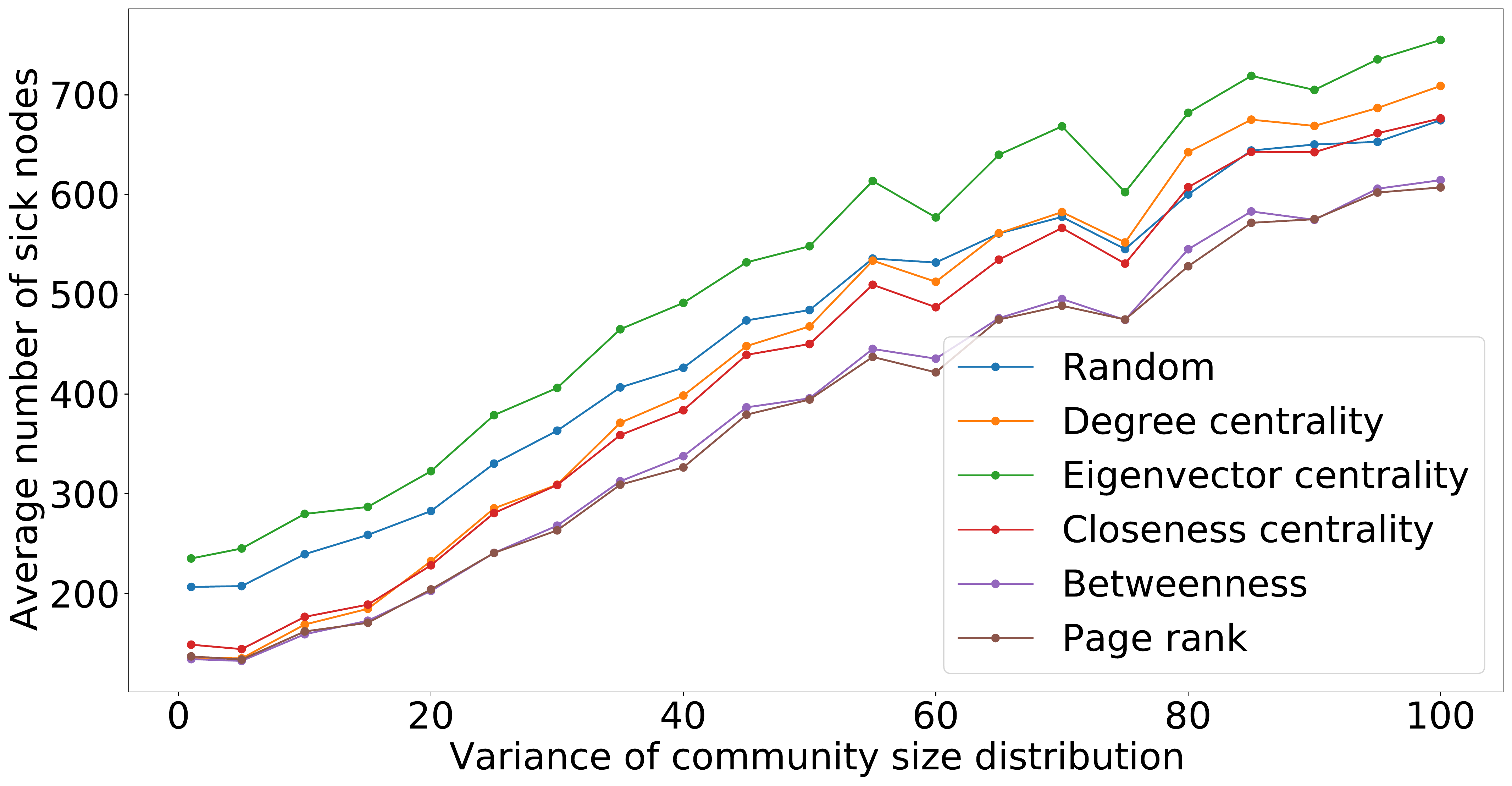}  
  \caption{ }
  \label{fig41_42_43_44:sub-second}
\end{subfigure}


\begin{subfigure}{.49\textwidth}
  \centering
  \includegraphics[width=.9\linewidth]{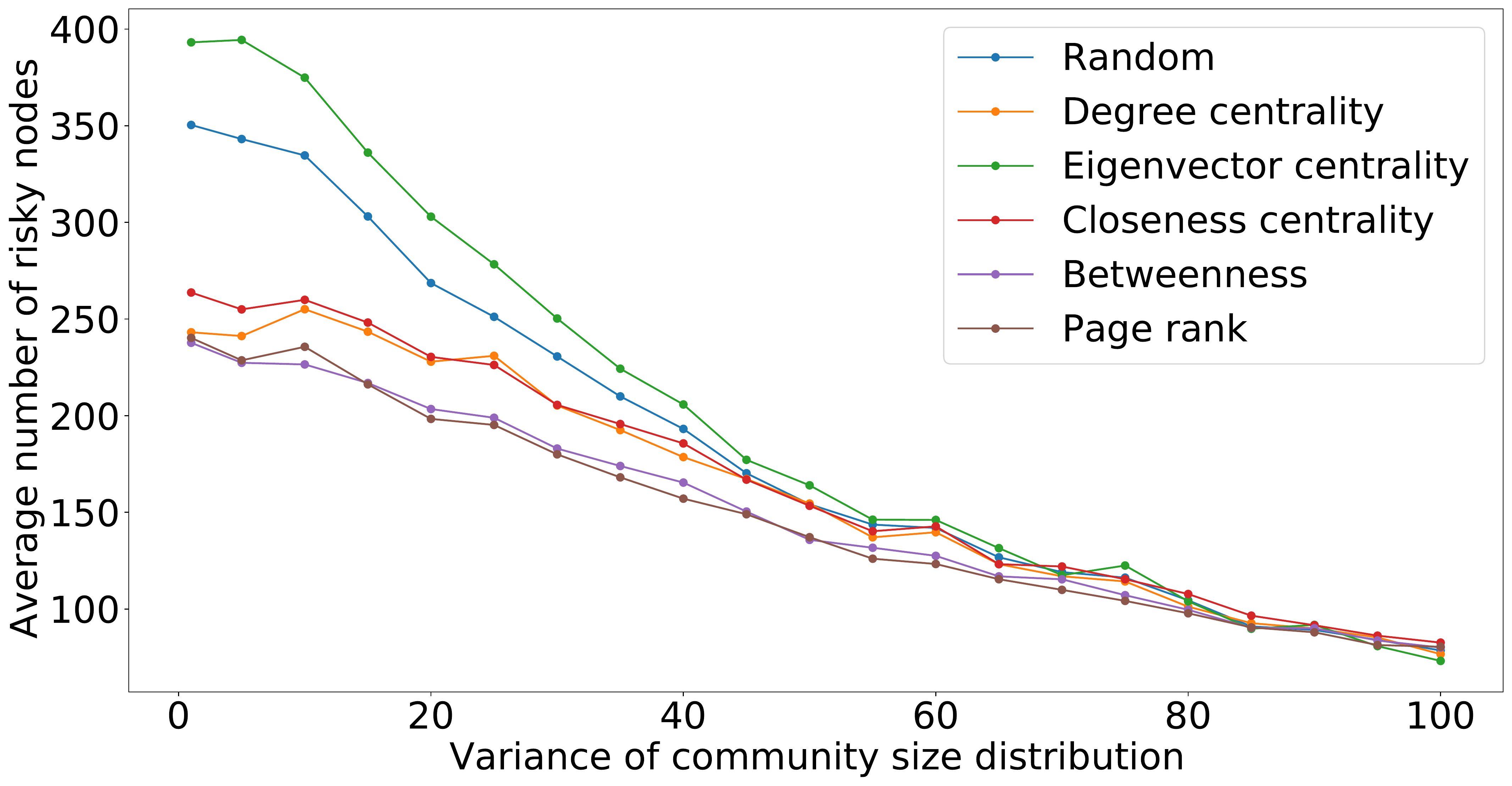}  
  \caption{ }
  \label{fig41_42_43_44:sub-third}
\end{subfigure}
\begin{subfigure}{.49\textwidth}
  \centering
  \includegraphics[width=.9\linewidth]{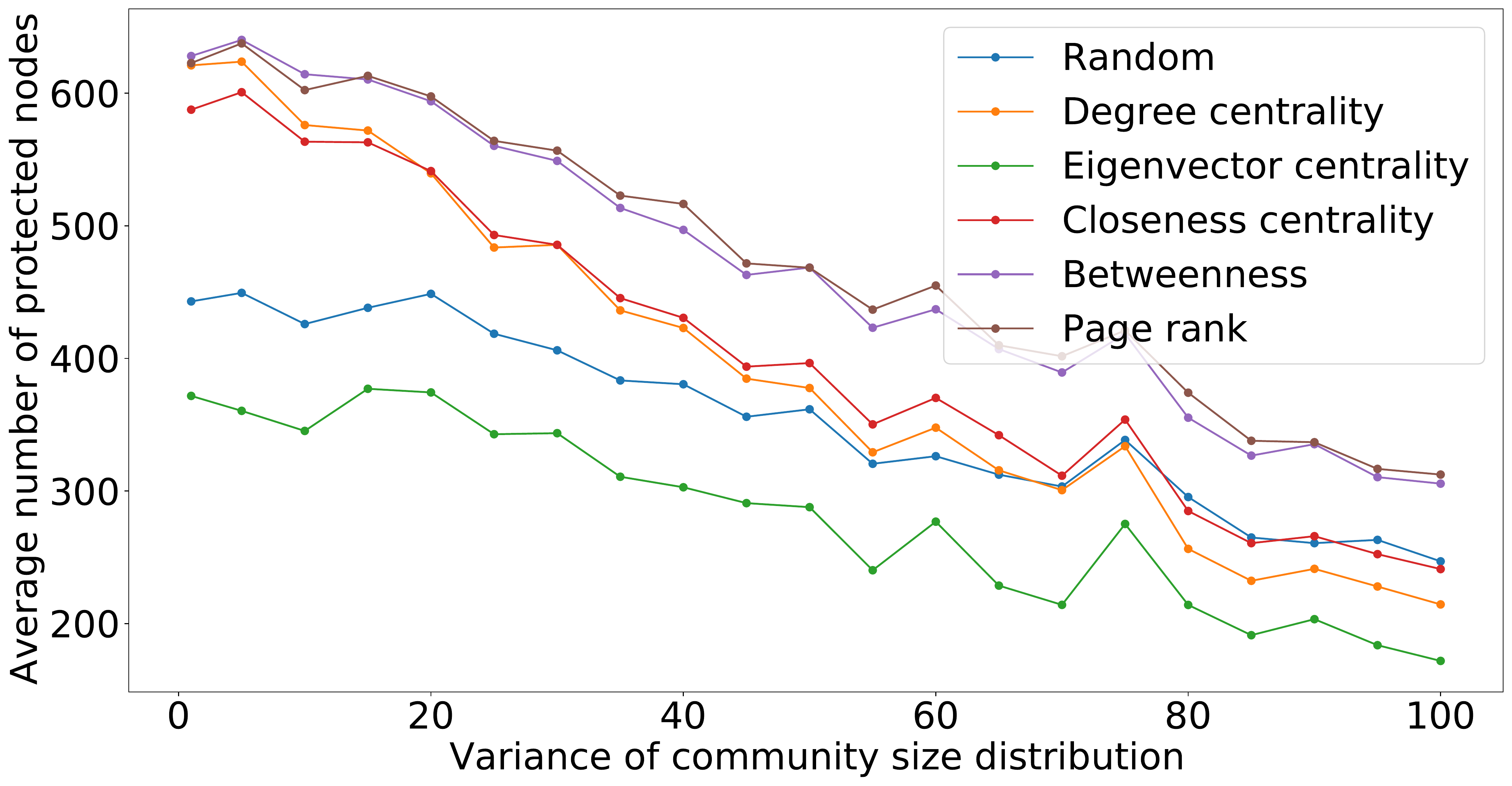}  
  \caption{ }
  \label{fig41_42_43_44:sub-fourth}
\end{subfigure}

\caption{Effect of variance in community size analysis for information interventions on (a) sum of $P_{IT}$; (b) number of infected nodes; (c) number of susceptible nodes; (d) number of protected nodes.}
\label{fig:fig41_42_43_44}
\end{figure*}

From results in Figures \ref{fig41_42_43_44:sub-first}, \ref{fig41_42_43_44:sub-second}, \ref{fig41_42_43_44:sub-third}, and \ref{fig41_42_43_44:sub-fourth}, when the variance of community size increases, the true information generally become less competitive. A Multi-community network with a higher variance will become closer to a one-community network. Also, as the edge existence probability within communities is higher than between communities, the density of the entire network will become larger with increasing community size, resulting in more “infected” nodes. Similarly, centrality based strategies have fewer advantages compared with random selection, with an increasing variance of community size.

\paragraph{\textbf{Information interventions analysis on LFR benchmark networks}}
Lastly, we apply our proposed combating information diffusion model on 50 LFR benchmark networks with parameters as ($n = 1000$, $false\_info\_starter = 3$, $true\_info\_starter = 10$, $P_F = 0.5$, $P_T = 0.4$, $T_D = 0.5$, $T_C = 0.1$, $tau1 = 3$, $tau2 = 1.5$, $\mu = 0.1$,  $average\_degree = 5$, $min\_community=50$) to analysis information intervention. Based on result in Figure \ref{fig:LFRtf} and Table \ref{table:LFRtf}, all the centrality based strategies except eigenvector out-perform random selection. The result is consistent with our previous finding, where betweenness and page rank centrality are still the optimal centrality measure of topology-based information intervention. The topology-based strategy is validated to be effectively empower the true information diffusion.

\section{Limitations and future works}
\label{sec:Limitations and future works}
The simulation of information diffusion depends significantly on the parameter selection, which makes our modeling highly data-driven. The centrality based strategies have optimal performance under graph with moderate density and diffusion with moderate transmission probability. Therefore, simulations on real-world network structure would improve the reliability of our simulation study and topological analysis.

Further works can be done for fine-tuning sets of parameters, targeting values closer to the real-world environment. A combination of studies of the real-world network’s topology structure with our simulation model could be applied. Additionally, transmitting probability on each edge would also be further extended to a set of trainable parameters, which could increase the flexibility of our proposed model.

\section{Conclusion}
\label{sec:Conclusion}
We have proposed a computationally efficient information diffusion model and information intervention model for combating information disorder, with a small number of parameters. We model the diffusion of information throughout independent boolean-valued outcomes, and further analyze and assess the effectiveness of centrality-based true news originator selection strategies. Centrality has been validated to be effective and efficient in both information diffusion and disorder interventions. According to the proposed information diffusion model, degree centrality leads to a deep information diffusion. However, considering the diffusion efficiency, betweenness and page rank would be the optimal centrality measurement for the information disorder intervention. Additionally, centrality-based strategies for both information diffusion and intervention have better performance in a moderately dense graph, which is similar to a real-world social media environment. 

\bibliographystyle{ACM-Reference-Format}
\bibliography{main}

\clearpage


\appendix
\section{{User Engagement on verified true vs false news}}
These list of user engagements were gathered from 3 widely used fact-checkers Snopes, Politifact and FactCheck.Each news ID labels one data record of the verified true news and the corresponding false news, collected from the fact-checkers. The data of user engagements are collected as a combination of times of shares, watches, likes and comments, as news being shared on social media has different forms.

\begin{table}[!htp]
\caption{User engagements of news on social media}
\label{table:engagement}
\begin{tabular}{lrr}

\label{tab:userengagement}
        & \multicolumn{2}{c}{User engagements of news on social media} \\
News ID & \multicolumn{1}{c}{True}     & \multicolumn{1}{r}{False}     \\
1       & 603                          & 37964                         \\
2       & 351                          & 3497                          \\
3       & 31                           & 20083196                      \\
4       & 69                           & 43221                         \\
5       & 123                          & 2810                          \\
6       & 4557                         & 24470                         \\
7       & 172                          & 48863                         \\
8       & 1167                         & 3592                          \\
9       & 610                          & 1136                          \\
10      & 555                          & 1604                          \\
11      & 1544                         & 981                           \\
12      & 730                          & 4464                          \\
13      & 1387                         & 1456                          \\
14      & 6456                         & 17139                         \\
15      & 3097                         & 17371                         \\
16      & 1743                         & 1516                          \\
17      & 791                          & 25630                         \\
18      & 8747                         & 40                            \\
19      & 662                          & 1116                          \\
20      & 2502                         & 140145                        \\
21      & 813                          & 546                           \\
22      & 5962                         & 3351                          \\
23      & 170                          & 2526                          \\
24      & 1973                         & 16000                         \\
25      & 4187                         & 16374                         \\
26      & 1364                         & 31800                         \\
27      & 2453                         & 2297                          \\
28      & 682                          & 71137                         \\
29      & 2690                         & 6575                          \\
30      & 3536                         & 1400                          \\
31      & 2264                         & 64197                         \\
32      & 687                          & 6170                          \\
33      & 1813                         & 3294                          \\
34      & 3905                         & 1060                          \\
35      & 487                          & 845262                        \\
36      & 2793                         & 497                           \\
37      & 3052                         & 4518                          \\
38      & 363                          & 1747                          \\
39      & 874                          & 2554                          \\
40      & 3999                         & 7700                          \\
41      & 5973                         & 1333                          \\
42      & 6394                         & 6148                          \\
43      & 2871                         & 10209                         \\
44      & 663                          & 1950                          \\
45      & 1185                         & 3373                          \\

\end{tabular}
\end{table}

\begin{table}[!htp]
\begin{tabular}{lrr}
        & \multicolumn{2}{c}{User engagements of news on social media} \\
News ID & \multicolumn{1}{c}{True}     & \multicolumn{1}{r}{False}     \\
46      & 814                          & 63100                         \\
47      & 596                          & 25575                         \\
48      & 9867                         & 7700                          \\
49      & 2399                         & 3926                          \\
50      & 1519                         & 3127                          \\
51      & 1223                         & 143854                        \\
52      & 92                           & 295                           \\
53      & 1614                         & 7757                          \\
54      & 3323                         & 37403                         \\
55      & 1497                         & 700                           \\
56      & 569                          & 4804                          \\
57      & 4833                         & 473                           \\
58      & 1008                         & 3627                          \\
59      & 714                          & 921                           \\
60      & 1977                         & 2131                          \\
61      & 5818                         & 1476                          \\
62      & 1867                         & 648                           \\
63      & 378                          & 2698                          \\
64      & 831                          & 16000                         \\
65      & 1585                         & 1588                          \\
66      & 3905                         & 5132                          \\
67      & 2292                         & 61013                         \\
68      & 3984                         & 18240                         \\
69      & 437                          & 1013                          \\
70      & 2037                         & 1096985                       \\
71      & 1070                         & 704                           \\
72      & 722                          & 315                           \\
73      & 1015                         & 824                           \\
74      & 1080                         & 155                           \\
75      & 2483                         & 2632                          \\
76      & 1429                         & 2513                          \\
77      & 1234                         & 72000                         \\
78      & 392                          & 4682                          \\
79      & 1718                         & 134201                        \\
80      & 24368                        & 100013                        \\
81      & 1013                         & 76013                         \\
82      & 2375                         & 1992                          \\
83      & 1333                         & 29868                         \\
84      & 864                          & 5833                          \\
85      & 1082                         & 1414                          \\
86      & 1222                         & 5335                          \\
87      & 1590                         & 20100                         \\
88      & 395                          & 337                           \\
89      & 293                          & 134533                        \\
90      & 869                          & 824                           \\

\end{tabular}
\end{table}

\begin{table}[!htp]
\begin{tabular}{lrr}
        & \multicolumn{2}{c}{User engagements of news on social media} \\
News ID & \multicolumn{1}{c}{True}     & \multicolumn{1}{r}{False}     \\
91      & 3909                         & 3523                          \\
92      & 1290                         & 843                           \\
93      & 167                          & 850                           \\
94      & 287                          & 147212                        \\
95      & 1795                         & 429                           \\
96      & 3599                         & 1986                          \\
97      & 1323                         & 539                           \\
98      & 3692                         & 4200                          \\
99      & 4676                         & 352                           \\
100     & 2630                         & 177                           \\
101     & 147                          & 16927                         \\
102     & 260                          & 4458                          \\
103     & 3681                         & 7366                          \\
104     & 260                          & 26835                         \\
105     & 2000                         & 20956                         \\
106     & 597                          & 4344                          \\
107     & 2421                         & 3262                          \\
108     & 2147                         & 8195                          \\
109     & 1721                         & 111156                        \\
110     & 6070                         & 20902                         \\
111     & 7054                         & 9849                          \\
112     & 10205                        & 1323                          \\
113     & 18683                        & 401824                        \\
114     & 3528                         & 2085                          \\
115     & 15683                        & 491423                        \\
116     & 3528                         & 2085                          \\
117     & 7950                         & 97679                         \\
118     & 8807                         & 143400                        \\
119     & 8081                         & 68900                         \\
120     & 5616                         & 1172                          \\
121     & 1705                         & 20000                         \\
122     & 6196                         & 12354                         \\
123     & 4450                         & 3017                          \\
124     & 4155                         & 5469                          \\
125     & 10914                        & 26210                         \\
126     & 390                          & 4668                          \\
127     & 22                           & 126180                        \\
128     & 60                           & 985                           \\
129     & 59                           & 75424                         \\
130     & 390                          & 4668                          \\
131     & 2110                         & 836                           \\
132     & 3136                         & 2953                          \\
133     & 297                          & 1917                          \\
134     & 1270                         & 70655                         \\
        &                              &                               \\
        & 2729 (average)               & 191316 (average)                       \\
        & 1587.5 (median)                   & 4461  (median)                       
\end{tabular}
\end{table}

\clearpage

\section{{Single diffusion supplementary results}}
\clearpage

\begin{table}
  \caption{$p$ values of information diffusion on 50 dense and sparse ER random graph}
  \label{table:table3}
  \begin{tabular}{*{5}{p{.165\linewidth}}}
    \toprule
     Centrality & iteration (dense) & sum of $P_I$ (dense) & iteration (sparse) & sum of $P_I$ (sparse)\\
    \midrule
        Degree & $6.535923*10^{-6}$ & $3.778465*10^{-10}$ & $0.000008$ & $3.778465*10^{-10}$ \\
        Eigenvector	& $6.535923*10^{-6}$ & $3.778465*10^{-10}$ & $0.000002$ & $3.778465*10^{-10}$\\
        Closeness & $2.866516*10^{-7}$ & $3.778465*10^{-10}$ & $0.000004$ & $3.778465*10^{-10}$\\
        Betweenness	& $1.868991*10^{-5}$ & $3.778465*10^{-10}$ & $0.000006$ & $3.778465*10^{-10}$\\
        Page rank & $3.167124*10^{-5}$ & $3.778465*10^{-10}$ & $0.000008$ & $3.778465*10^{-10}$\\
  \bottomrule
\end{tabular}
\end{table}

\begin{figure*}[!htb]
\begin{subfigure}{.35\textwidth}
  \centering
  \includegraphics[width=.9\linewidth]{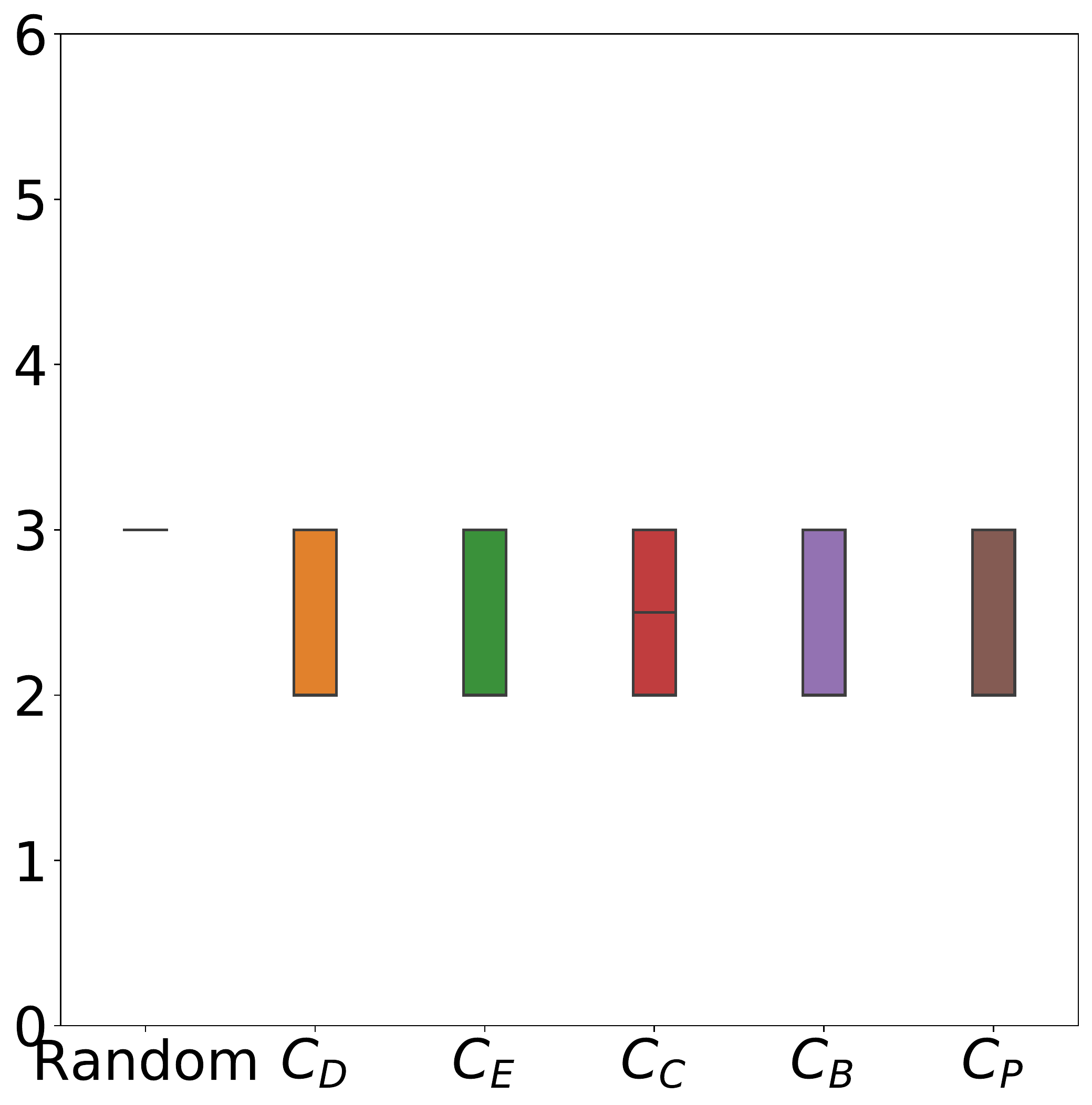}  
  \caption{ }
  \label{fig06_07_08_09:sub-first}
\end{subfigure}
\begin{subfigure}{.35\textwidth}
  \centering
  \includegraphics[width=.9\linewidth]{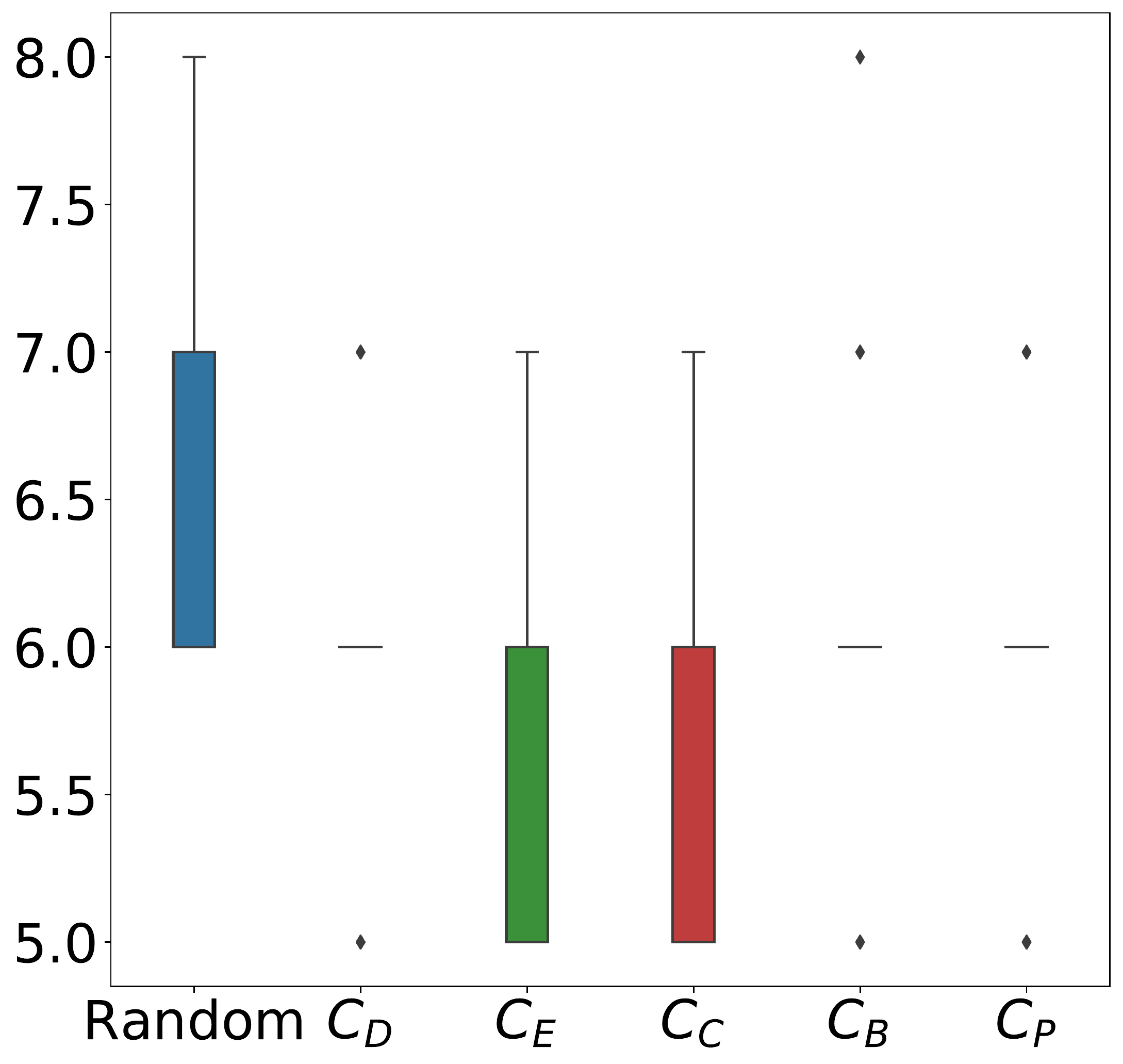}  
  \caption{ }
  \label{fig06_07_08_09:sub-second}
\end{subfigure}


\begin{subfigure}{.35\textwidth}
  \centering
  \includegraphics[width=.9\linewidth]{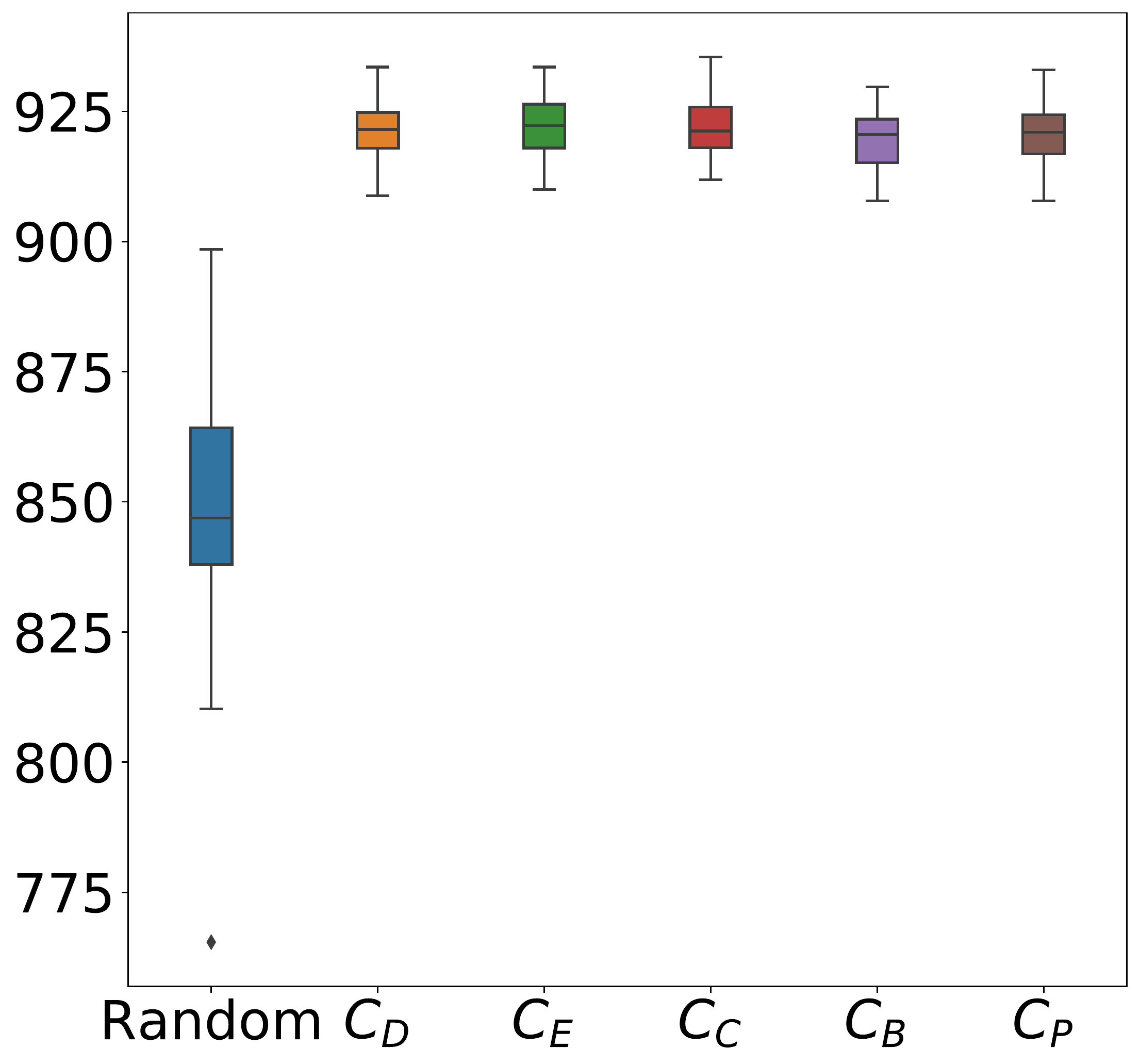}  
  \caption{ }
  \label{fig06_07_08_09:sub-third}
\end{subfigure}
\begin{subfigure}{.35\textwidth}
  \centering
  \includegraphics[width=.9\linewidth]{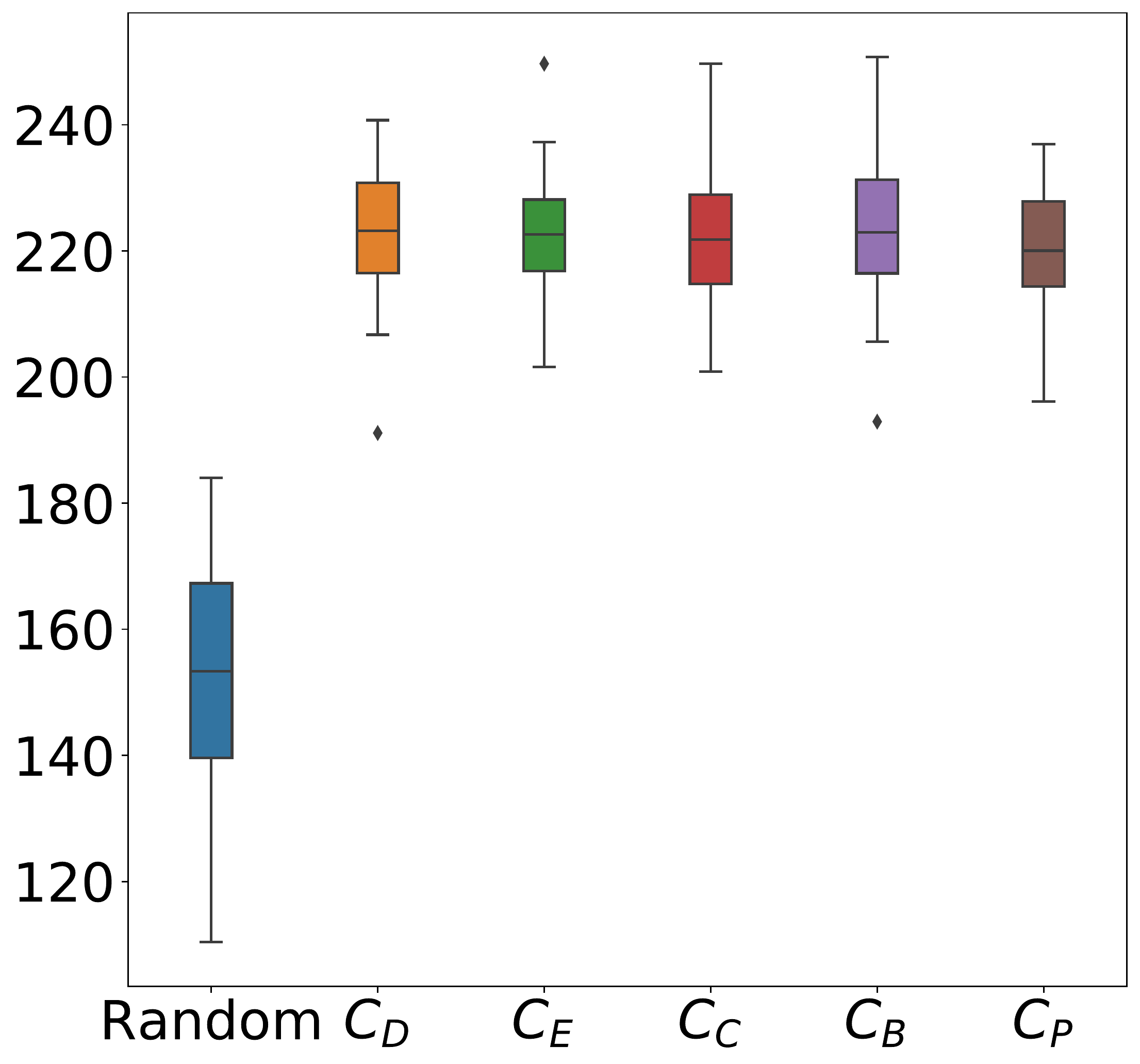}  
  \caption{ }
  \label{fig06_07_08_09:sub-fourth}
\end{subfigure}

\caption{Information diffusion analysis by box plots of (a) iterations on 50 one-community (dense) networks; (b) iterations on 50 one-community (sparse) networks; (c) sum of $P_I$ on 50 one-community (dense) networks; (d) sum of $P_I$ on 50 one-community (sparse) networks. $C_D$: Degree centrality, $C_E$: Eigenvector centrality, $C_C$: Closeness centrality, $C_B$: Betweenness, $C_P$: Page rank.}
\label{fig:fig06_07_08_09}
\end{figure*}

\begin{table}
  \caption{$p$ values of information diffusion on 50 clustered Gaussian random partition graph with similar and varying community size}
  \label{table:table5}
  \begin{tabular}{*{5}{p{.165\linewidth}}}
    \toprule
     Centrality & iteration (similar) & sum of $P_I$ (similar) & iteration (varying) & sum of $P_I$ (varying)\\
    \midrule
    Degree & $6.566666*10^{-9}$ & $3.778465*10^{-10}$ & $2.850821*10^{-6}$ &  $3.778465*10^{-10}$\\
    Eigenvector	& $1.960530*10^{-3}$ & $1.443198*10^{-7}$ & $8.175611*10^{-3}$ &  $3.778465*10^{-10}$\\
    Closeness & $1.207344*10^{-7}$ & $3.778465*10^{-10}$ & $4.915502*10^{-6}$ &  $4.015545*10^{-10}$\\
    Betweenness	& $3.611943*10^{-8}$ & $3.778465*10^{-10}$ & $3.750258*10^{-6}$ &  $4.267113*10^{-10}$\\
    Page rank & $4.405628*10^{-7}$ & $3.778465*10^{-10}$ & $1.782999*10^{-7}$ & $4.267113*10^{-10}$\\
  \bottomrule
\end{tabular}
\end{table}

\begin{figure*}[!htb]
\begin{subfigure}{.35\textwidth}
  \centering
  \includegraphics[width=.9\linewidth]{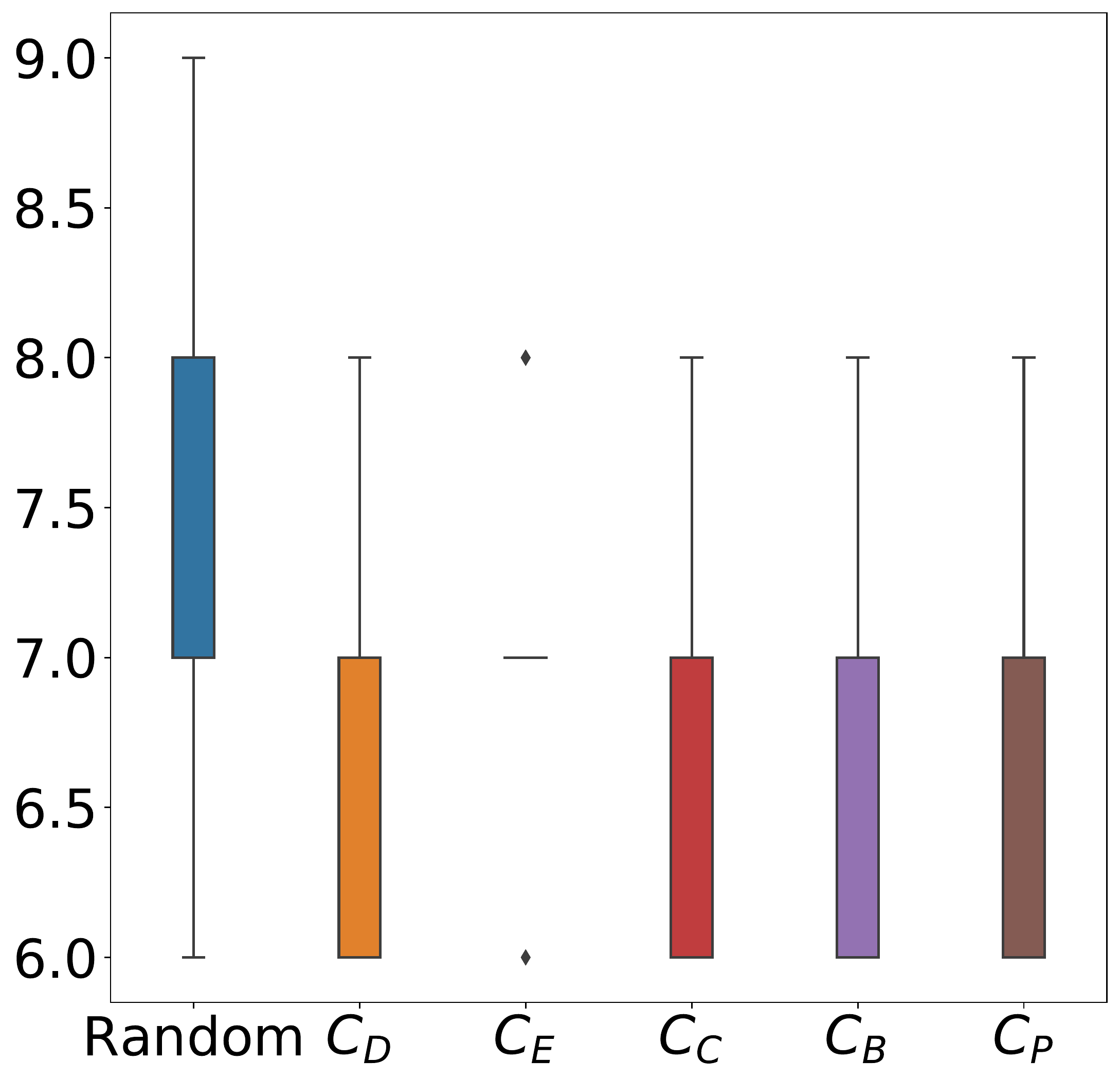}  
  \caption{ }
  \label{fig10_11_12_13:sub-first}
\end{subfigure}
\begin{subfigure}{.35\textwidth}
  \centering
  \includegraphics[width=.9\linewidth]{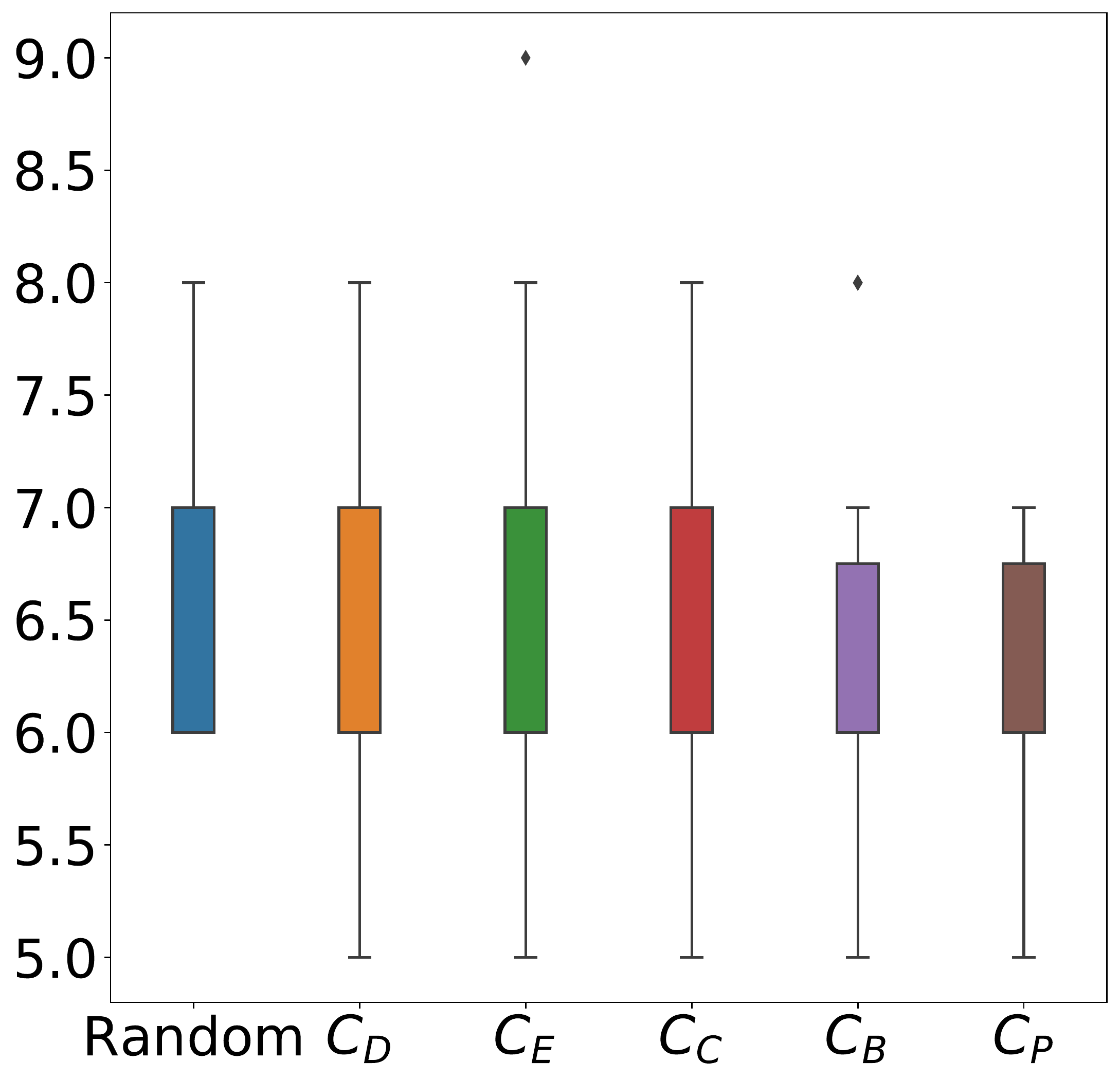}  
  \caption{ }
  \label{fig10_11_12_13:sub-second}
\end{subfigure}


\begin{subfigure}{.35\textwidth}
  \centering
  \includegraphics[width=.9\linewidth]{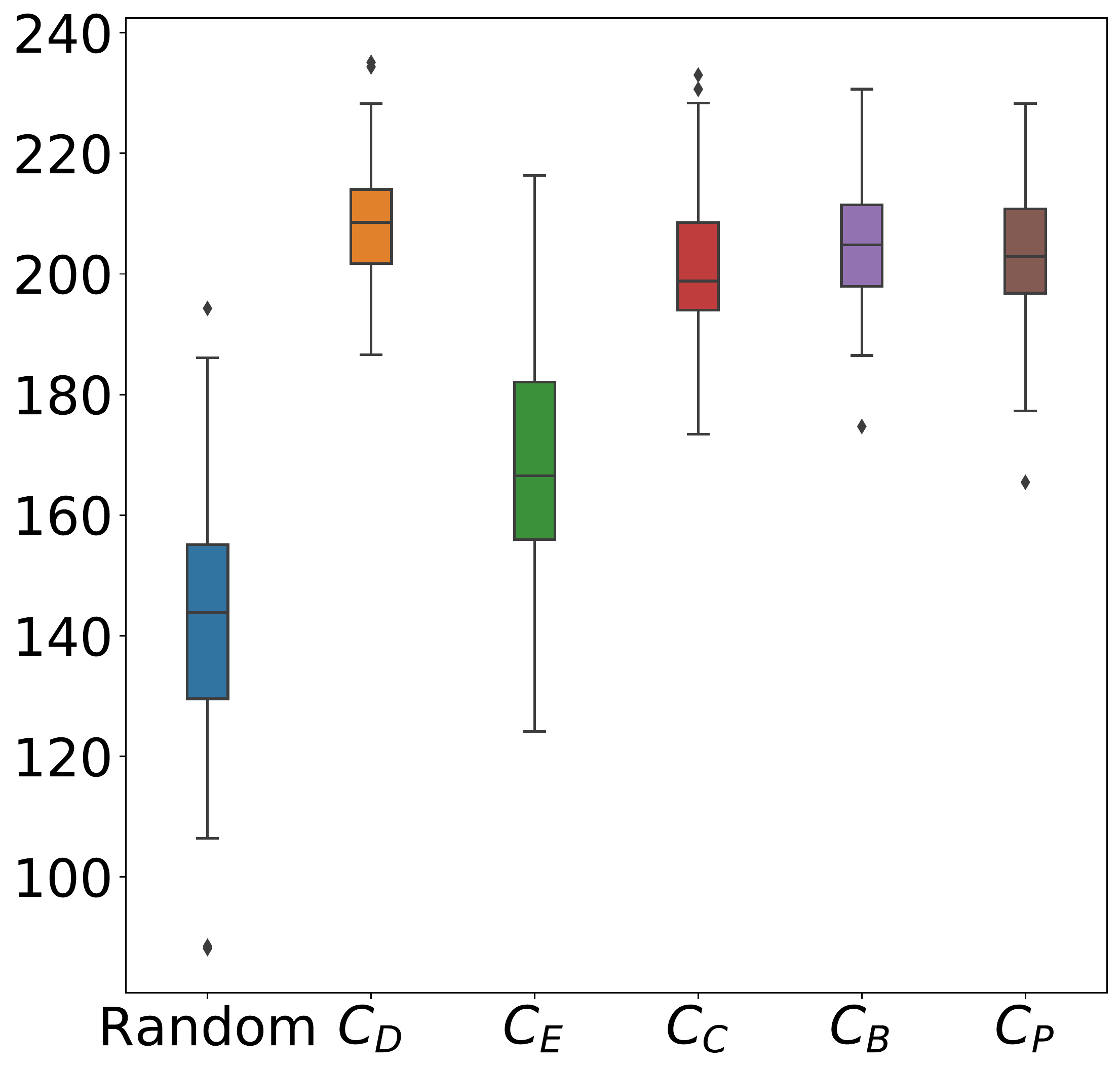}  
  \caption{ }
  \label{fig10_11_12_13:sub-third}
\end{subfigure}
\begin{subfigure}{.35\textwidth}
  \centering
  \includegraphics[width=.9\linewidth]{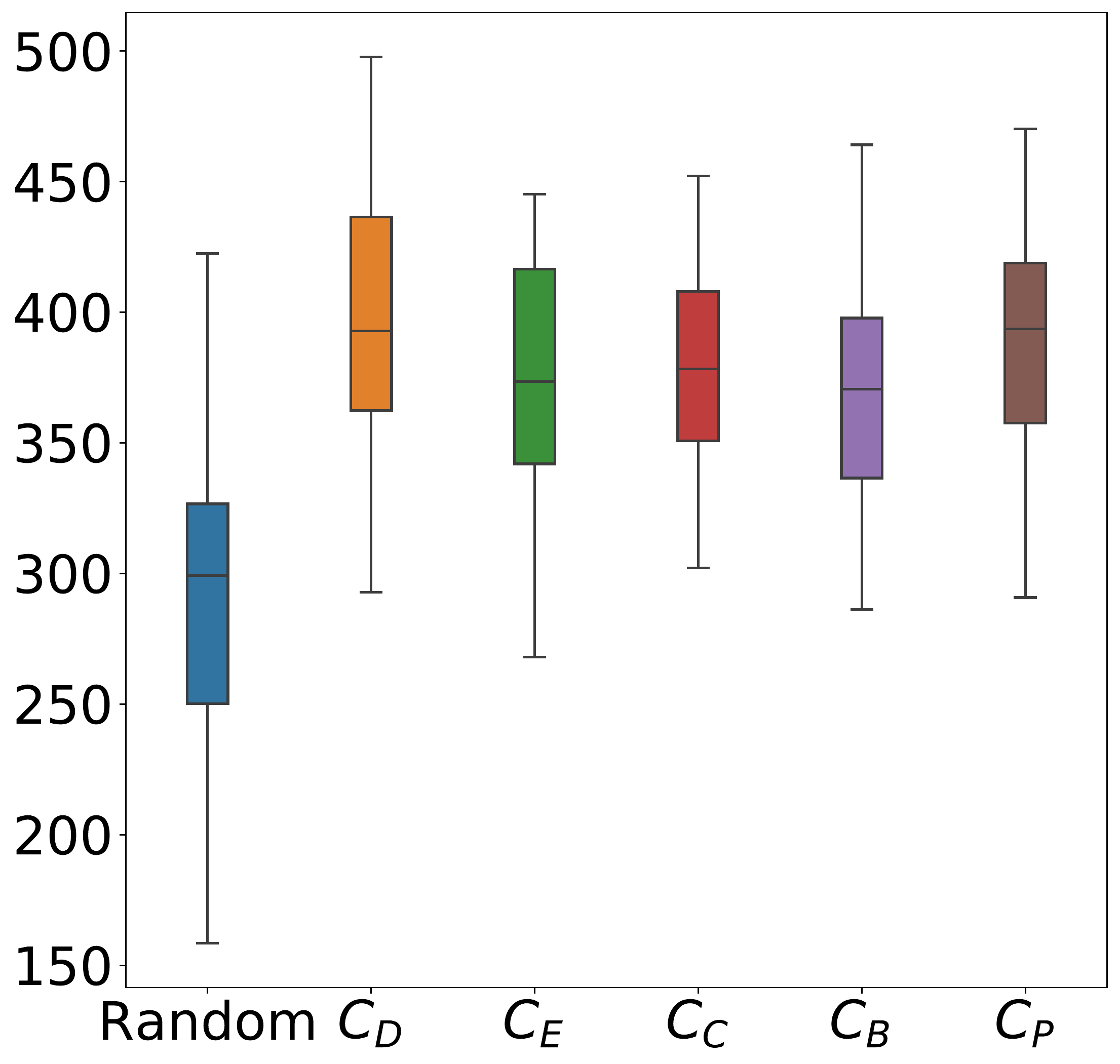}  
  \caption{ }
  \label{fig10_11_12_13:sub-fourth}
\end{subfigure}

\caption{Information diffusion analysis by box plots of (a) iterations on 50 multi-communities networks with similar community size; (b) iterations on 50 multi-communities networks with varying community size; (c) sum of $P_I$ on 50 multi-communities networks with similar community size; (d) sum of $P_I$ on 50 multi-communities networks with varying community size. $C_D$: Degree centrality, $C_E$: Eigenvector centrality, $C_C$: Closeness centrality, $C_B$: Betweenness, $C_P$: Page rank.}
\label{fig:fig10_11_12_13}
\end{figure*}

\begin{figure*}[!htb]

\begin{subfigure}{.35\textwidth}
  \centering
  \includegraphics[width=.9\linewidth]{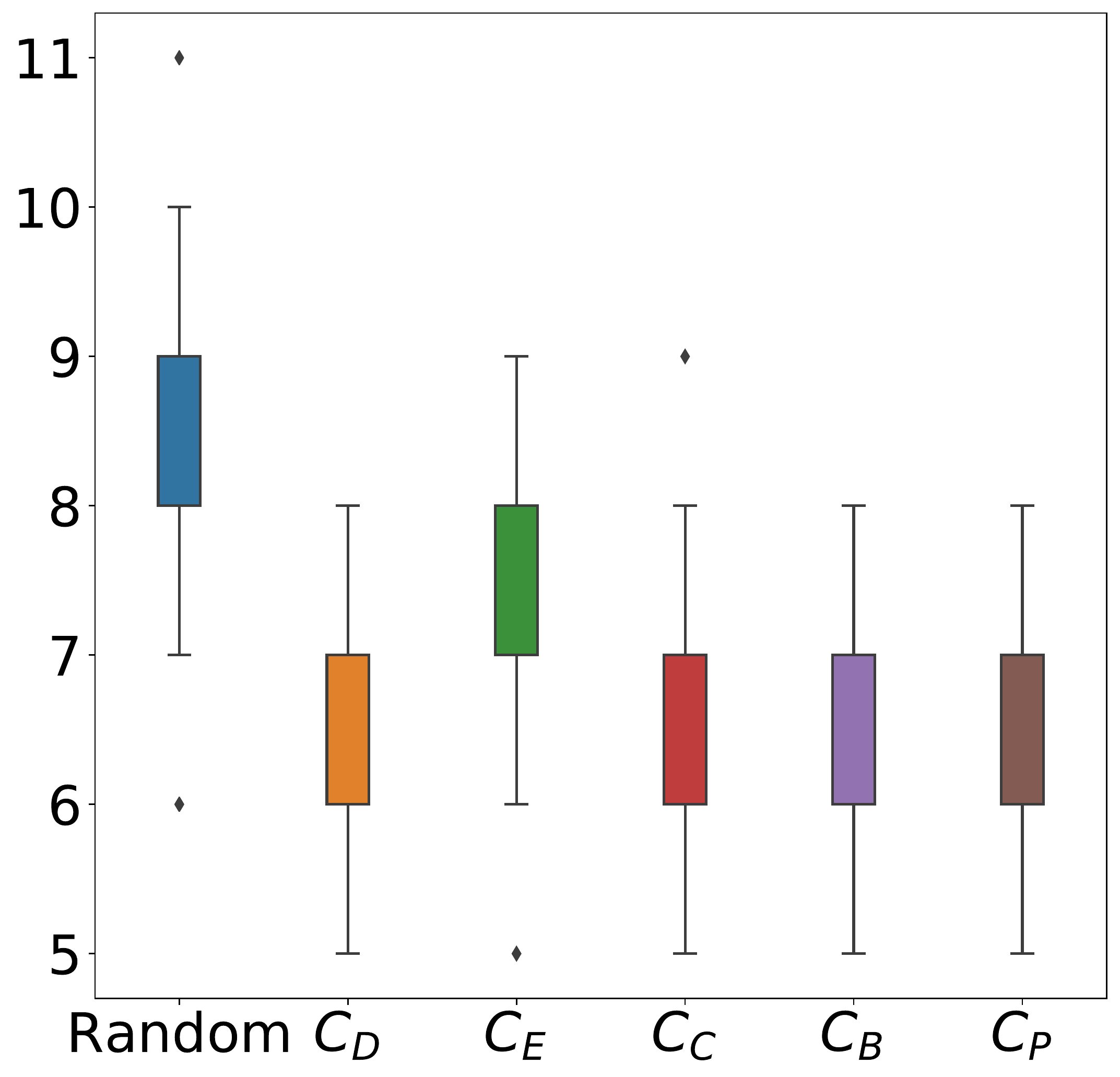}  
  \caption{ }
  \label{figLFRsingle:sub-first}
\end{subfigure}
\begin{subfigure}{.35\textwidth}
  \centering
  \includegraphics[width=.9\linewidth]{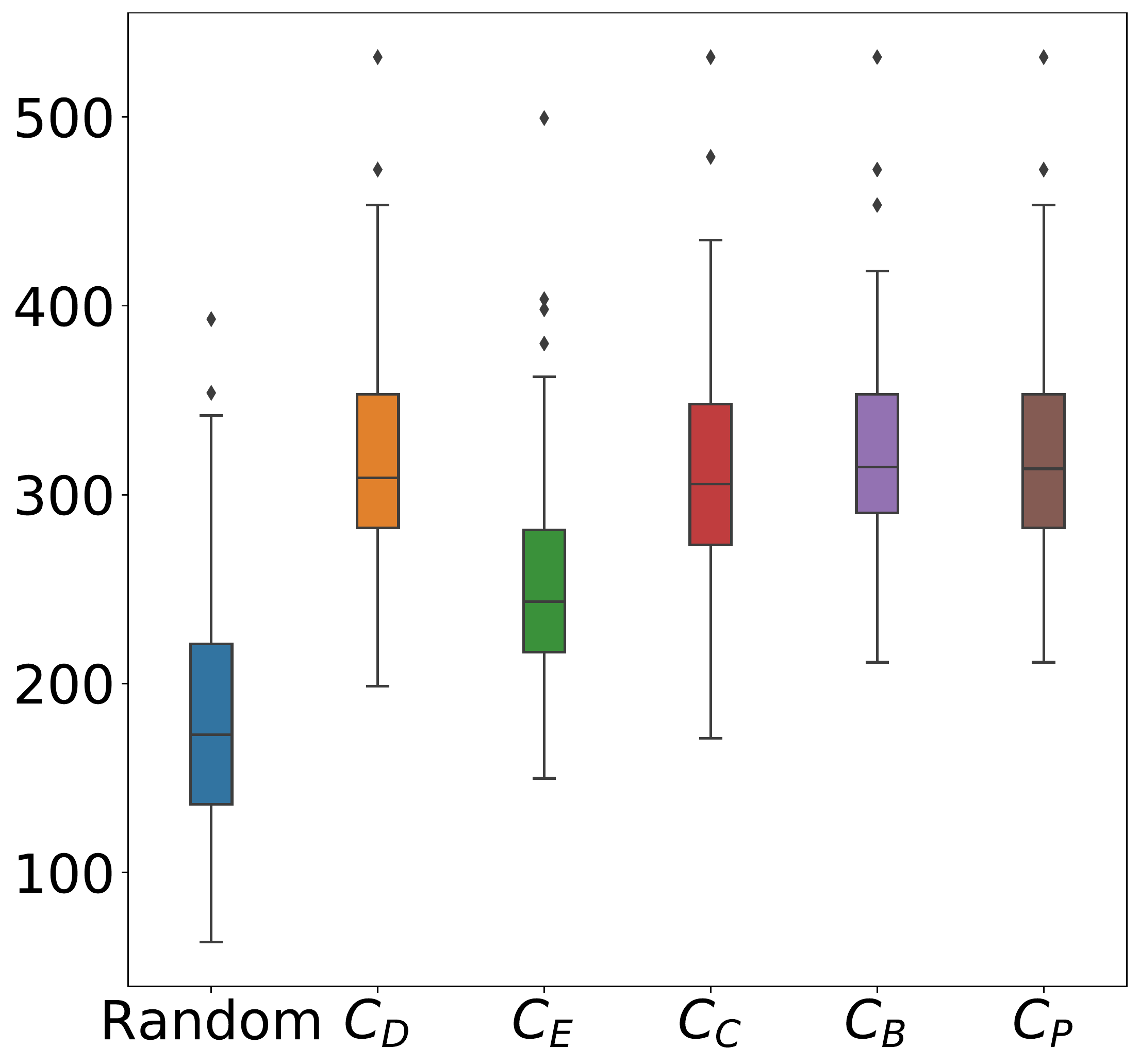}  
  \caption{ }
  \label{figLFRsingle:sub-second}
\end{subfigure}
\caption{Information diffusion analysis by box plots of (a) iterations on 50 LFR benchmark graph; (b) sum of $P_I$ on 50 50 LFR benchmark graph. $C_D$: Degree centrality, $C_E$: Eigenvector centrality, $C_C$: Closeness centrality, $C_B$: Betweenness, $C_P$: Page rank.}
\label{fig:LFRsingle}
\end{figure*}

\begin{table}
  \caption{$p$ values of information diffusion on 50 LFR benchmark graph}
  \label{table:LFRsingle}
  \begin{tabular}{lrr}
    \toprule
     Centrality & iteration & sum of $P_I$ \\
    \midrule
        Degree & $2.820804*10^{-10}$ & $3.778465*10^{-10}$\\
        Eigenvector	& $3.811027*10^{-6}$ & $1.496882*10^{-9}$\\
        Closeness & $5.730597*10^{-10}$ & $3.778465*10^{-10}$\\
        Betweenness	& $1.363267*10^{-8}$ & $3.778465*10^{-10}$\\
        Page rank & $3.104716*10^{-10}$ & $3.778465*10^{-10}$\\
  \bottomrule
\end{tabular}
\end{table}

\clearpage

\section{Combating diffusion supplementary results}
\clearpage

\begin{table}
  \caption{$p$ values of information interventions on 50 dense ER random graph}
  \label{table:table6}
  \begin{tabular}{*{5}{p{.165\linewidth}}}
    \toprule
     Centrality & sum of $P_{IT}$ & "infected" nodes & "susceptible" nodes & "protected" nodes \\
    \midrule
        Degree & $1.048948*10^{-8}$ & $5.347646*10^{-7}$ & $0.992581$ & $0.000002$\\
        Eigenvector	& $2.399432*10^{-7}$	& $2.302321*10^{-6}$	& $0.997225$ & $0.000031$\\
        Closeness & $5.081552*10^{-8}$ & $5.500584*10^{-7}$ & $0.977221$ & $0.000006$\\
        Betweenness	& $6.329399*10^{-9}$ & $2.612913*10^{-7}$ & $0.968295$ & $0.000001$\\
        Page rank & $3.139597*10^{-8}$ & $1.907753*10^{-6}$ & $0.986730$ & $0.000012$\\
  \bottomrule
\end{tabular}
\end{table}

\begin{figure*}[!htb]
\begin{subfigure}{.35\textwidth}
  \centering
  \includegraphics[width=.9\linewidth]{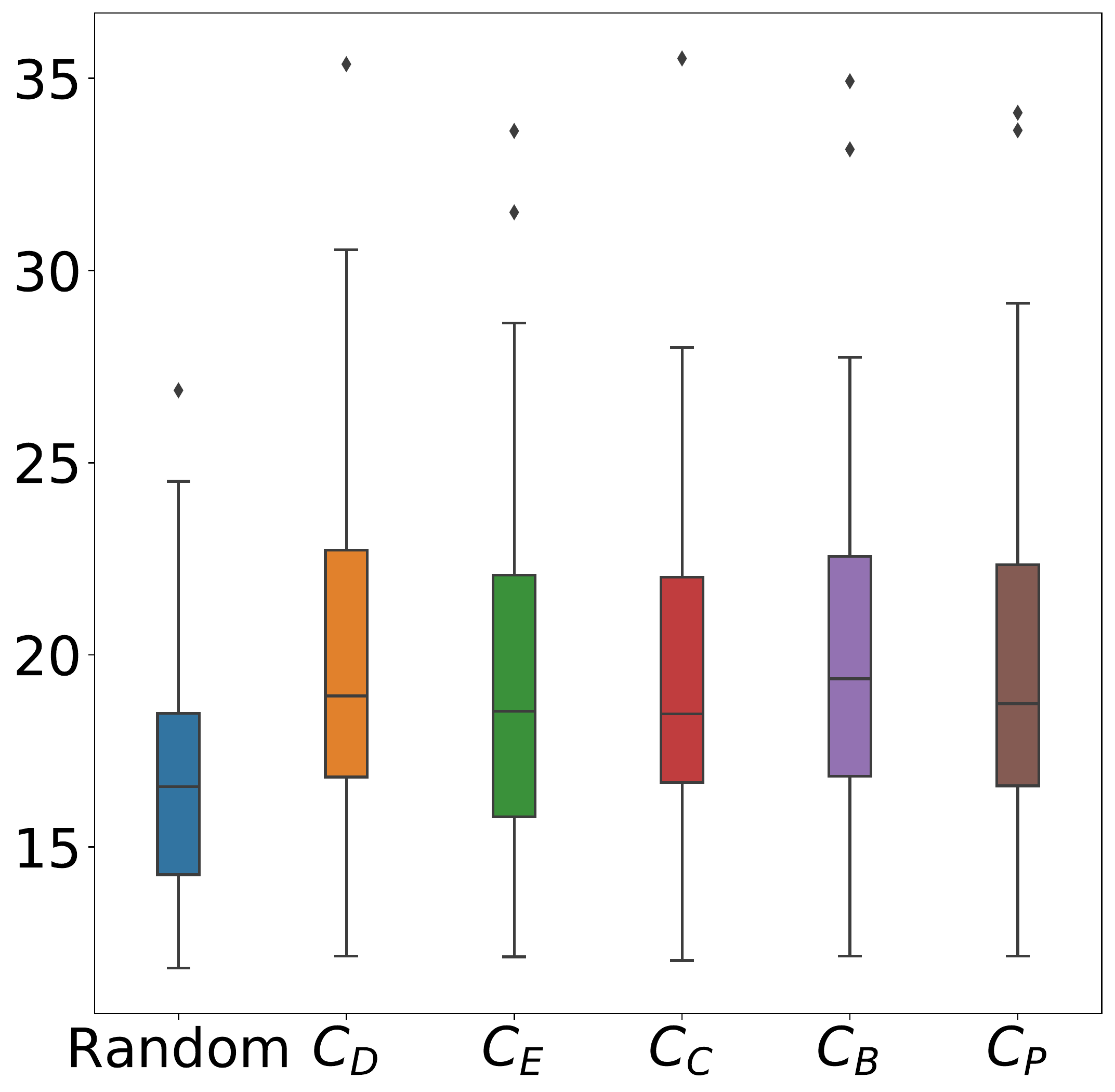}  
  \caption{ }
  \label{fig21_22_23_24:sub-first}
\end{subfigure}
\begin{subfigure}{.35\textwidth}
  \centering
  \includegraphics[width=.9\linewidth]{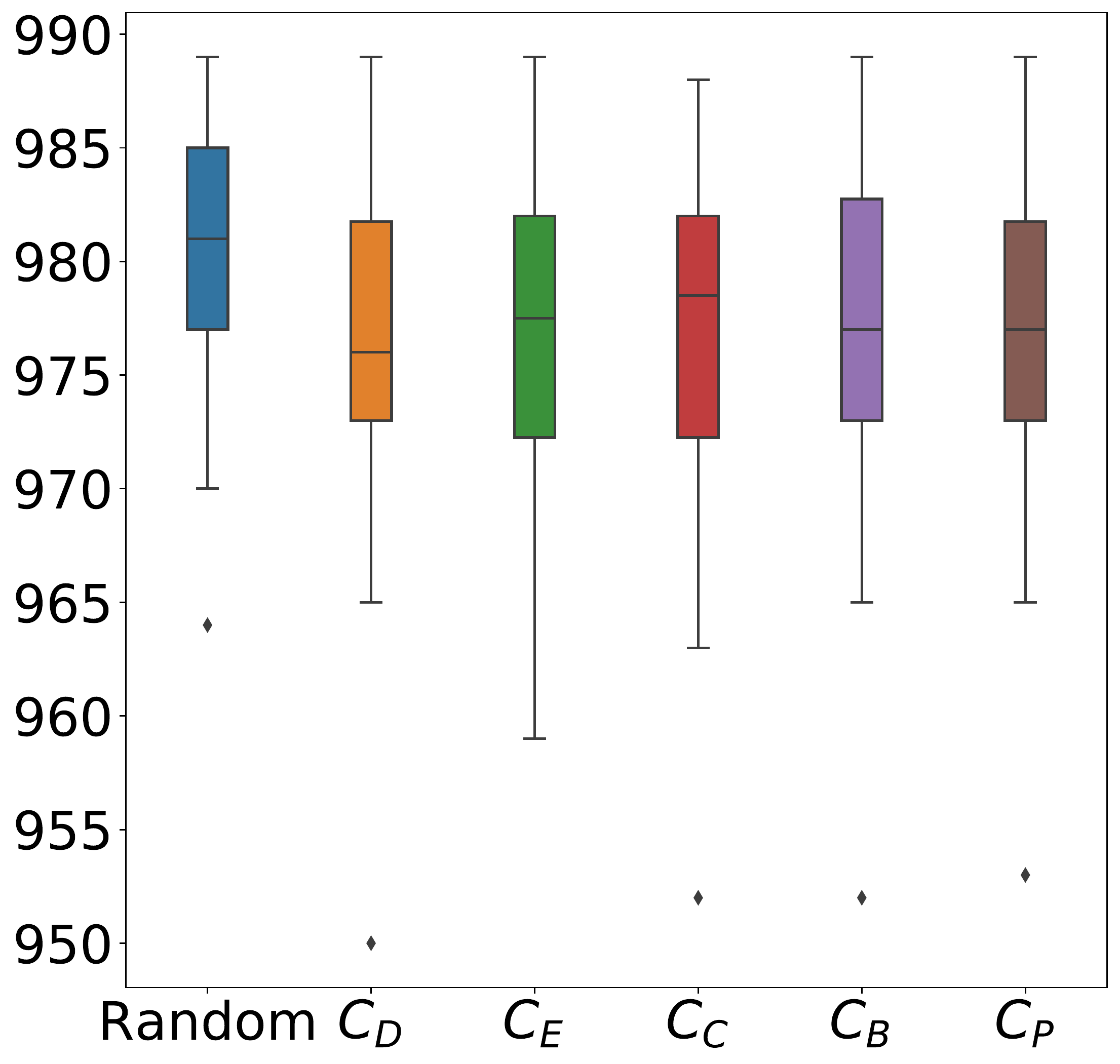}  
  \caption{ }
  \label{fig21_22_23_24:sub-second}
\end{subfigure}


\begin{subfigure}{.35\textwidth}
  \centering
  \includegraphics[width=.9\linewidth]{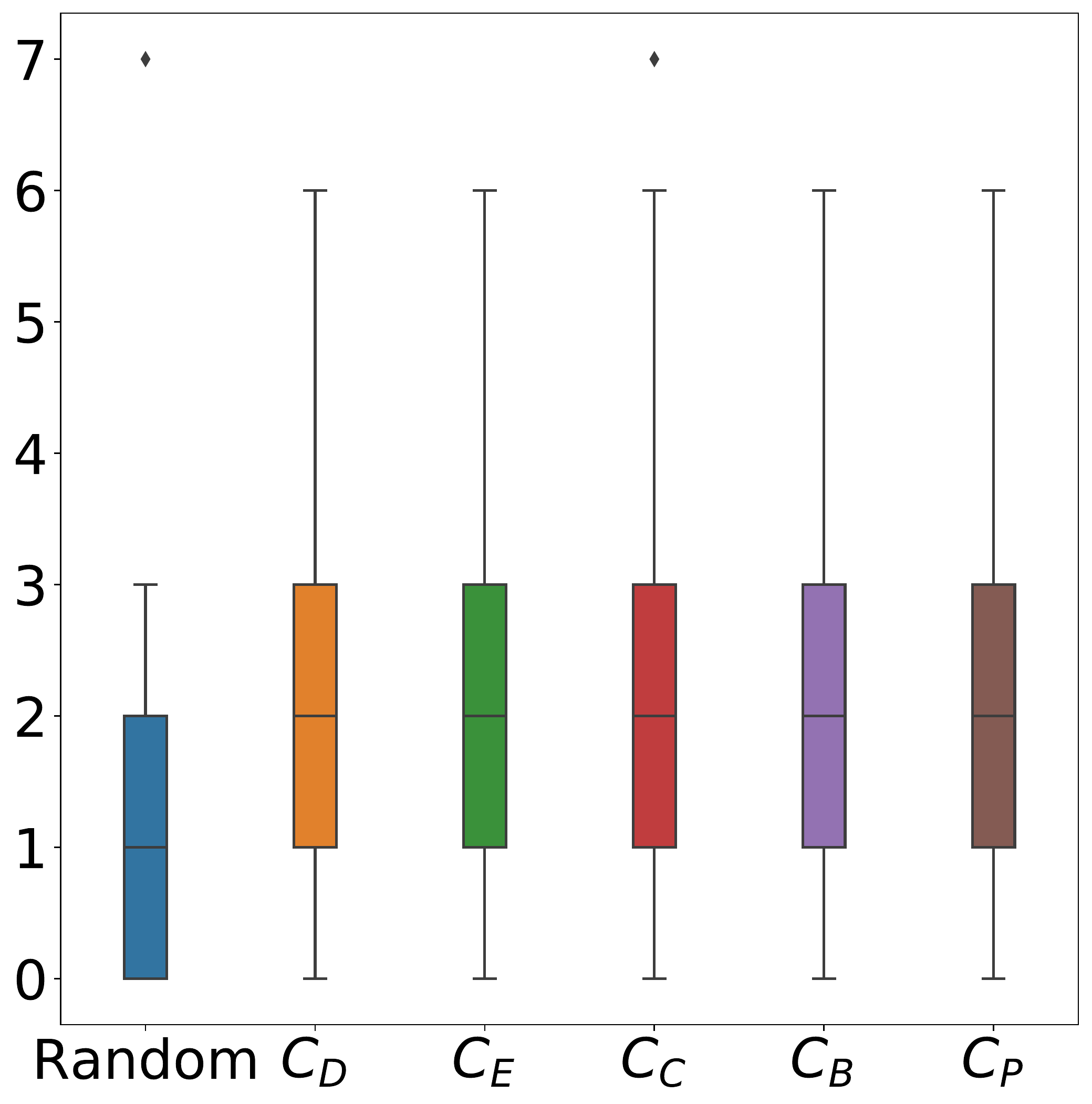}  
  \caption{ }
  \label{fig21_22_23_24:sub-third}
\end{subfigure}
\begin{subfigure}{.35\textwidth}
  \centering
  \includegraphics[width=.9\linewidth]{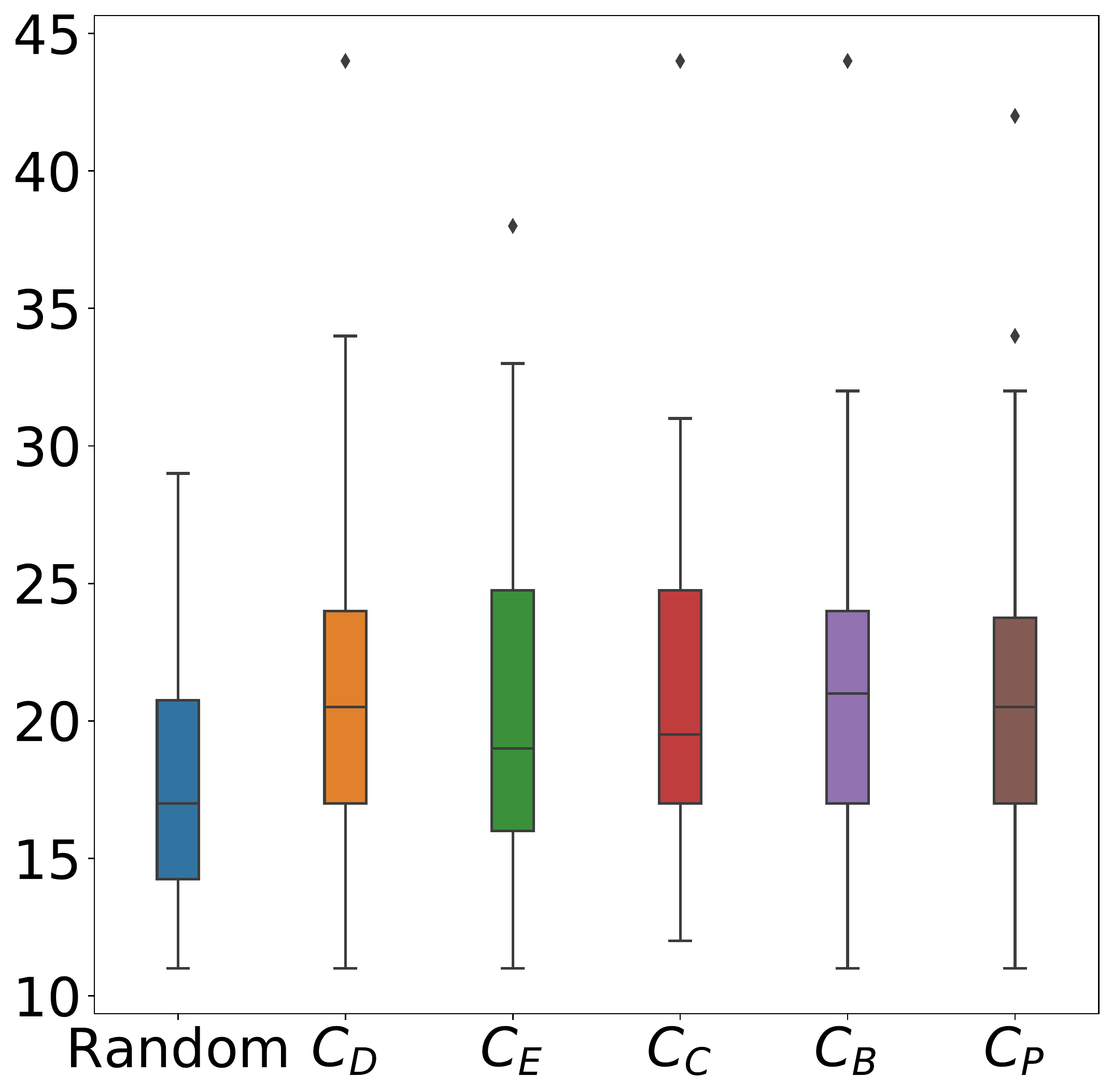}  
  \caption{ }
  \label{fig21_22_23_24:sub-fourth}
\end{subfigure}

\caption{Information interventions analysis on 50 one-community (dense) networks by box plots of (a) sum of $P_{IT}$; (b) number of infected nodes; (c) number of susceptible nodes; (d) number of protected nodes.}
\label{fig:fig21_22_23_24}
\end{figure*}

\begin{table}
  \caption{$p$ values of information interventions on 50 sparse ER random graph}
  \label{table:table7}
  \begin{tabular}{*{5}{p{.165\linewidth}}}
    \toprule
     Centrality & sum of $P_{IT}$ & "infected" nodes & "susceptible" nodes & "protected" nodes \\
    \midrule
        Degree & $3.778465*10^{-10}$ & $3.768225*10^{-10}$ & $3.772488*10^{-10}$ & $3.770782*10^{-10}$\\
        Eigenvector	& $3.778465*10^{-10}$ & $3.772488*10^{-10}$ & $3.771635*10^{-10}$ & $3.760561*10^{-10}$\\
        Closeness & $3.778465*10^{-10}$ & $3.769930*10^{-10}$ & $3.767372*10^{-10}$ & $3.772488*10^{-10}$\\
        Betweenness	& $3.778465*10^{-10}$ & $3.764817*10^{-10}$ & $3.769930*10^{-10}$ & $3.770782*10^{-10}$\\
        Page rank & $3.778465*10^{-10}$ & $3.764817*10^{-10}$ & $3.773342*10^{-10}$ & $3.772488*10^{-10}$\\
        
  \bottomrule
\end{tabular}
\end{table}

\begin{figure*}[!htb]
\begin{subfigure}{.35\textwidth}
  \centering
  \includegraphics[width=.9\linewidth]{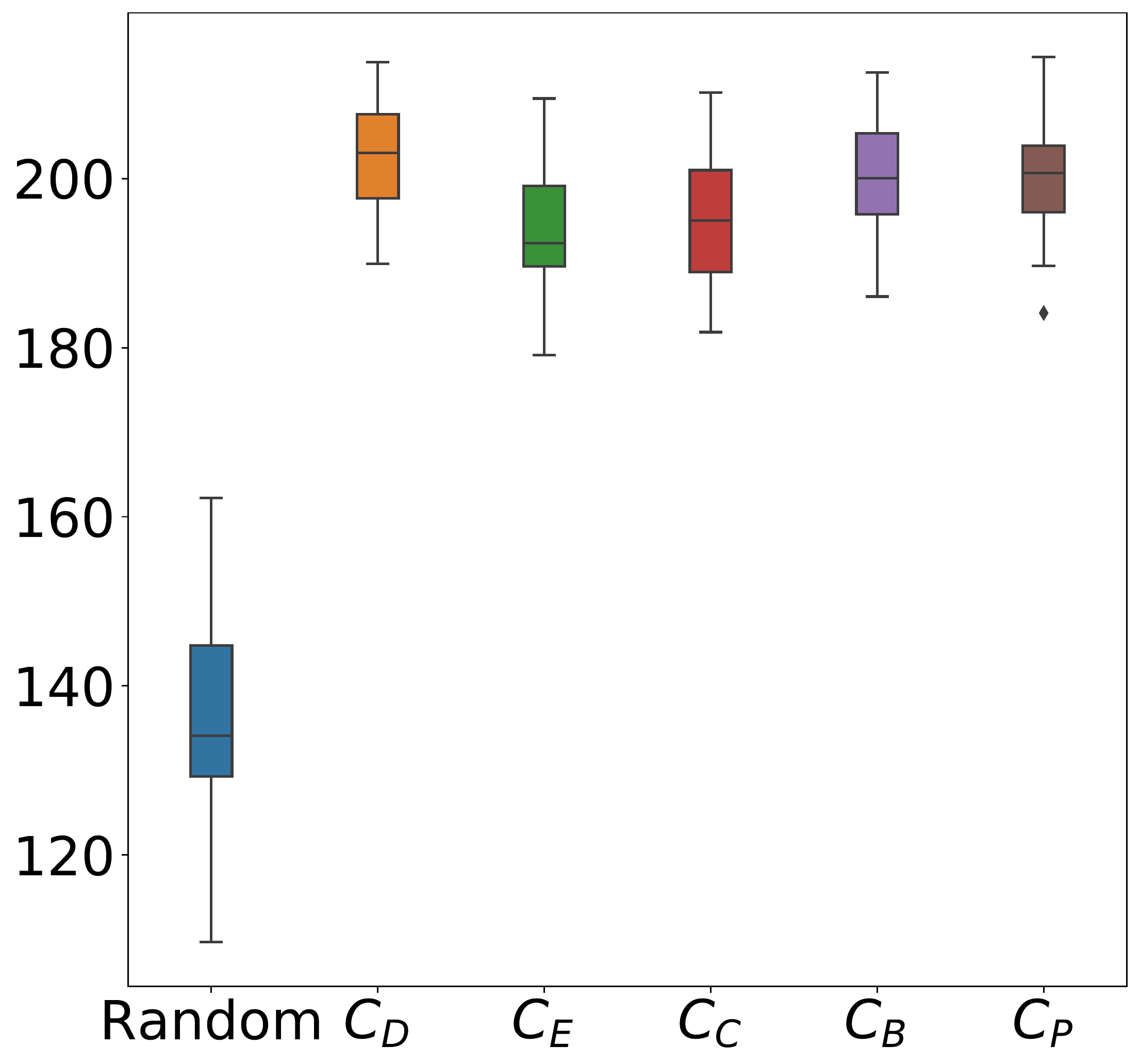}  
  \caption{ }
  \label{fig25_26_27_28:sub-first}
\end{subfigure}
\begin{subfigure}{.35\textwidth}
  \centering
  \includegraphics[width=.9\linewidth]{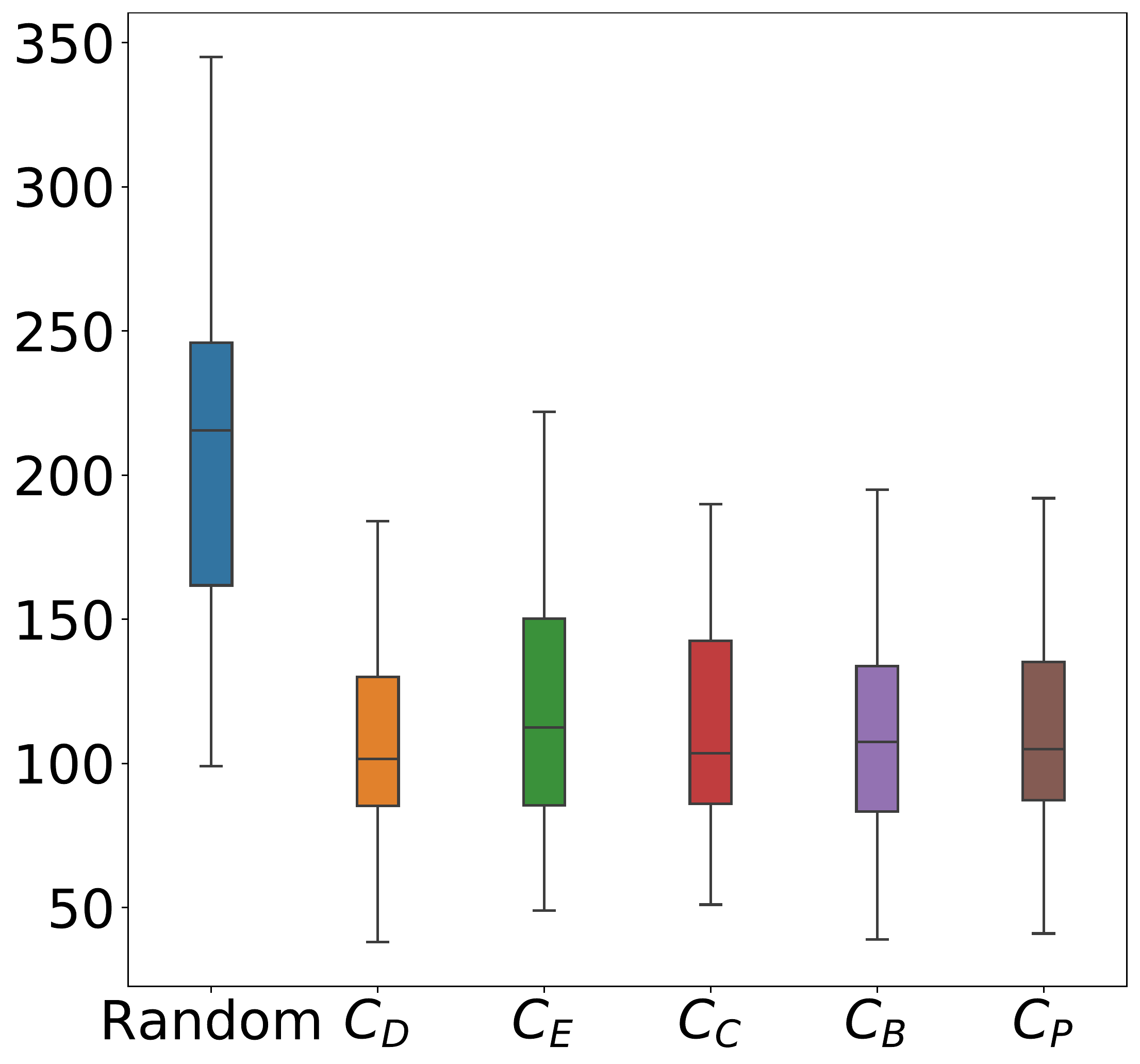}  
  \caption{ }
  \label{fig25_26_27_28:sub-second}
\end{subfigure}


\begin{subfigure}{.35\textwidth}
  \centering
  \includegraphics[width=.9\linewidth]{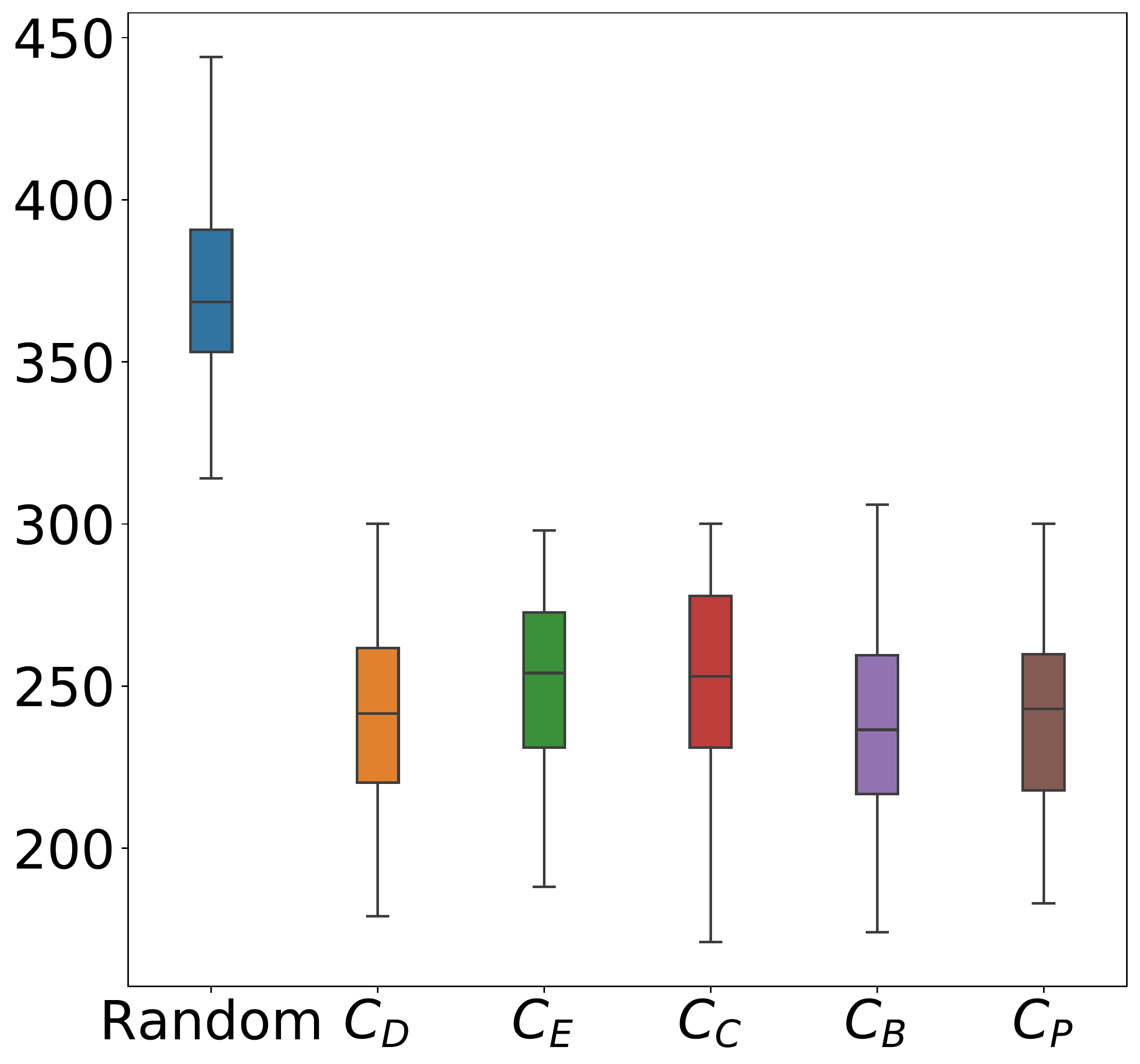}  
  \caption{ }
  \label{fig25_26_27_28:sub-third}
\end{subfigure}
\begin{subfigure}{.35\textwidth}
  \centering
  \includegraphics[width=.9\linewidth]{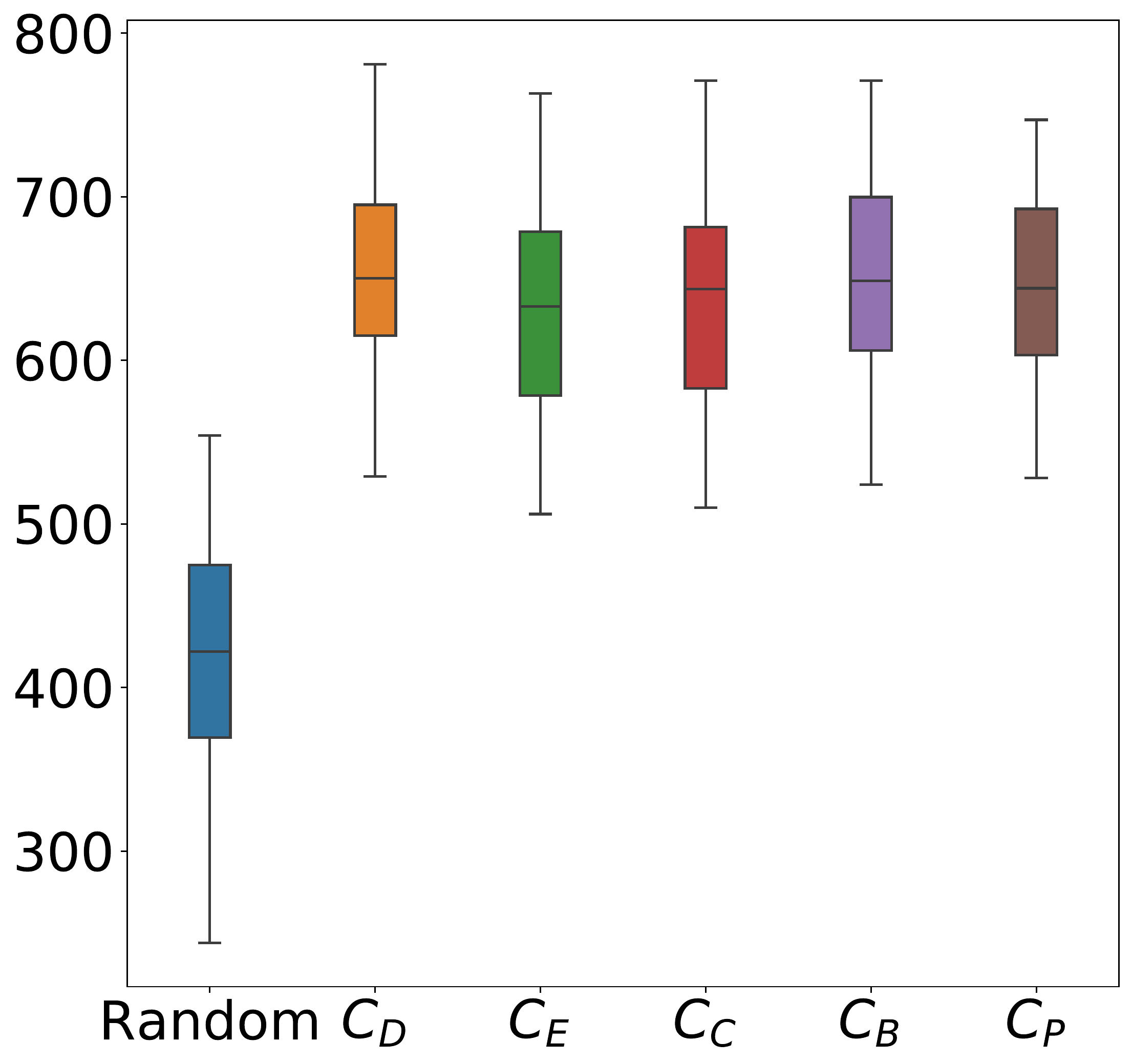}  
  \caption{ }
  \label{fig25_26_27_28:sub-fourth}
\end{subfigure}

\caption{Intervention analysis on 50 one-community networks (sparse) by box plots of (a) sum of $P_{IT}$; (b) number of infected nodes; (c) number of susceptible nodes; (d) number of protected nodes.}
\label{fig:fig25_26_27_28}
\end{figure*}

\begin{table}
  \caption{$p$ values of information interventions on 50 Gaussian random partition graph with similar community size}
  \label{table:table8}
  \begin{tabular}{*{5}{p{.165\linewidth}}}
    \toprule
     Centrality & sum of $P_{IT}$ & "infected" nodes & "susceptible" nodes & "protected" nodes \\
    \midrule
        Degree & $3.778465*10^{-9}$ & $3.770782*10^{-10}$ & $3.763114*10^{-10}$ & $3.775049*10^{-10}$\\
        Eigenvector	& $9.964106*10^{-1}$ & $9.990294*10^{-1}$ & $9.999458*10^{-1}$ & $9.999414*10^{-1}$\\
        Closeness & $4.015545*10^{-10}$ & $3.763965*10^{-10}$ & $4.006503*10^{-10}$ & $4.009214*10^{-10}$\\
        Betweenness	& $3.778465*10^{-10}$ & $4.007406*10^{-10}$ &	$3.769077*10^{-10}$ & $3.774195*10^{-10}$\\
        Page rank & $3.778465*10^{-10}$ & $3.769930*10{-10}$	& $3.775049*10{-10}$ & $3.771635*10^{-10}$\\
        
  \bottomrule
\end{tabular}
\end{table}

\begin{figure*}[!htb]
\begin{subfigure}{.35\textwidth}
  \centering
  \includegraphics[width=.9\linewidth]{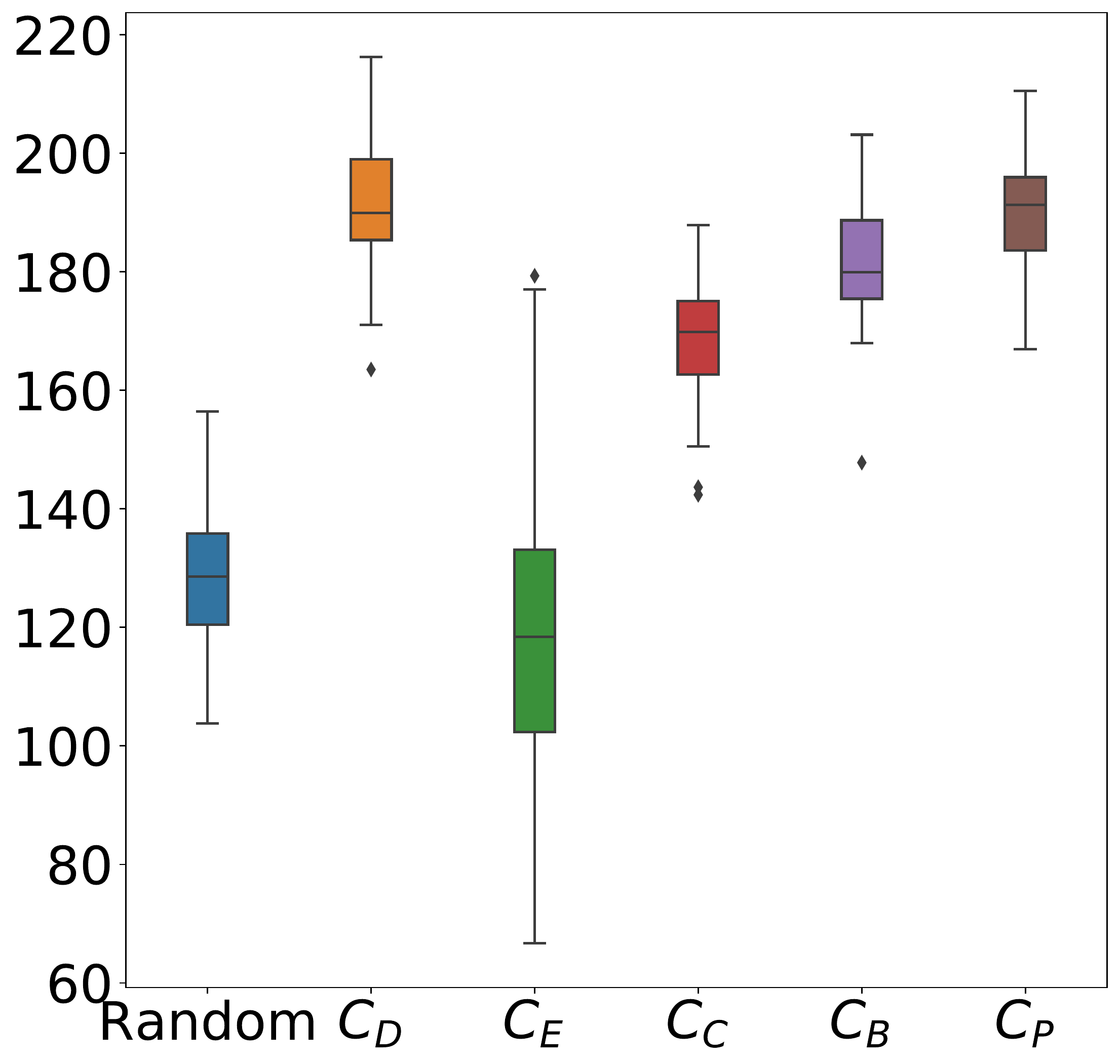}  
  \caption{ }
  \label{fig29_30_31_32:sub-first}
\end{subfigure}
\begin{subfigure}{.35\textwidth}
  \centering
  \includegraphics[width=.9\linewidth]{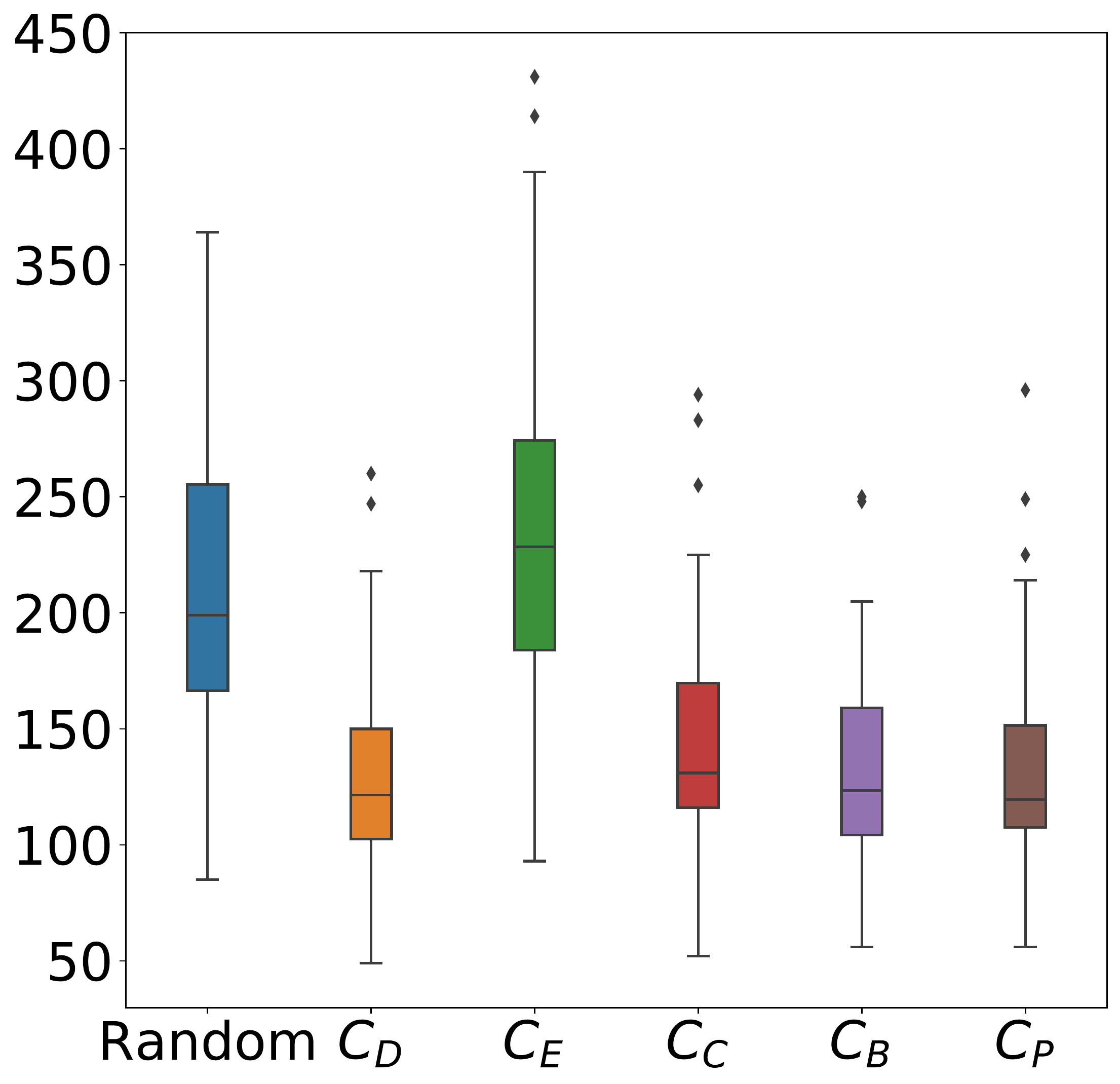}  
  \caption{ }
  \label{fig29_30_31_32:sub-second}
\end{subfigure}


\begin{subfigure}{.35\textwidth}
  \centering
  \includegraphics[width=.9\linewidth]{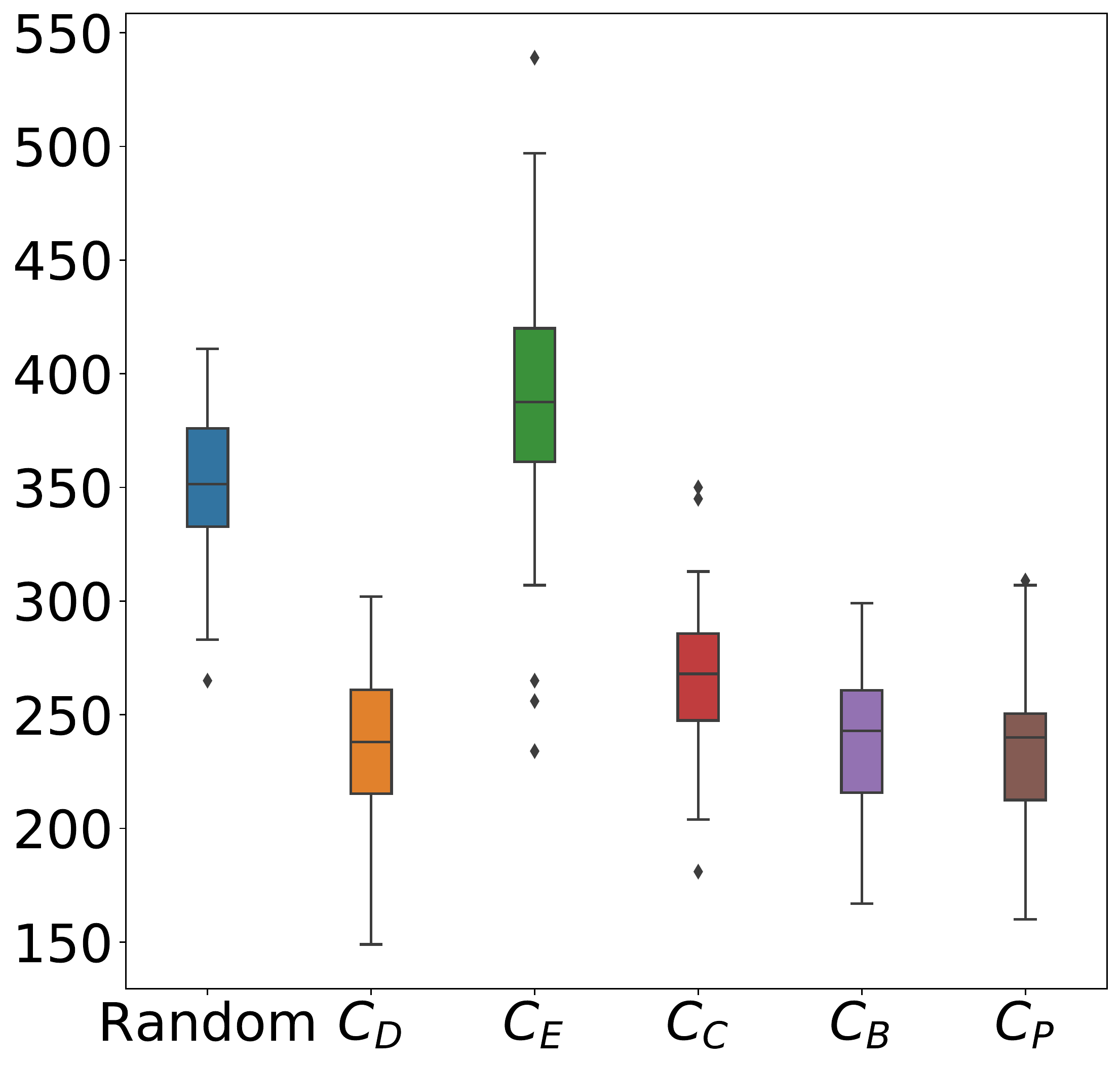}  
  \caption{ }
  \label{fig29_30_31_32:sub-third}
\end{subfigure}
\begin{subfigure}{.35\textwidth}
  \centering
  \includegraphics[width=.9\linewidth]{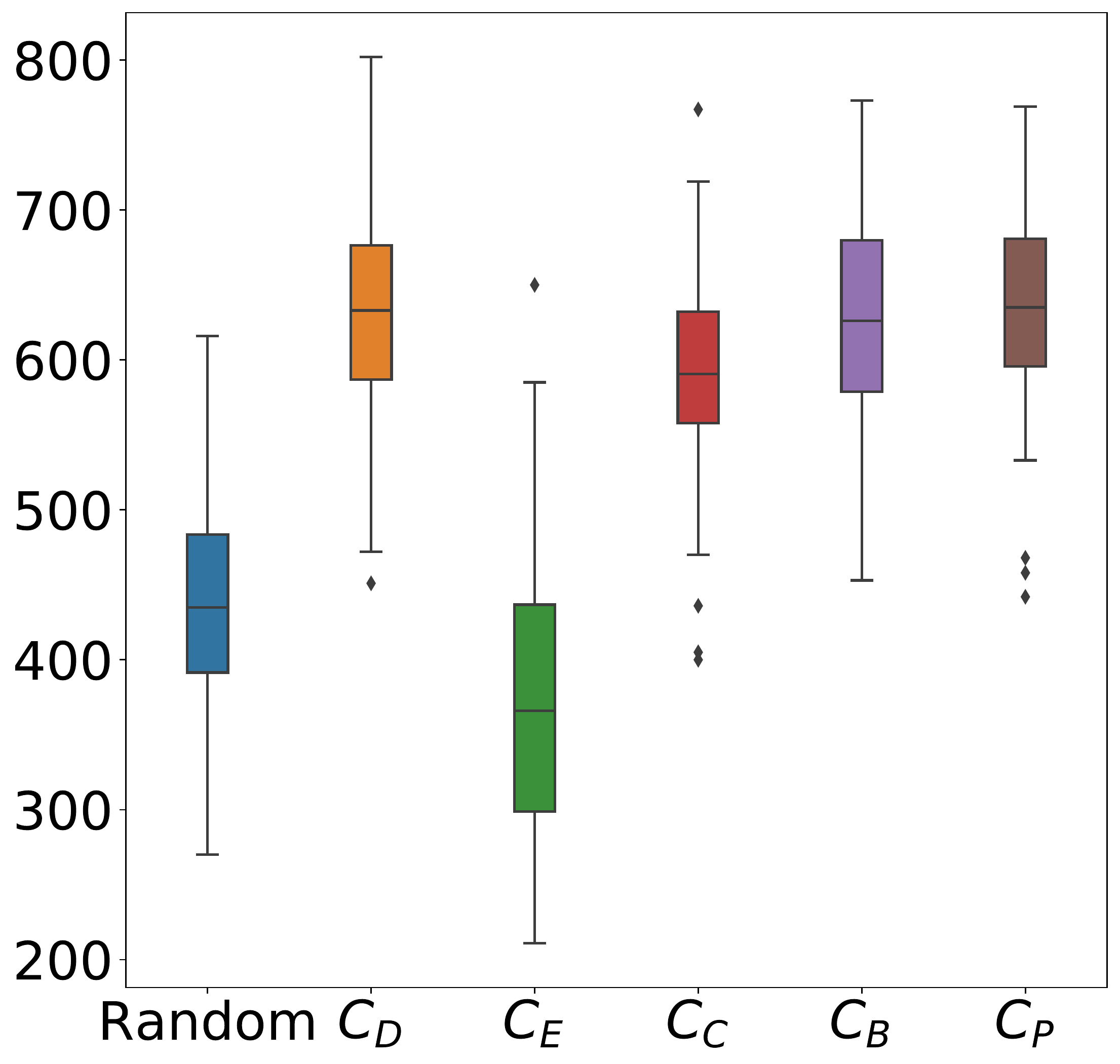}  
  \caption{ }
  \label{fig29_30_31_32:sub-fourth}
\end{subfigure}

\caption{Intervention analysis on 50 multi-communities networks with similar community size by box plots of (a) sum of $P_{IT}$; (b) number of infected nodes; (c) number of susceptible nodes; (d) number of protected nodes.}
\label{fig:fig29_30_31_32}
\end{figure*}

\begin{table}[!htb]
  \caption{$p$ values of information interventions on 50 Gaussian random partition graph with varying community size}
  \label{table:table9}
  \begin{tabular}{*{5}{p{.165\linewidth}}}
    \toprule
     Centrality & sum of $P_{IT}$ & "infected" nodes & "susceptible" nodes & "protected" nodes \\
    \midrule
        Degree & $5.294732*10^{-2}$ & $9.021268*10^{-1}$ & $0.999187$ & $9.976993*10^{-1}$\\
        Eigenvector	& $9.999984*10^{-1}$  & $1.000000$ & $0.999988$ & $1.000000$\\
        Closeness & $1.938026*10^{-1}$  & $2.988948*10^{-1}$   & $0.998233$ & $9.775570*10^{-1}$\\
        Betweenness	& $7.929826*10^{-9}$  & $2.830820*10^{-8}$   & $0.000146$ & $4.311838*10^{-9}$\\
        Page rank  & $3.778465*10^{-10}$ & $4.622548*10^{-10}$   & $0.000002$ & $4.495575*10^{-10}$\\
        
  \bottomrule
\end{tabular}
\end{table}

\begin{figure*}[!htb]
\begin{subfigure}{.35\textwidth}
  \centering
  \includegraphics[width=.9\linewidth]{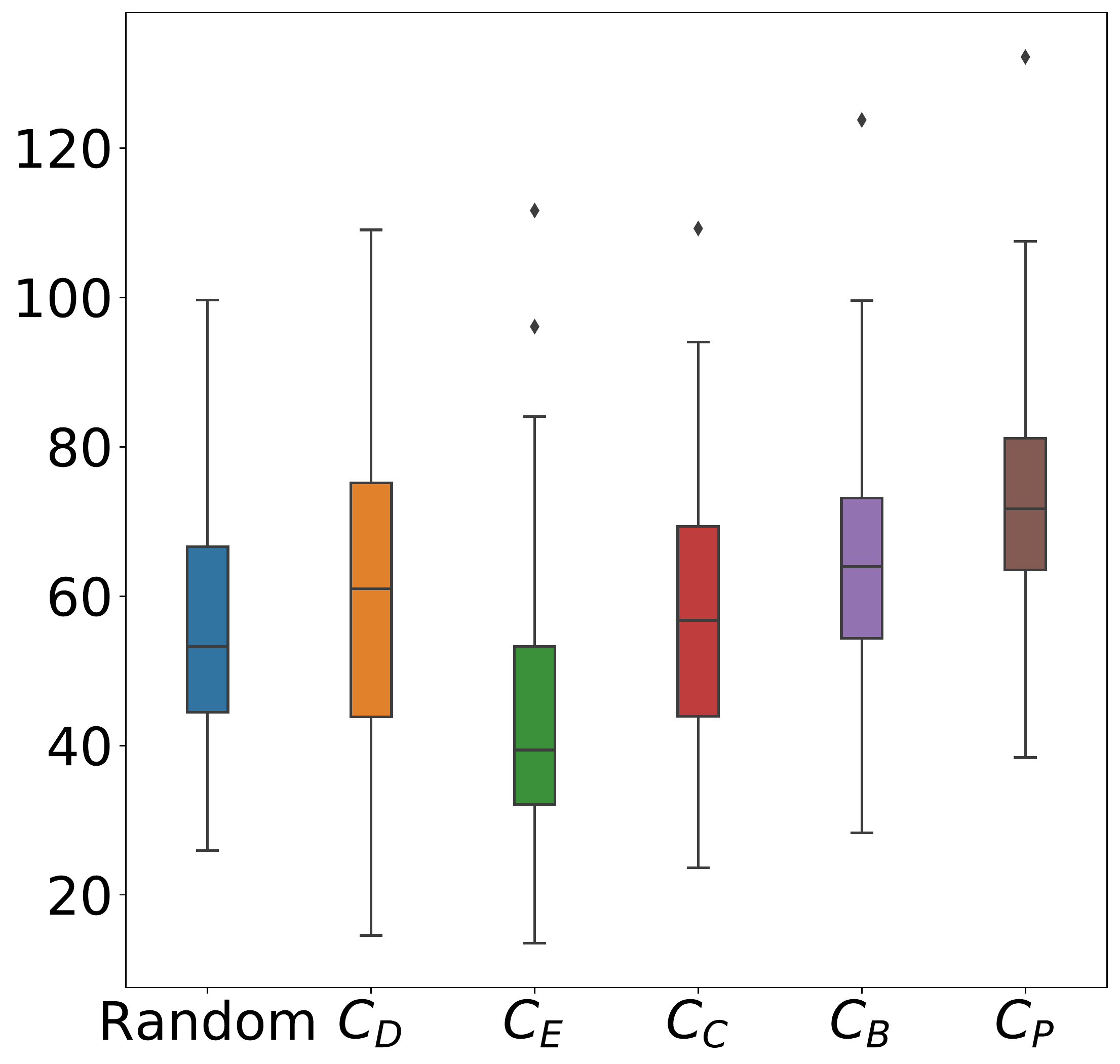}  
  \caption{ }
  \label{fig33_34_35_36:sub-first}
\end{subfigure}
\begin{subfigure}{.35\textwidth}
  \centering
  \includegraphics[width=.9\linewidth]{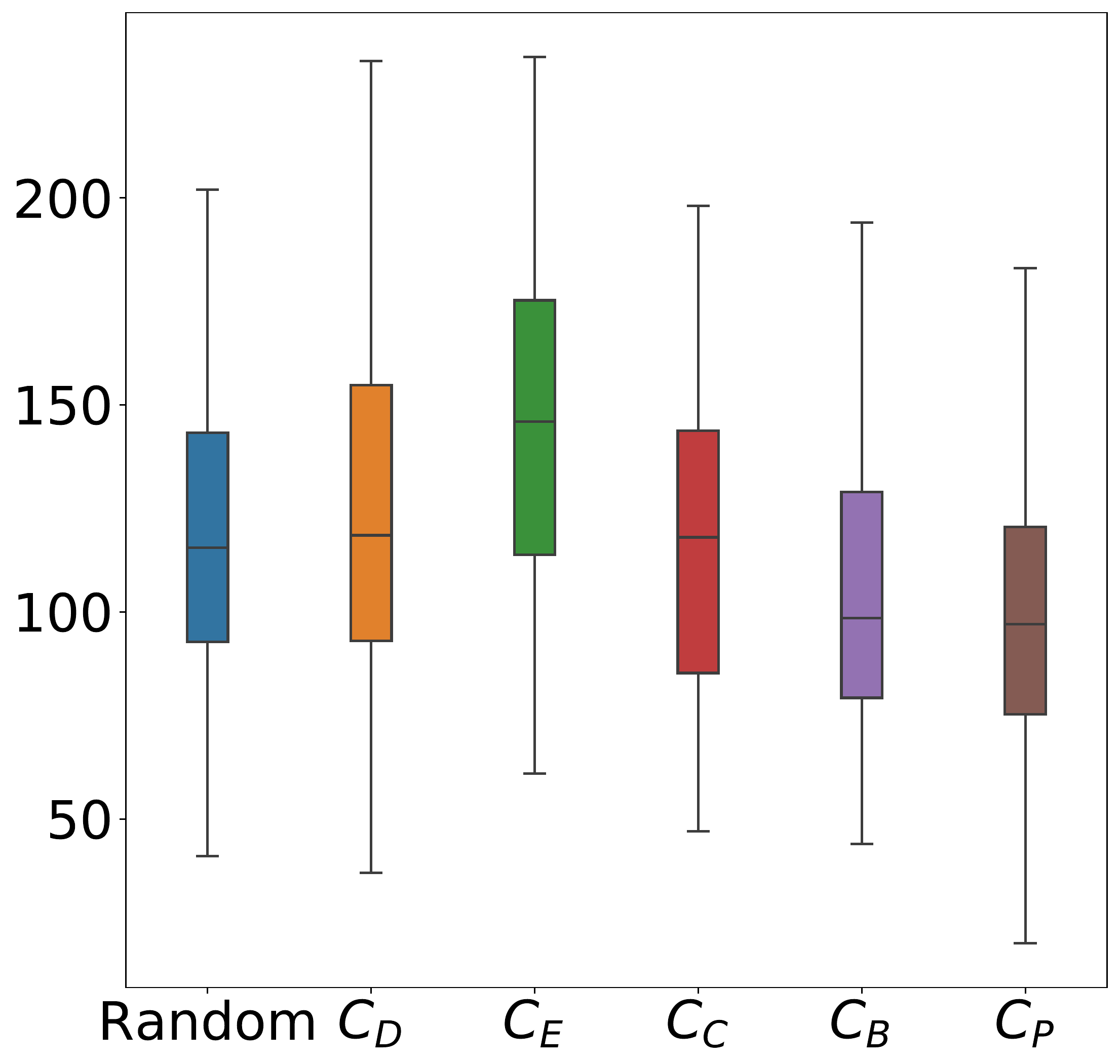}  
  \caption{ }
  \label{fig33_34_35_36:sub-second}
\end{subfigure}


\begin{subfigure}{.35\textwidth}
  \centering
  \includegraphics[width=.9\linewidth]{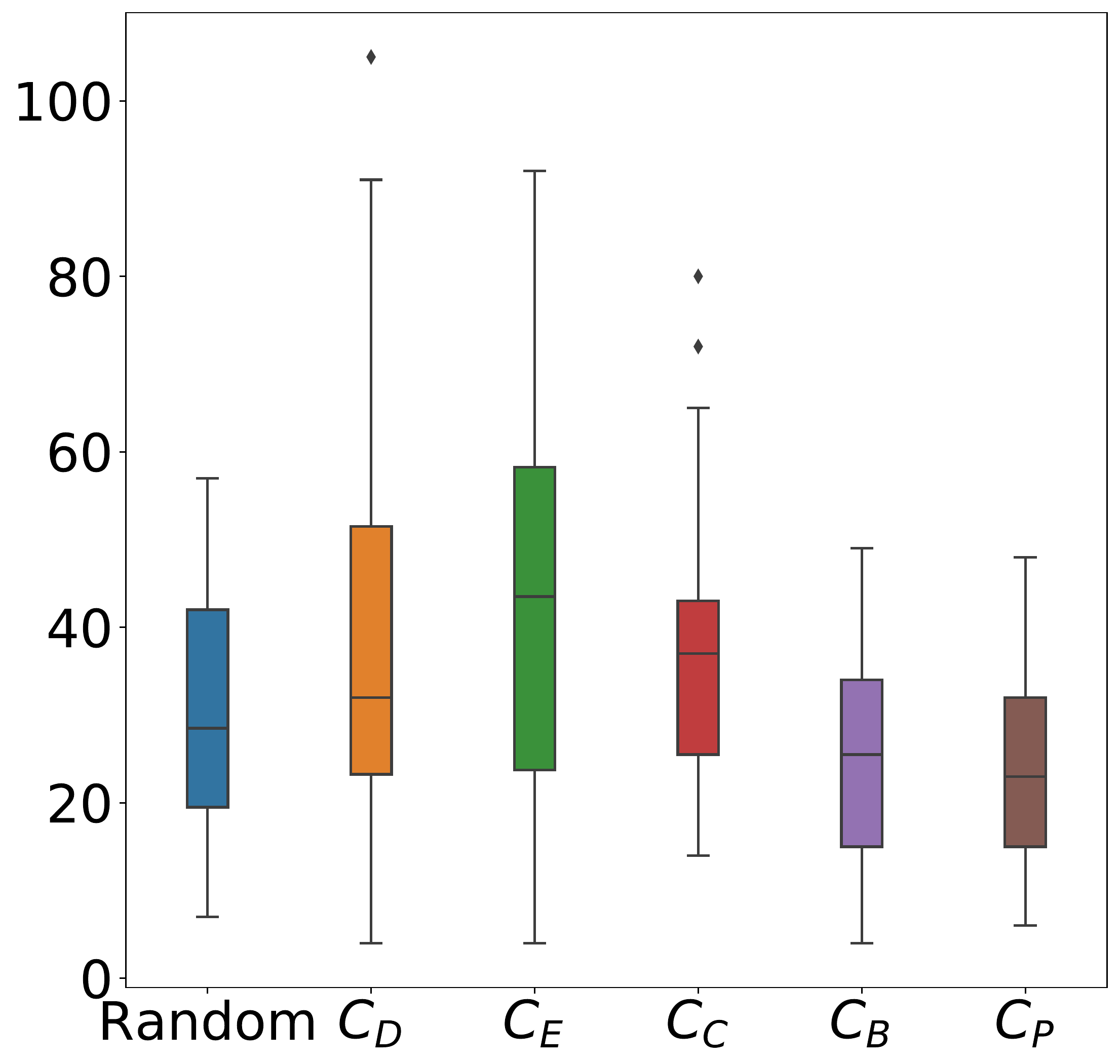}  
  \caption{ }
  \label{fig33_34_35_36:sub-third}
\end{subfigure}
\begin{subfigure}{.35\textwidth}
  \centering
  \includegraphics[width=.9\linewidth]{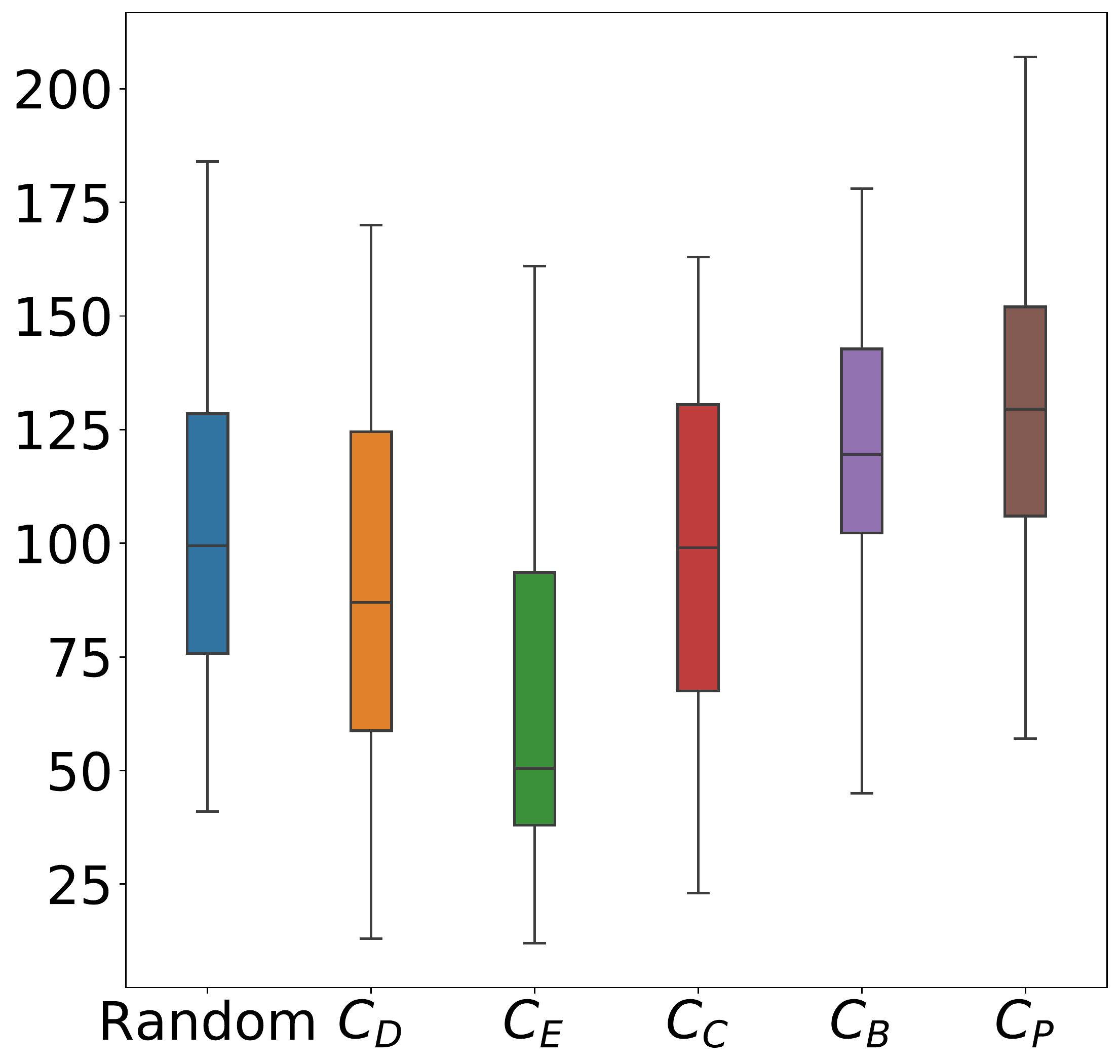}  
  \caption{ }
  \label{fig33_34_35_36:sub-fourth}
\end{subfigure}

\caption{Intervention analysis on 50 multi-communities networks with varying community size by box plots of (a) sum of $P_{IT}$; (b) number of infected nodes; (c) number of susceptible nodes; (d) number of protected nodes.}
\label{fig:fig33_34_35_36}
\end{figure*}

\begin{figure*}[!htb]
\begin{subfigure}{.35\textwidth}
  \centering
  \includegraphics[width=.9\linewidth]{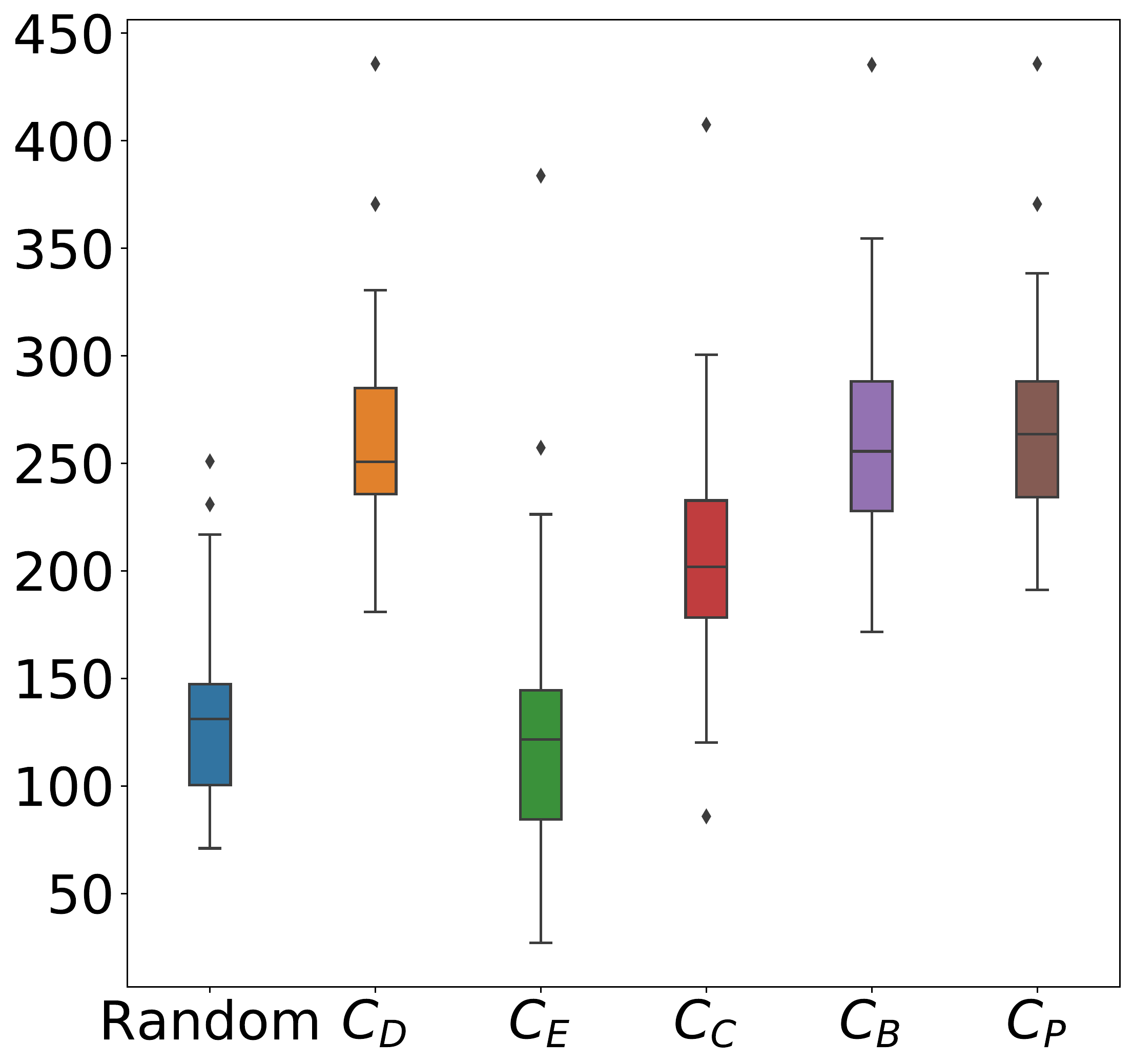}  
  \caption{ }
  \label{LFRtf:sub-first}
\end{subfigure}
\begin{subfigure}{.35\textwidth}
  \centering
  \includegraphics[width=.9\linewidth]{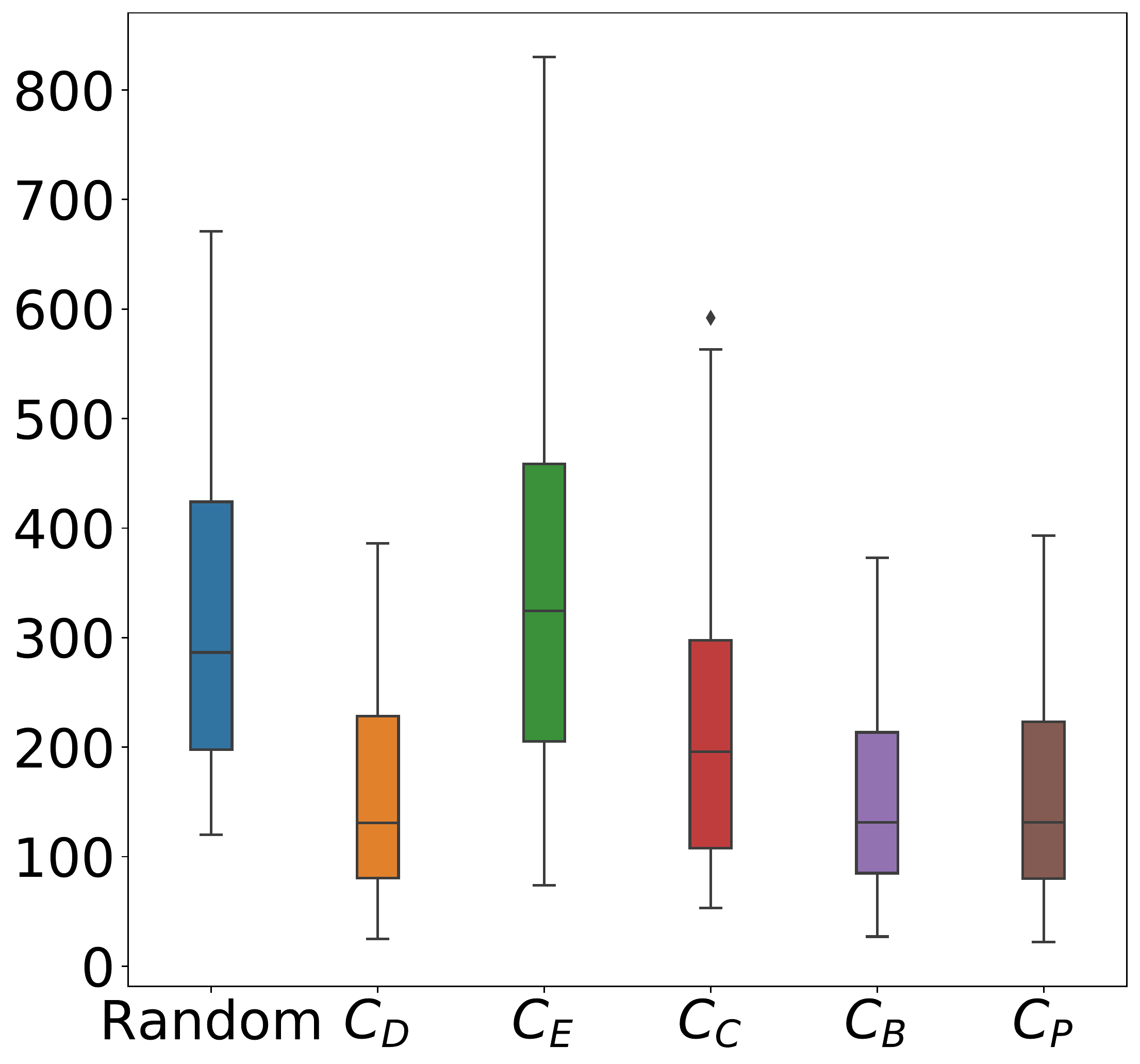} 
  \caption{ }
  \label{LFRtf:sub-second}
\end{subfigure}


\begin{subfigure}{.35\textwidth}
  \centering
  \includegraphics[width=.9\linewidth]{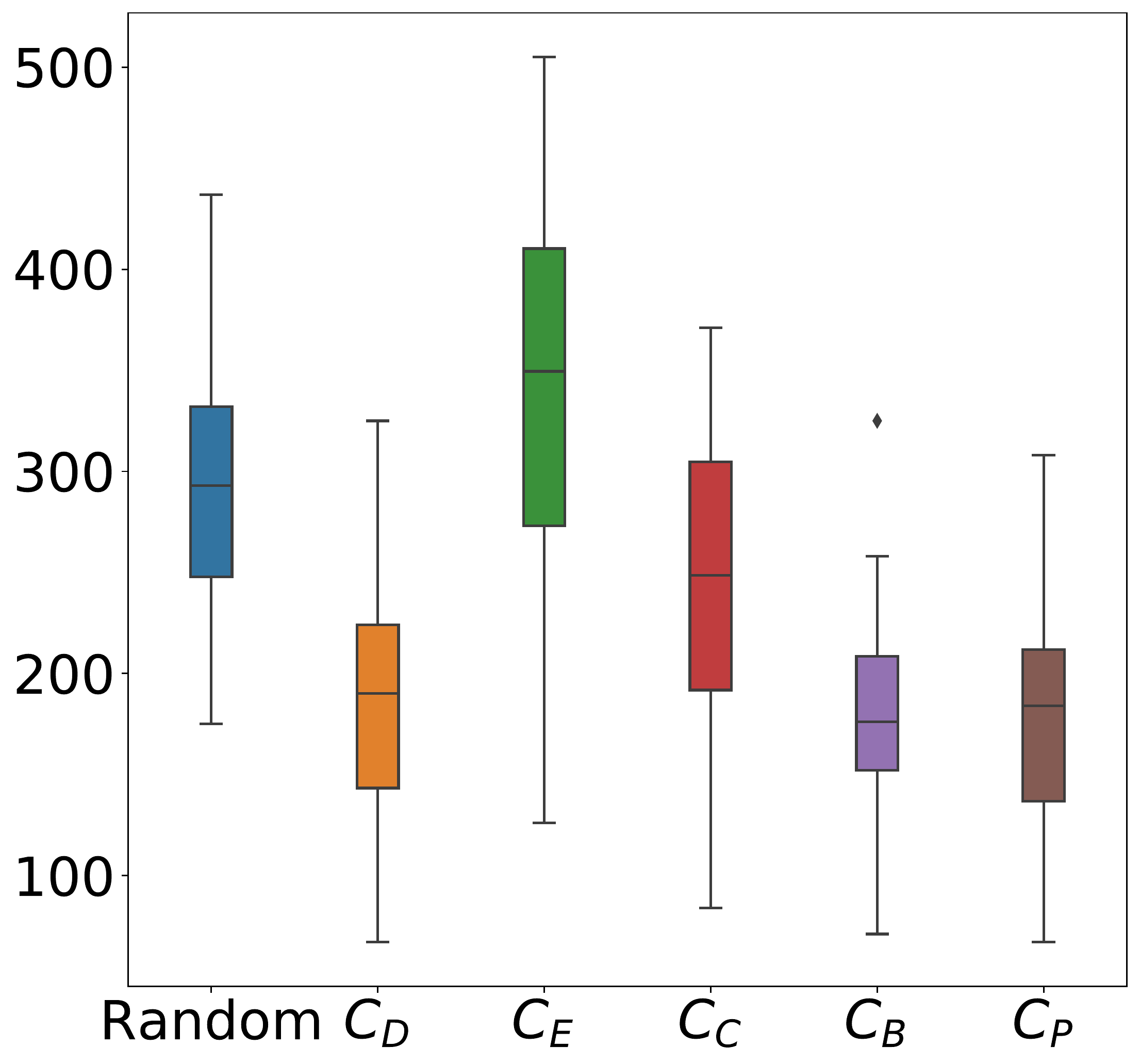}
  \caption{ }
  \label{LFRtf:sub-third}
\end{subfigure}
\begin{subfigure}{.35\textwidth}
  \centering
  \includegraphics[width=.9\linewidth]{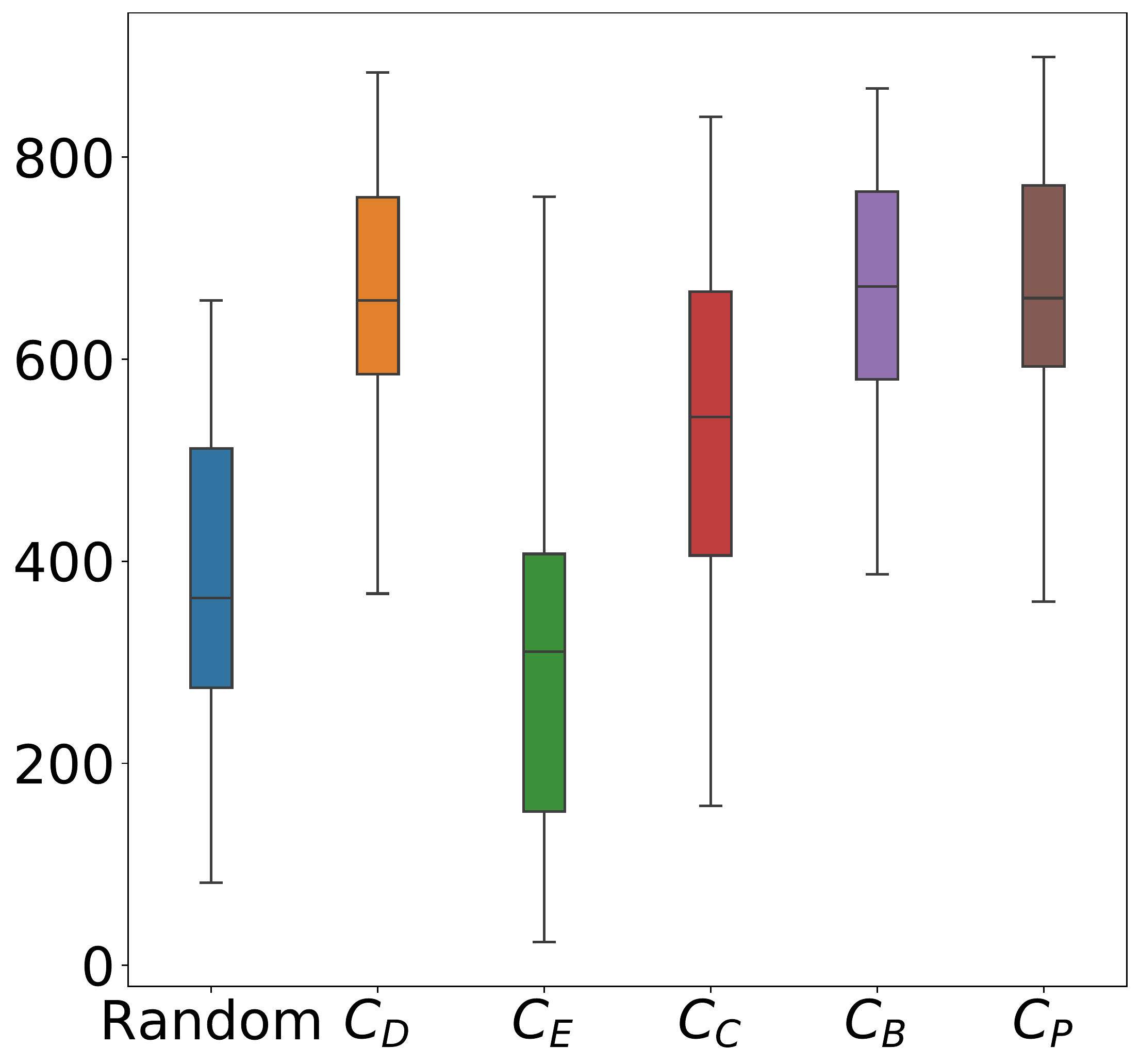}
  \caption{ }
  \label{LFRtf:sub-fourth}
\end{subfigure}

\caption{Information intervention analysis on 50 LFR benchmark graph by box plots of (a) sum of $P_{IT}$; (b) number of infected nodes; (c) number of susceptible nodes; (d) number of protected nodes.}
\label{fig:LFRtf}
\end{figure*}

\begin{table}[!htb]
  \caption{$p$ values of information intervention on 50 LFR benchmark graph}
  \label{table:LFRtf}
  \begin{tabular}{*{5}{p{.165\linewidth}}}
    \toprule
     Centrality & sum of $P_{IT}$ & "infected" nodes & "susceptible" nodes & "protected" nodes \\
    \midrule
        Degree & $3.778465*10^{-10}$  & $3.774195*10^{-10}$	 & $5.934975*10^{-10}$ & $3.776756*10^{-10}$\\
        Eigenvector	& $9.924046*10^{-1}$  & $9.934516*10^{-1}$ & $9.999421*10^{-1}$ & $9.999918*10^{-1}$\\
        Closeness & $3.778465*10^{-10}$  & $3.769930*10^{-10}$   & $1.469963*10^{-5}$ & $4.013735*10^{-10}$\\
        Betweenness	& $3.778465*10^{-10}$  & $3.774195*10^{-10}$   & $4.525909*10^{-10}$ & $3.775049*10^{-10}$\\
        Page rank  & $3.778465*10^{-10}$ & $3.775902*10^{-10}$ & $6.471115*10^{-10}$ & $3.775902*10^{-10}$\\
        
  \bottomrule
\end{tabular}
\end{table}

\end{document}